\documentstyle[preprint,aps]{revtex}
\begin{document}
\voffset -5mm
\draft
\preprint{
\vbox{
\halign{&##\hfil\cr
        &July 1999,  hep-ph/9908295\cr}}
}
\title{The minimum supersymmetric standard model without
R-parity through various lepton-number violations}
\author{Chao-hsi Chang$^{1,2}$, 
Tai-fu Feng$^{2}$}
\address{$^1$CCAST (World Laboratory), P.O.Box 8730,
Beijing 100080, China\thanks{Not post mail address}}
\address{$^2$Institute of Theoretical Physics, 
Chinese Academy of Sciences, P.O.Box 2735,
Beijing 100080, China}
\maketitle
\begin{center}
{\large\bf Abstract}
\end{center}
{\small  
The minimum supersymmetric standard model (MSSM) 
without R-parity through various 
lepton-number violations is investigated systematically.
All kinds of possible mixing in the model
are formulated precisely. The remarkable issue that 
the lightest `Higgs' $H_1^0$ may be heavier than the 
weak-boson $Z$ at tree level is kept, as the special R-parity
violation MSSM where only bilinear violation of the lepton
numbers is allowed. It is also shown explicitly that there 
is a freedom $U(n+1)$ ($n$ is the number of the broken 
lepton-numbers) in re-defining the lepton and Higgs superfields. 
Feynman rules relevant to the R-parity violations are given 
precisely in $'$t Hooft-Feynman gauge. With further 
assumptions and concerning 
all available experimental constraints, spectrum for 
interest sectors is computed numerically.}
\vspace{4mm}

\section{Introduction}
It is being increasingly realized by those engaged in
search for supersymmetry (SUSY) that the principle of R-parity
conservation\cite{s1,s13}, assumed to be sacrosanct in 
the prevalent search strategies, is not in practice 
inviolable\cite{s2}. The R-parity of a particle is defined by 
$R=(-1)^{2S+3B+L}$ and can be violated either by baryon-number 
(B) breaking or by lepton-number (L) breaking\cite{s2}. 
Proton decay experiments have set stringent 
restrictions on the violations of the first and 
the second generations of the baryon-numbers, but 
the existent experiment data do not impose 
so stringent restriction neither on lepton-number violations
nor on the third generations of baryon-numbers. In particular, 
for some cosmology models to explain the baryongenesis, 
it requests lepton-numbers not to be conserved, that 
the intensive studies of the supersymmetry models without R-parity,
through lepton-number violations and/or the violation 
for the third generation of baryon-number, have attracted quite a lot 
of attentions recently[3-22]. As far as the 
literature is concerned, besides those the so-called 
basis-independent studies of the R-parity violations\cite{s7}, 
the models with lepton-numbers being broken are characterized 
by certain `special' Lagrangian which has 
`bilinear'\cite{s21,s11,chf,s5,s55} and/or 
trilinear\cite{s5,s55,s3,s4,s6,s10} R-parity 
violations explicitly in
superpotential and/or the SUSY soft-breaking terms, 
and by the violations spontaneously generated 
by nonzero vacuum expectation values (VEVs) of sneutrinos[3-9]. 
The supersymmetry models without R-parity can also be arranged 
well that there will be no contradiction with all the existent 
experimental data[9-22]. Respect to the very general case, 
the MSSM without R-parity through various possible 
lepton-number violations simultaneously has not 
been investigated thoroughly.

In general, the minimal supersymmetric standard model (MSSM) 
(R-parity is conserved) has the following 
general form for the superpotential in terms of superfields:
\begin{eqnarray}
{\cal W}_{MSSM} &=& \mu\varepsilon_{ij} \hat{H}_{i}^{1} \hat{H}_{j}^{2} 
+ l_{I}\varepsilon_{ij} \hat{H}_{i}^{1} \hat{L}_{j}^{I} \hat{R}^{I} -
u_{I}(\hat{H}_{1}^{2} C^{JI*}\hat{Q}_{2}^{J} - \hat{H}_{2}^{2} 
\hat{Q}_{1}^{I}
)\hat{U}^{I}  \nonumber \\
 & & - d_{I}(\hat{H}_{1}^{1}\hat{Q}_{2}^{I} - \hat{H}_{2}^{1} C^{IJ}
\hat{Q}_{1}^{J})\hat{D}^{I}.
\label{eq-1}
\end{eqnarray}
Here $\hat{H}^{1}$, $\hat{H}^{2}$ are Higgs superfields; 
$\hat{Q}^{I}$ and $\hat{L}^{I}$ being quark and lepton superfields
(I=1, 2, 3 is the index of generation), all are  
in doublet of the weak SU(2) respectively. The rest superfields:
$\hat{U}^{I}$ and $\hat{D}^{I}$ being quark superfields 
and $\hat{R}^{I}$ charged lepton ones, but in singlet 
of the weak SU(2). Here the indices i, j are contracted 
in a general way for the SU(2) group, and $C^{IJ}$ $(I, J=1, 2, 3)$ 
are the elements of the CKM matrix. 

When R-breaking interactions are incorporated, 
the superpotential will be modified as the 
follows:
\begin{equation}
{\cal W} = {\cal W}_{MSSM} + {\cal W}_{L} + {\cal W}_{B}
\label{eq-2}
\end{equation}
with 
\begin{eqnarray}
{\cal W}_{L} &=& \varepsilon_{ij} [ \lambda_{IJK} \hat{L}_{i}^{I} 
\hat{L}_{j}^{J}
\hat{R}^{K} + \lambda_{IJK}^{\prime}\hat{L}_{i}^{I} \hat{Q}_{j}^{J} \hat{D}
^{K} + \epsilon_{I} \hat{H}_{i}^{2} \hat{L}_{j}^{I} ]  \nonumber \\
{\cal W}_{B} &=& \lambda_{IJK}^{\prime\prime} 
\hat{U}^{I} \hat{D}^{J} \hat{D}^{K}.
\label{eq-3}
\end{eqnarray}

Considering the stringent constraint by proton decay experiments 
on the violations for the first and the second generation 
baryon-numbers, many authors would like to focus on the third 
generation baryon-number\cite{s5,s55,s24,s22}, but the other 
authors would like to examine the effects of the broken 
lepton-numbers. Here we will suppress ${\cal W}_{B}$ for all 
generations totally. The first two terms in ${\cal W}_{L}$ in 
Eq.\ (\ref{eq-3}) have received a lot of
consideration, and many restrictions on them have been
derived from existing experimental
data\cite{s7,s3,s4,s6,s25,se1}. 
However, the term $\epsilon_{I}\varepsilon_{ij}
\hat{H}_{i}^{2}\hat{L}_{j}^{J}$ is also a viable agent for 
R-parity breaking. It is particularly interesting because it 
with proper SUSY soft-breaking terms
can result in observable factors that cannot
be effected by the trilinear terms alone. One of these 
distinctive effects which we would like to mention
here is that, at tree level the
lightest neutralino can decay invisibly into three neutrinos, 
which is not possible if only the trilinear terms in 
${\cal W}_{L}$ are presented. In addition, such as
non-zero vacuum expectation values (VEVs) 
of some sneutrinos or/and bilinear violation terms 
will cause `fresh' mixing and different phenomenology etc,
the interesting results are obtained\cite{s21,s11,chf,s5,s55}.
Whereas what happens to the most 
general case where all possible lepton-number violations 
(not only from superpotential but also from the terms 
causing SUSY soft breaking) are simultaneously involved, 
is still an interesting problem to be investigated, thus 
we would like to turn to the problem in the paper. 
Indeed in this kind model there is a freedom in principle for
defining the superfields, thus remarkable problems, how big
of the freedom and how to recognize two different 
`parameterization' manners of the same model among
this kind of models, emerge. Therefore we will start 
with the general SUSY version 
and keep all the possible R-parity violation terms in the 
superpotential ${\cal W}_{L}$ and in the SUSY soft-breaking
Lagrangian properly, then to work out the Lagrangian in
`component version'\footnote{In fact, we may consider 
the resultant effective theory as a SUSY superfield version, 
renormalized at the energy-scale of SUSY breaking.
Thus the full renormalization at weak-interaction energy-scale
on the parameters in the effective Lagrangian should be made 
only `further' in component version.}. 
All possible mixing and Feynman rules for 
further precise studies of the phenomenology of 
the R-parity violation effective theory will be given 
precisely. The straightforward deductions for these
pursers are omitted and lengthy formulas are put into 
Appendices. Moreover we put the problem aside that the effective 
theory may have a more fundamental origin at a comparatively 
high energy scale, although it is interesting and
the effective theory may be helpful to trace out some
clue on the problem\footnote{Some parameters maybe vanish 
due to higher symmetries if one relates the theory to 
a specific GUT model\cite{s9}, whereas the investigation here 
is still applicable as long as to set the
corresponding parameters to vanish.}. As for the freedom
for defining the superfields, we would also take one section
to demonstrate how big it is precisely and make some suggestions
on it for later conveniences in later applications.

The paper is organized as follows: in Sect.II, we describe the 
basic ingredient of the SUSY without R-parity through various
lepton-number breaking. The mass matrices of the CP-even Higgs, 
CP-odd Higgs and charged Higgs are derived precisely. As an 
important result, relations for CP-even and CP-odd Higgs masses 
as those in the bilinear case\cite{chf}, are recovered.
For completeness, we also give the mixing matrix of charginos and
charged leptons, and that of neutralinos and neutrinos.
In Sect.III, the Feynman rules for the interactions 
relevant to R-parity violation, i.e., those of 
the Higgs bosons (sleptons) with the gauge bosons, and the 
charginos, neutralinos with gauge bosons or Higgs bosons 
(sleptons) are presented. The self interactions of the Higgs 
and the interactions of chargino (neutralino)-squark-quark 
are also given. In Sect.IV, we examine the 
freedom\cite{s7,s25,se1} for re-defining the Higgs 
superfield and the $n-$lepton superfields which are 
relevant to the R-parity violation. We precisely show the
equivalence for two superficially different parameterizations 
generated by two sets of the $n+1$ superfields, if the 
two sets of the $n+1$ superfields may be connected by a $U(n+1)$
transformation exactly, hence the $U(n+1)$ transformation can be 
understood as a freedom for redefining the Higgs and lepton 
superfields at very beginning. In Sect.V, we try to consider
the comparatively interesting `particle spectrum' 
numerically under a few further reasonable assumptions on
the parameter space of the model partly for simplifying the
practical calculations.

\section{The physical masses in the MSSM without R-parity}

Generally the lepton-number violations in a MSSM not only cause
R-parity broken but also make quite a lot of fresh 
and interesting mixings between particles and sparticles.
Let us examine the subject for the model with various
lepton-number violations in this section.
Since those parts, such as gauge, matter and the 
gauge-matter interactions etc, in the model are the same 
as the MSSM, thus we will omit them in the paper everywhere
except special needs.

As stated above, we are to consider the superpotential,
(to combine Eq.(1) and Eq.(2)):
\begin{eqnarray}
{\cal W} &=& \mu \varepsilon_{ij} \hat{H}_{i}^{1} \hat{H}_{j}^{2} 
+ l_{I}\varepsilon_{ij}\hat{H}_{i}^{1}\hat{L}_{j}^{I}\hat{R}^{I} - 
u^{I}(\hat{H}_{1}^{2} C^{JI*}\hat{Q}_{2}^{J} - 
\hat{H}_{2}^{2}\hat{Q}_{1}^{I})\hat{U}^{I}  \nonumber \\
 & & - d^{I}(\hat{H}_{1}^{1}\hat{Q}_{2}^{I} - \hat{H}_{2}^{1} C^{IJ} 
\hat{Q}_{1}^{J})\hat{D}^{I} +
\varepsilon_{ij} [ \lambda_{IJK} \hat{L}_{i}^{I} \hat{L}_{j}^{J}
\hat{R}^{K}  + \epsilon_{I} \hat{H}_{i}^{2} \hat{L}_{j}^{I}] \nonumber \\
&& + \lambda_{IJK}^{\prime}(\hat{L}_{1}^{I}\hat{Q}_{2}^{J}\delta_{JK} - 
\hat{L}_{2}^{I} C^{JK} 
\hat{Q}_{1}^{K})\hat{D}^{J}
\label{eq-4}
\end{eqnarray}
with $\mu$, $\epsilon_{I}$ are the parameters with units of mass, 
$u^{I}$, $d^{I}$ and $l^{I}$ are the Yukawa couplings as in the 
MSSM with R-parity, and the parameters $\lambda_{IJK}$,
$\lambda_{IJK}^{\prime}$ describe the trilinear R-parity violation.
Since now we consider the case with three families thus the
subscripts $I,J,K=1,2,3$.
To break the SUSY so as to have a correct phenomenology,
the general soft SUSY-breaking terms are introduced 
accordingly:
\begin{eqnarray}
{\cal L}_{soft} & = & -m_{H^{1}}^{2}H_{i}^{1*}H_{i}^{1} - 
m_{H^{2}}^{2} H_{i}^{2*}
H_{i}^{2}-m_{L^{I}}^{2} \tilde{L}_{i}^{I*} \tilde{L}_{i}^{I} - 
m_{R^{I}}^{2}\tilde{R}^{I*} 
\tilde{R}^{I}  \nonumber  \\
&& -\sum\limits_{I}m_{HL^{I}}^{2}(H_{i}^{1*}\tilde{L}_{i}^{I}
+ H_{i}^{1}\tilde{L}_{i}^{I*}) - \sum\limits_{I\neq J}
m_{L^{IJ}}^{2}\tilde{L}_{i}^{I*}\tilde{L}_{i}^{J} - \sum\limits_{I\neq J}
m_{R^{IJ}}^{2}\tilde{R}^{I*}\tilde{R}^{J} \nonumber  \\
 &  & -m_{Q^{I}}^{2} \tilde{Q}_{i}^{I*} \tilde{Q}_{i}^{I} - m_{D^{I}}^{2}
\tilde{D}
 ^{I*} \tilde{D}^{I} - m_{U^{I}}^{2}\tilde{U}^{I*} \tilde{U}^{I} + (m_{1}
\lambda_{B}
\lambda_{B}  \nonumber  \\
 &  & + m_{2}\lambda_{A}^{i}\lambda_{A}^{i} + 
 m_{3} \lambda_{G}^{a}\lambda_{G}^{a} + h.c.) + 
 \{ B \varepsilon_{ij}H_{i}^{1}H_{j}^{2} + 
 B_{I}\varepsilon_{ij}H_{i}^{2}\tilde{L}_{j}^{I}  \nonumber \\
 & & + \varepsilon_{ij} l_{sI} H_{i}^{1}\tilde{L}_{j}^{I}\tilde{R}^{I} +
  d_{sI} (-H_{1}^{1}\tilde{Q}_{2}^{I} + C^{IK}H_{2}^{1}\tilde{Q}_{1}^{K})
 \tilde{D}^{I}  \nonumber  \\
 & &+ u_{sI} (-C^{KI*}H_{1}^{2}\tilde{Q}_{2}^{I} + 
 H_{2}^{2}\tilde{Q}_{1}^{I})\tilde{U}^{I} + 
 \varepsilon_{ij}\lambda_{IJK}^{S}\tilde{L}_{i}^{I}\tilde{L}_{j}^{J}
\tilde{R}^{K}  \nonumber \\
& & + \lambda_{IJK}^{S^{\prime}}(\tilde{L}_{1}^{I}\tilde{Q}_{2}^{J}
\delta^{JK} - \tilde{L}_{2}^{I}C^{JK}\tilde{Q}_{1}^{J})\tilde{D}^{K} + h.c.\}
 \label{eq-5}
 \end{eqnarray}
where $m_{H^{1}}^{2}$, $m_{H^{2}}^{2}$, $m_{L^{I}}^{2}$, $m_{R^{I}}^{2}$, 
$m_{Q^{I}}^{2}$, 
$m_{D^{I}}^{2}$, $m_{U^{I}}^{2}$, $B$ and $B_{I}$ are the 'bare" mass 
parameters while $m_{3}$, $m_{2}$, $m_{1}$ denote the masses of 
$\lambda_{G}^{a}$, $\lambda_{A}^{i}$ and $\lambda_{B}$, 
the $SU(3)\times SU(2) \times U(1)$ gauginos.  
$d_{sI}$, $u_{sI}$, $l_{sI}$ $(I=1,2,3)$ and $\lambda_{IJK}^{S}$, 
$\lambda_{IJK}^{S^{\prime}}$
are the soft breaking 
parameters that make necessary mass splitting between the quarks, 
leptons and their supersymmetric partners. To correspond to the
superpotential in Eq.\ (\ref{eq-4}), all the possible lepton number 
violation terms for breaking SUSY softly are involved in
Eq.\ (\ref{eq-5}).

In general, the scalar potential of the model can be written as
\begin{eqnarray}
V &=& \sum_{i} |\frac{\partial{\cal W}}{\partial A_{i}}|^{2} + V_{D} + 
V_{soft}  \nonumber \\
 &=& V_{F} + V_{D} + V_{soft}
\label{eq-6}
\end{eqnarray}
where $A_{i} (i=\cdots)$ denote scalar components, $V_{D}$ 
the usual D-terms, $V_{soft}$ just the SUSY soft breaking terms given in 
Eq.\ (\ref{eq-5}). Using the superpotential Eq.\ (\ref{eq-4}) and the soft
breaking terms Eq.\ (\ref{eq-5}), we can write down the scalar 
potential precisely with the following forms:
\begin{eqnarray}
V_{F} &=& |\frac{\partial{\cal W}}{\partial H^{1}}|^{2} + |\frac{\partial
{\cal W}}{\partial H^{2}}|^{2} + |\frac{\partial{\cal W}}{\partial 
\tilde{L}^{I}
}|^{2} + |\frac{\partial{\cal W}}{\partial \tilde{R}^{I}}|^{2} +
|\frac{\partial{\cal W}}{\partial \tilde{Q}^{I}}|^{2} + 
|\frac{\partial{\cal W}}{\partial \tilde{U}^{I}}|^{2} +
|\frac{\partial{\cal W}}{\partial \tilde{D}^{I}}|^{2} \; . \nonumber \\
\end{eqnarray}

As MSSM but more general, the electroweak symmetry in the model
is broken spontaneously if the two Higgs 
doublets $H^{1}$, $H^{2}$, and the sleptons as well
acquire nonzero vacuum expectation values (VEVs):
\begin{equation}
H^{1}=
\left( 
\begin{array}{c}
\frac{1}{\sqrt{2}}(\chi_{1}^{0} + \upsilon_{1} + i\phi_{1}^{0}) \\
H_{2}^{1}  \end{array}  \right)
\label{H1-vacuum}
\end{equation}
\begin{equation}
H^{2}=
\left(
\begin{array}{c}
H_{1}^{2}  \\
\frac{1}{\sqrt{2}}(\chi_{2}^{0} + \upsilon_{2} + i\phi_{2}^{0})  
\end{array}  \right)
\label{H2-vacuum}
\end{equation}
and
\begin{equation}
\tilde{L}^{I} = 
\left(
\begin{array}{c}
\frac{1}{\sqrt{2}}(\chi_{\tilde{\nu}_{I}}^{0} + 
\upsilon_{\tilde{\nu}_{I}} + i\phi_{\tilde{\nu}_{I}}^{0})  \\
\tilde{L}_{2}^{I}    \end{array}  \right)
\label{L3-vacuum}
\end{equation}
where $\tilde{L}^{I}$ denote the slepton SU(2) 
doublets, and $I=e$, $\mu$, $\tau$, i.e. the indices of the
three families for leptons. From 
Eqs.\ (\ref{eq-6},\ref{H1-vacuum},\ref{H2-vacuum},\ref{L3-vacuum})
it is easy to find the scalar potential includes the linear terms as
follows:
\begin{equation}
V_{tadpole} = t_{1}^{0}\chi_{1}^{0} + t_{2}^{0} \chi_{2}^{0} 
+ t_{\tilde{\nu}_{e}}^{0} \chi_{\tilde{\nu}_{e}}^{0} + 
t_{\tilde{\nu}_{\mu}}^{0} \chi_{\tilde{\nu}_{\mu}}^{0} + 
t_{\tilde{\nu}_{\tau}}^{0} \chi_{\tilde{\nu}_{\tau}}^{0}
\label{tadpole0}
\end{equation}
where
\begin{eqnarray}
t_{1}^{0} &=& \frac{1}{8}(g^{2} + 
g^{\prime^{2}})\upsilon_{1}(\upsilon_{1}^{2} - \upsilon_{2}^{2} + 
\sum_{I}\upsilon_{\tilde{\nu}_{I}}^{2})
+ |\mu|^{2}\upsilon_{1} + m_{H^{1}}^{2}\upsilon_{1} \nonumber \\ 
&& +\sum\limits_{I}m_{HL^{I}}^{2}\upsilon_{\tilde{\nu}_{\tau}}- 
B \upsilon_{2} -
\sum_{I} \mu \epsilon_{I}\upsilon_{\tilde{\nu}_{I}},  \nonumber \\
t_{2}^{0} &=& -\frac{1}{8}(g^{2} + 
g^{\prime^{2}})\upsilon_{2}(\upsilon_{1}^{2} - 
\upsilon_{2}^{2} + \sum_{I}\upsilon_{\tilde{\nu}_{I}}^{2})
+ |\mu|^{2}\upsilon_{2} + m_{H^{2}}^{2}\upsilon_{2} - 
B \upsilon_{1} + \sum_{I}\epsilon_{I}^{2}\upsilon_{2} \nonumber \\
 & & +\sum_{I} B_{I}\upsilon_{\tilde{\nu}_{I}},
\nonumber \\
t_{\tilde{\nu}_{I}}^{0} &=& \frac{1}{8}(g^{2} + 
g^{\prime^{2}})\upsilon_{\tilde{\nu}_{I}}(\upsilon_{1}^{2} - 
\upsilon_{2}^{2} + 
\sum_{I}\upsilon_{\tilde{\nu}_{I}}^{2})
+m_{L^{I}}^{2}\upsilon_{\tilde{\nu}_{I}} + 
\epsilon_{I}\sum_{J}\epsilon_{J}\upsilon_{\tilde{\nu}_{J}}  \nonumber \\
 & & - \mu\epsilon_{I}
\upsilon_{1} + B_{I}\upsilon_{2} + m_{HL^{I}}^{2}\upsilon_{1}
+\sum\limits_{J\neq I}m_{L^{IJ}}^{2}\upsilon_{\tilde{\nu}_{J}}.
\label{tadpole1}
\end{eqnarray}
Here $t_{i}^{0}$ ($i=1$, $2$, $\tilde{\nu}_{e}$, $\tilde{\nu}_{\mu}$, 
$\tilde{\nu}_{\tau}$) 
are tadpoles at `tree level', thus the true VEVs
of the neutral scalar fields should satisfy the condition $t_{i}^{0}=0$, 
($i=1$, $2$, $\tilde{\nu}_{e}$, $\tilde{\nu}_{\mu}$, 
$\tilde{\nu}_{\tau}$), therefore we obtain:
\begin{eqnarray}
m_{H^{1}}^{2} &=& -\Bigg(|\mu|^{2} - \sum_{I}\bigg(\epsilon_{I}
\mu - m_{HL^{I}}^{2}\bigg)\frac{\upsilon_{\tilde{\nu}_{I}}}{\upsilon_{1}} - 
B\frac{\upsilon_{2}}{\upsilon_{1}}
 + \frac{1}{8}\bigg(g^{2} + g^{\prime^{2}}\bigg)
\bigg(\upsilon_{1}^{2} - \upsilon_{2}^{2} 
 + \sum_{I}\upsilon_{\tilde{\nu}_{I}}^{2}\bigg)\Bigg),  \nonumber \\
m_{H^{2}}^{2} &=& -\Bigg(|\mu|^{2} + \sum_{I}\epsilon_{I}^{2} + 
\sum_{I}B_{I}\frac{\upsilon_{\tilde{\nu}_{I}}}{\upsilon_{2}} - 
B\frac{\upsilon_{1}}{\upsilon_{2}} - 
\frac{1}{8}\bigg(g^{2} + g^{\prime^{2}}\bigg)\bigg(\upsilon_{1}^{2} 
- \upsilon_{2}^{2} + 
\sum_{I}\upsilon_{\tilde{\nu}_{I}}^{2}\bigg)\Bigg),   \nonumber \\
m_{L^{I}}^{2} &=& -\Bigg(\frac{1}{8}\bigg(g^{2} + g^{\prime^{2}}\bigg)
\bigg(\upsilon_{1}^{2} 
- \upsilon_{2}^{2} + \sum_{I}\upsilon_{\tilde{\nu}_{I}}^{2}\bigg)
+ \epsilon_{I}\sum_{J}\epsilon_{J}\frac{\upsilon_{\tilde{\nu}_{J}}}
{\upsilon_{\tilde{\nu}_{I}}} - 
\epsilon_{I}\mu\frac{\upsilon_{1}}{\upsilon_{\tilde{\nu}_{I}}}   \nonumber \\
 & & +B_{I}\frac{\upsilon_{2}}{
\upsilon_{\tilde{\nu}_{I}}} + m_{HL^{I}}^{2}\frac{\upsilon_{1}}
{\upsilon_{\tilde{\nu}_{I}}} + \sum\limits_{J\neq I}m_{L^{IJ}}^{2}
\frac{\upsilon_{\tilde{\nu}_{J}}}{\upsilon_{\tilde{\nu}_{I}}}
\Bigg), \hspace{5mm} (I=e, \mu, \tau). \nonumber \\
\label{masspara}
\end{eqnarray}
For convenience, later on we will call all of these scalar bosons ($H^{1}$, 
$H^{2}$ and $\tilde{L}^{I}$) as `Higgs'. 
As for the scalar sector, the Higgs-boson mass-matrices
squared may be obtained by:
\begin{equation}
{\cal M}_{ij}^{2} = \frac{\partial^{2} V}{\partial \phi_{i} 
\phi_{j}}|_{minimum},
\label{getsmatrix}
\end{equation}
here `minimum' means to evaluate the values at $<H_{1}^{1}> = 
\frac{\upsilon_{1}}{\sqrt{2}}$, $<H_{2}^{2}> = 
\frac{\upsilon_{2}}{\sqrt{2}}$, 
$<\tilde{L}_{1}^{I}> = \frac{\upsilon_{\tilde{\nu}_{I}}}{\sqrt{2}}$ and 
$<A_{i}>=0$ ($A_{i}$ represent all the other scalar fields).
Note that the matrices of the CP-even and the CP-odd scalar 
bosons both are $5\times 5$ as we have the sneutrinos
which correspond to three left-handed neutrinos;, 
whereas the matrix of the charged Higgs is $8\times 8$
as we have the charged sleptons which correspond to
three left-handed and three right-handed charged leptons.
 
Now let us summarize the results and try to 
make the matrices diagonal in the following subsections.

\subsection{The mass matrices for Higgs}

The mass terms of the
CP-even Higgs from the scalar potential Eq.\ (\ref{eq-6}):
\begin{equation}
{\cal L}_{m}^{even} = - \Phi_{even}^{\dag} {\cal M}_{even}^{2} \Phi_{even}
\label{l-add1}
\end{equation}
are obtained with the interaction CP-even Higgs fields 
$\Phi_{even}^{T} = (\chi_{1}^{0}$, $\chi_{2}^{0}$,
$\chi_{\tilde{\nu}_{e}}^{0}$, $\chi_{\tilde{\nu}_{\mu}}^{0}$, 
$\chi_{\tilde{\nu}_{\tau}}^{0})$ 
as basis, and the corresponding mass matrix is
\begin{equation}
{\cal M}_{even}^{2} = \left(
\begin{array}{ccccc}
r_{11} & -e_{12} - B  &  e_{13} - \mu\epsilon_{1}  & 
e_{14} - \mu\epsilon_{2} & e_{15} - \mu\epsilon_{3} \\
-e_{12} - B & r_{22} & -e_{23} + B_{1} & -e_{24} + 
B_{2} & -e_{25} + B_{3} \\
e_{13} - \mu\epsilon_{1} & -e_{23} + B_{1} & r_{33} & 
e_{34} + \epsilon_{1}\epsilon_{2} & e_{35} + \epsilon_{1}\epsilon_{3} \\
e_{14} - \mu\epsilon_{2} & -e_{24} + B_{2} & e_{34} + 
\epsilon_{1}\epsilon_{2} & r_{44} & e_{45} + \epsilon_{2}\epsilon_{3} \\
e_{15} - \mu\epsilon_{3} & -e_{25} + B_{3} & e_{35} + 
\epsilon_{1}\epsilon_{3} & e_{45} + \epsilon_{2}\epsilon_{3} & r_{55} 
\end{array}
\right)\; .
\label{matrix-even}
\end{equation}
The parameters appearing in the matrix elements are defined in Appendix A.
Note that when obtaining the above mass matrix, the Eq.\ (\ref{masspara})
is used. 

The physical CP-even `Higgs' $H_i^0$ (eigenvalues 
and corresponding eigenstates) are obtained 
by means of a standard method to make the matrix 
Eq.\ (\ref{matrix-even}) diagonal.
Namely we may find a unitary matrix $Z_{even}$:
\begin{equation}
H_{i}^{0} = \sum_{j=1}^{5}Z_{even}^{ij}\chi_{j}^{0}\;, 
\label{masseven}
\end{equation}
where $Z^{ij}_{even}$ ($i$, $j=1$, $2$, $3$, $4$, $5$) are the 
elements of the matrix that converts the mass matrix 
Eq.\ (\ref{matrix-even}) into a diagonal one: 
$${\cal L}_{m}^{even} = - \Phi_{even}^{\dag} 
\cdot {\cal M}_{even}^{2} \cdot \Phi_{even} =
- {\cal H}^{0 \dag} \cdot {\cal M'}_{even}^{2} \cdot {\cal H}^0\;,$$
where 
$$ {\cal M'}_{even}^2 = Z_{even} \cdot {\cal M}_{even}^2 \cdot
Z_{even}^\dag = diag (m_{H^0_1}^2, m_{H^0_2}^2, m_{H^0_3}^2,
m_{H^0_5}^2, m_{H^0_1}^2)\; .$$
Thus $\Phi_{even}$ are the `interaction fields' and ${\cal 
H}^0$ are the physical fields (the eigenstates of the mass matrix).

As for the CP-odd sector of the Higgs, with the `interaction' 
basis $\Phi_{odd}^{T} =
(\phi_{1}^{0}$, $\phi_{2}^{0}$, $\phi_{\tilde{\nu}_{e}}^{0}$,
$\phi_{\tilde{\nu}_{\mu}}^{0}$, $\phi_{\tilde{\nu}_{\tau}}^{0})$, 
the mass matrix for the CP-odd `Higgs' can be written as:
\begin{equation}
{\cal M}_{odd}^{2} =
\left(
\begin{array}{ccccc}
s_{11} & B & -\mu\epsilon_{1}+ m_{HL^{1}}^{2} & 
-\mu\epsilon_{2}+ m_{HL^{2}}^{2} & -\mu\epsilon_{3}+ m_{HL^{3}}^{2}  \\
B & s_{22} & -B_{1} & 
-B_{2} & -B_{3} \\
-\mu\epsilon_{1}+ m_{HL^{1}}^{2} & -B_{1} & 
s_{33} & \epsilon_{1}\epsilon_{2}+ m_{L^{12}}^{2} & 
\epsilon_{1}\epsilon_{3}+ m_{L^{13}}^{2} \\
 -\mu\epsilon_{2}+ m_{HL^{2}}^{2} & -B_{2} & 
 \epsilon_{1}\epsilon_{2}+ m_{L^{12}}^{2} & s_{44} & 
\epsilon_{2}\epsilon_{3}+ m_{L^{23}}^{2}  \\
-\mu\epsilon_{3}+ m_{HL^{3}}^{2} & -B_{3} & 
\epsilon_{1}\epsilon_{3}+ m_{L^{13}}^{2} & 
\epsilon_{2}\epsilon_{3}+ m_{L^{23}}^{2} & s_{55}
\end{array}
\right)\; .  \label{massodd}
\end{equation}
The parameters appearing in the matrix elements 
are defined precisely in Appendix A.

To be different from the CP-even sector,
it is easy,  as Ref. \cite{s11} 
from Eq.\ (\ref{massodd}), to find a neutral 
Goldstone boson (with zero eigenvalue):
\begin{eqnarray}
H_{6}^{0} &=& \sum_{i=1}^{5} Z_{odd}^{1i} \phi_{i}^{0} \nonumber \\
  &=& \frac{1}{\upsilon}(\upsilon_{1}\phi_{1}^{0} - 
  \upsilon_{2}\phi_{2}^{0} + 
  \upsilon_{\tilde{\nu}_{e}}\phi_{\tilde{\nu}_{e}}^{0}
 + \upsilon_{\tilde{\nu}_{\mu}}\phi_{\tilde{\nu}_{\mu}}^{0} + 
 \upsilon_{\tilde{\nu}_{\tau}}\phi_{\tilde{\nu}_{\tau}}^{0}),
\label{ngold}
\end{eqnarray}
which is indispensable for spontaneously breaking the EW gauge symmetry. 
Here the $\upsilon=\sqrt{\upsilon_{1}^{2} + \upsilon_{2}^{2} + 
\sum\limits_{I}\upsilon_{\tilde{\nu}_{I}}^{2}}$ and similar
to the R-parity conserved MSSM, the mass of $Z$-boson
$m_{Z} = \frac{\sqrt{g^{2} + g^{\prime^{2}}}}{2}\upsilon$ is kept.
The other four massive neutral bosons can be written as:
\begin{equation}
H_{5+i}(i=2, 3, 4, 5)=\sum_{j=1}^{5}Z_{odd}^{ij}\phi_{j}^{0}
\label{oddhiggs}
\end{equation}
where again $Z^{ij}_{odd}$ ($i$, $j=1$, $2$, $3$, $4$, $5)$ is 
the matrix elements and the matrix converts the interaction fields
into the physical ones.

From the eigenvalue equations for CP-even and CP-odd 
`Higgs' and the identities of the model, similar to 
the case of Ref.\cite{chf},
it is not very difficult to find two independent 
relations for the eigenvalues as follows:
\begin{eqnarray}
  \sum_{i=1}^{5}m_{H_{i}}^{2}  &=& \sum_{i=2}^{5}m_{H_{5+i}}^{2} 
  + m_{Z}^{2}, \nonumber \\
  \prod_{i=1}^{5}m_{H_{i}}^{2}  &=& \Bigg[\frac{\upsilon_{1}^{2} - 
  \upsilon_{2}^{2} + 
\sum\limits_{I=1}^{3}\upsilon_{\tilde{\nu}_{I}}^{2}}
{\upsilon^{2}}\Bigg]^{2} m_{Z}^{2} \prod_{i=2}^{5}m_{H_{5+i}}^{2}. 
\label{massrelation}
\end{eqnarray}
The first relation of Eq.\ (\ref{massrelation}) 
is obtained by relating the traces of the two
neutral Higgs mass matrices (CP-even and CP-odd)
and the second is relating the determinants of the
mass matrices. Note: there is a Goldstone in CP-odd 
sector, thus to obtain the second relation of 
Eq.\ (\ref{massrelation}), the Goldstone mode must 
have been taken away already, and it is reason why 
the multi-product in the r.h.s. of the equation is 
start from $2$ (According to the convention here the 
number $1$ corresponds to the Goldstone). 
If we introduce the following notations:
\begin{eqnarray}
 & & \upsilon_{1} = \upsilon\cos\beta\cos\theta_{\upsilon} \; ,  
\nonumber \\
 & & \upsilon_{2} = \upsilon\sin\beta \; ,  \nonumber \\
 & & \sqrt{\sum\limits_{I=1}^{3}\upsilon_{\nu_{I}}^{2}} = 
\upsilon\cos\beta\sin\theta_{\upsilon} \; ,
\label{defineang}
\end{eqnarray}
the second relation of Eq.\ (\ref{massrelation}) becomes:
\begin{equation}
\prod_{i=1}^{5}m_{H_{i}}^{2} = \cos^{2}2\beta m_{Z}^{2} 
\prod_{i=2}^{5}m_{H_{5+i}}^{2}\;.
\label{massrelation1}
\end{equation}
The first relation of Eq.\ (\ref{massrelation}) was obtained in
Ref.\cite{s12} firstly. The second relation of Eq.\ (\ref{massrelation})
was obtained in Ref.\cite{chf} firstly in special case of the bilinear 
R-parity violating. 

The two equations are independent, and restrict the masses of 
the neutral `Higgs' bosons substantially. As discussed in Ref.\cite{chf},
for instance, with Eq.\ (\ref{massrelation}),
Eq.\ (\ref{massrelation1}) and simple algebra reduction,
the upper limit on the mass of the lightest Higgs at tree level
\begin{equation}
m_{H_{1}}^{2} \leq 
m_{H_{n}}^{2}\Bigg(\frac{m_{Z}^{2}\cos^{2}2\beta}{m_{H_{n}}^{2}}
\Bigg)^{\frac{1}{n-1}}
\frac{1-\frac{1}{n-1}\frac{m_{Z}^{2}}{m_{H_{n}}^{2}}}{1-\frac{1}{n-1}
\bigg(\frac{m_{Z}^{2}\cos^{2}2\beta}{m_{H_{n}}^{2}}\bigg)^{\frac{1}{n-1}}},
\label{bound-masshiggs}
\end{equation}
is obtained straightforwardly.
Here $n \geq 2$ is the number of the CP-even `Higgs' (the `original'
Higgs and the sneutrinos), $m_{H_{1}}$ is the mass of the lightest 
one among them, whereas $m_{H_{n}}$ is the heaviest one. 

On Eq.\ (\ref{bound-masshiggs}) two points
should be noted: 
\begin{itemize}
  \item    When $n=2$ or $m_{H_{1}}^{2} = \cdots = m_{H_{n}}^{2} = 
m_{H_{n+2}}^{2}=\cdots =
  m_{H_{n+n}}^{2}=m_{Z}^{2}$, and $\cos^{2}2\beta=1$, the symbol "=" 
is established.
  \item    In the case of MSSM with R-parity i.e. n=2,
  $m_{H_{1}}=m_{Z}^{2}\cos^{2}2\beta\frac{1-\frac{m_{Z}^{2}}{m_{H_{2}}
  ^{2}}}{1-\frac{m_{Z}^{2}}{m_{H_{2}}^{2}}\cos^{2}2\beta} 
  \leq m_{Z}^{2}\cos^{2}2\beta $ is recovered. 
\end{itemize}
In present case when $n > 2$, such a strong constraint on the
lightest Higgs mass $m_{H_{1}}$ at tree level as that for the
R-parity conserved MSSM\cite{s13,s19} 
  $$m_{H_{1}}^{2} \leq m_{Z}^{2}\cos^{2} 2\beta \leq m_{Z}^{2}$$
cannot be obtained.

In the MSSM with R-parity, the radiative corrections make
the mass of the lightest Higgs larger than that of tree level
when completing one-loop corrections and leading two-loop 
corrections of ${\cal O}(\alpha\alpha_{s})$
are included\cite{sg1}. For instance the
Ref.\cite{sg2} by precise loop calculations sets the 
limit on the lightest Higgs 
mass: $m_{H_{1}^{0}}\leq 132$GeV.
In the MSSM without R-parity, as indicated here there is no such a 
stringent restriction on the lightest Higgs at tree level
as R-parity conserved one, hence one can quite be sure that the 
`theoretical' bound on the lightest Higgs mass must be loosened 
a lot (i.e. it can be heavier than the restriction from MSSM with
R-parity conservation), especially, when loop corrections are 
involved . In Section V we will
show the indications of Eqs.(\ref{massrelation},
\ref{massrelation1}) on the lightest Higgs mass more precisely
numerically.

\subsection{The mass matrix for charged Higgs\label{app13}}

With the interaction basis $\Phi_{c}=(H_{2}^{1*}$, $H_{1}^{2}$,
$\tilde{L}_{2}^{1*}$, $\tilde{L}_{2}^{2*}$, $\tilde{L}_{2}^{3*}$,
$\tilde{R}^{1}$, $\tilde{R}^{2}$, $\tilde{R}^{3})$
\footnote{The model which we are considering here is that there
is no right-handed neutrinos at all thus there are neutrinos'
SUSY partners corresponding to the lift-handed ones, but as for
charged leptons, there are not only left-handed ones but
also right-handed ones, thus the numbers of `charged Higgs'
are 8 instead of 5 for the `neutral Higgs' (more than 3 in
three generations of leptons.} 
and Eq.\ (\ref{eq-6}),
it is easy to obtain the following mass terms for charged
`Higgs':
\begin{equation}
{\cal L}_{m}^{C} = -\Phi_{c}^{\dag}{\cal M}_{c}^{2}\Phi_{c},
\label{eq-20}
\end{equation}
the symmetric matrix ${\cal M}_{c}^{2}$ is given as Appendix A.

Making the mass matrix diagonal, a zero mass Goldstone boson state:
\begin{eqnarray}
H_{1}^{+} &=& \sum_{i=1}^{8}Z_{c}^{1i}\Phi_{c}^{i} \nonumber \\
 &=& \frac{1}{\upsilon}(\upsilon_{1}H_{2}^{1*} - \upsilon_{2}H_{1}^{2} + 
 \upsilon_{\tilde{\nu}_{e}}\tilde{L}_{2}^{1*} + 
 \upsilon_{\tilde{\nu}_{\mu}}\tilde{L}_{2}^{2*} + 
 \upsilon_{\tilde{\nu}_{\tau}}\tilde{L}_{2}^{3*})
\label{cgold}
\end{eqnarray}
is obtained. Together with its charge conjugate state $H_{1}^{-}$ 
are needed to break electroweak symmetry and give $W^{\pm}$ bosons 
masses. With the transformation matrix $Z_{c}^{ij}$ (to convert 
the interaction fields into 
the physical eigenstates), the other seven physical eigenstates
$H_{i}^{+}$ $(i=2$, $3$, $4$, $5$, $6$, $7$, $8)$ can be expressed as:
\begin{equation}
H_{i}^{+} = \sum_{j=1}^{8}Z_{c}^{ij}\Phi^{c}_{j} \hspace{0.2in}(i, j=1,
\cdots, 8).
\label{charhiggs}
\end{equation}

\subsection{The mixing of neutralinos and neutrinos:}

Due to the lepton number violations in the model, fresh and
interesting mixing of neutralinos-neutrinos and charginos-charged leptons
may happen. We devote two subsections to outline the mixing and
solve them numerically late in Sect.V. The piece of Lagrangian 
responsible for the mixing of neutralinos and neutrinos is:
\begin{eqnarray}
{\cal L}_{\chi_{i}^{0}}^{mass} &=& \{ 
ig\sqrt{2}T_{ij}^{a}\lambda^{a}\psi_{j}A_{i}^{*} - 
\frac{1}{2}\frac{\partial^{2} 
{\cal W}}{\partial A_{i} \partial A_{j}}\psi_{i}\psi_{j} + h.c. \} + 
m_{1}(\lambda_{B}\lambda_{B} + h.c.) + \nonumber \\
 & & m_{2}(\lambda_{A}^{i}\lambda_{A}^{i} + h.c.)
\label{massneutra}
\end{eqnarray}
where ${\cal W}$ is given by Eq.\ (\ref{eq-4}). $T^{a}$ are the 
generators of the SU(2)$\times$U(1) gauge group and $\psi$, 
$A_{i}$ stand for generic two-component fermion and scalar fields. 
Writing down the Eq.\ (\ref{massneutra}) explicitly, we obtain:
\begin{equation}
{\cal L}_{\chi_{i}^{0}}^{mass} = -\frac{1}{2}(\Phi^{0})^{T}
{\cal M}_{N}\Phi^{0} + h.c.
\label{massneutra1}
\end{equation}
with the interaction basis $(\Phi^{0})^{T} = (-i\lambda_{B}$, 
$-i\lambda_{A}^{3}$, 
$\psi_{H^{1}}^{1}$, $\psi_{H^{2}}^{2}$, $\nu_{e_{L}}$, $\nu_{\mu_{L}}$,
$\nu_{\tau_{L}})$ and
\begin{equation}
{\cal M}_{N} = \left(
\begin{array}{ccccccc}
2m_{1} & 0 & -\frac{1}{2}g^{\prime}\upsilon_{1} & 
\frac{1}{2}g^{\prime}\upsilon_{2} & 
-\frac{1}{2}g^{\prime}\upsilon_{\tilde{\nu}_{e}} &
-\frac{1}{2}g^{\prime}\upsilon_{\tilde{\nu}_{\mu}} & 
-\frac{1}{2}g^{\prime}\upsilon_{\tilde{\nu}_{\tau}} \\
0 & 2m_{2} & \frac{1}{2}g\upsilon_{1} & -\frac{1}{2}g\upsilon_{2} & 
\frac{1}{2}g\upsilon_{\tilde{\nu}_{e}} 
& \frac{1}{2}g\upsilon_{\tilde{\nu}_{\mu}} & 
\frac{1}{2}g\upsilon_{\tilde{\nu}_{\tau}}  \\
-\frac{1}{2}g^{\prime}\upsilon_{1} & \frac{1}{2}g\upsilon_{1} & 0 & 
-\frac{1}{2}\mu & 0  & 0 & 0\\
\frac{1}{2}g^{\prime}\upsilon_{2} & -\frac{1}{2}g\upsilon_{2} & 
-\frac{1}{2}\mu 
& 0 & \frac{1}{2}\epsilon_{1} 
& \frac{1}{2}\epsilon_{2} & \frac{1}{2}\epsilon_{3}   \\
-\frac{1}{2}g^{\prime}\upsilon_{\tilde{\nu}_{e}} & 
\frac{1}{2}g\upsilon_{\tilde{\nu}_{e}} & 0 & \frac{1}{2}\epsilon_{1} 
& 0 & 0 & 0\\
-\frac{1}{2}g^{\prime}\upsilon_{\tilde{\nu}_{\mu}} & 
\frac{1}{2}g\upsilon_{\tilde{\nu}_{\mu}} & 0 & \frac{1}{2}\epsilon_{2} 
& 0 & 0 & 0\\
-\frac{1}{2}g^{\prime}\upsilon_{\tilde{\nu}_{\tau}} & 
\frac{1}{2}g\upsilon_{\tilde{\nu}_{\tau}} & 0 & \frac{1}{2}\epsilon_{3} 
& 0 & 0 & 0 
\end{array} \right)\; .
\label{neutralino-matrix}
\end{equation}
The mixing has the formulation:
\begin{eqnarray}
& & -i\lambda_{B} = Z_{N}^{1i}\tilde{\chi}_{i}^{0}, 
\hspace{5mm}
-i\lambda_{A}^{3} = Z_{N}^{2i}\tilde{\chi}_{i}^{0}, 
\hspace{5mm}
\psi_{H^{1}}^{1} = Z_{N}^{3i}\tilde{\chi}_{i}^{0}, \nonumber \\
& &\psi_{H^{2}}^{2} = Z_{N}^{4i}\tilde{\chi}_{i}^{0}, 
\hspace{9mm}
\nu_{e_{L}} = Z_{N}^{5i}\tilde{\chi}_{i}^{0}, 
\hspace{9mm}
\nu_{\mu_{L}} = Z_{N}^{6i}\tilde{\chi}_{i}^{0}, \nonumber \\
& &\nu_{\tau_{L}} = Z_{N}^{7i}\tilde{\chi}_{i}^{0}
\label{mixing-neutralino}
\end{eqnarray}
and the transformation matrix $Z_{N}$ has the property
\begin{eqnarray}
Z_{N}^{T}{\cal M}_{N}Z_{N} &=& diag(m_{\tilde{\kappa}_{1}^{0}}, 
m_{\tilde{\kappa}_{2}^{0}}, 
m_{\tilde{\kappa}_{3}^{0}}, m_{\tilde{\kappa}_{4}^{0}}, 
m_{\nu_{e}},m_{\nu_{\mu}},m_{\nu_{\tau}}).
\label{define-ZN}
\end{eqnarray}
For convenience as in Ref.\cite{s13}, we formulate all the neutral
fermions into four component Majorana spinors as follows:
\begin{equation}
\nu_{e}=
\left(  \begin{array}{c}
\tilde{\chi}_{5}^{0} \\
\bar{\tilde{\chi}}_{5}^{0}  \end{array}  
\right),  \label{define-eneutrino}
\end{equation}
\begin{equation}
\nu_{\mu}=
\left(  \begin{array}{c}
\tilde{\chi}_{6}^{0} \\
\bar{\tilde{\chi}}_{6}^{0}  \end{array}  
\right),  \label{define-muneutrino}
\end{equation}
\begin{equation}
\nu_{\tau}=
\left(  \begin{array}{c}
\tilde{\chi}_{7}^{0} \\
\bar{\tilde{\chi}}_{7}^{0}  \end{array}  
\right),  \label{define-tauneutrino}
\end{equation}
\begin{equation}
\kappa_{i}^{0}(i=1, 2, 3, 4)=
\left(  \begin{array}{c}
\tilde{\chi}_{i}^{0} \\
\bar{\tilde{\chi}}_{i}^{0}  \end{array}  
\right).  \label{define-neutralino}
\end{equation}
It is easy from Eq.\ (\ref{neutralino-matrix}) 
to find that only one type 
of neutrinos obtains mass from the mixing
at tree level, as pointed out by Ref. \cite{se6} firstly,
and we will assume it is $\tau$-neutrino naively.
One of the stringent restrictions comes from the bound 
that the mass of $\tau$-neutrino should be less than
20 MeV\cite{s14}. Late on for convenience, we will call the
mixtures of neutralinos and
neutrinos as `neutralinos' shortly as long as there is no confusion.

\subsection{The mixing of charginos and charged leptons}

Similar to the mixing of neutralinos and neutrino, charginos 
mix with the charged leptons and form a set of physical
charged fermions: $e^{-}$, $\mu^{-}$, $\tau^{-}$, $\kappa_{1}^{\pm}$, 
$\kappa_{2}^{\pm}$. In the interaction basis, $\Psi^{+T}
=(-i\lambda^{+}$, $\tilde{H}_{2}^{1}$, $e_{R}^{+}$, $\mu_{R}^{+}$, 
$\tau_{R}^{+})$ and $\Psi^{-T}=(-i\lambda^{-}$, $\tilde{H}_{1}^{2}$,
$e_{L}^{-}$, $\mu_{L}^{-}$, $\tau_{L}^{-})$, the charged fermion mass terms 
of the Lagrangian have a general formulation\cite{s15}:
\begin{equation}
{\cal L}_{\chi_{i}^{\pm}}^{mass} = -\Psi^{-T}{\cal M}_{C} \Psi^{+} + h.c. 
\label{masscharg1}
\end{equation}
and the mass matrix:
\begin{equation}
{\cal M}_{C} = \left(
\begin{array}{ccccc}
2m_{2} & \frac{e\upsilon_{2}}{\sqrt{2}S_{W}} & 0  & 0 & 0 \\
\frac{e\upsilon_{1}}{\sqrt{2}S_{W}} & \mu  & 
\frac{l_{1}\upsilon_{\tilde{\nu}_{e}}}{\sqrt{2}} & 
\frac{l_{2}\upsilon_{\tilde{\nu}_{\mu}}}{\sqrt{2}} & \frac{l_{3}
\upsilon_{\tilde{\nu}_{\tau}}}{\sqrt{2}} \\
\frac{e\upsilon_{\tilde{\nu}_{e}}}{\sqrt{2}S_{W}} & 
-\epsilon_{1} & \frac{l_{1}\upsilon_{1}}{\sqrt{2}}+
\frac{1}{\sqrt{2}}\sum\limits_{I}\lambda_{I11}\upsilon_{\tilde{\nu}_{I}} & 
\frac{1}{\sqrt{2}}\sum\limits_{I}\lambda_{I12}\upsilon_{\tilde{\nu}_{I}} & 
\frac{1}{\sqrt{2}}\sum\limits_{I}\lambda_{I13}\upsilon_{\tilde{\nu}_{I}} \\
\frac{e\upsilon_{\tilde{\nu}_{\mu}}}{\sqrt{2}S_{W}} & -\epsilon_{2} & 
\frac{1}{\sqrt{2}}\sum\limits_{I}\lambda_{I21}\upsilon_{\tilde{\nu}_{I}} & 
\frac{l_{2}\upsilon_{1}}{\sqrt{2}} +\frac{1}{\sqrt{2}}\sum\limits_{I}
\lambda_{I22}
\upsilon_{\tilde{\nu}_{I}} & \frac{1}{\sqrt{2}}\sum\limits_{I}\lambda_{I23}
\upsilon_{\tilde{\nu}_{I}} \\
\frac{e\upsilon_{\tilde{\nu}_{\tau}}}{\sqrt{2}S_{W}} & -\epsilon_{3} & 
\frac{1}{\sqrt{2}}
\sum\limits_{I}\lambda_{I31}\upsilon_{\tilde{\nu}_{I}} & 
\frac{1}{\sqrt{2}}\sum\limits_{I}\lambda_{I32}\upsilon_{\tilde{\nu}_{I}} & 
\frac{l_{3}\upsilon_{1}}{\sqrt{2}} + 
\frac{1}{\sqrt{2}}\sum\limits_{I}\lambda_{I33}
\upsilon_{\tilde{\nu}_{I}}
\end{array} \right)\; .
\label{chargino-matrix}
\end{equation}
Here $S_{W}=\sin\theta_{W}$ and $\lambda^{\pm} = \frac{\lambda_{A}^{1} 
\mp i\lambda_{A}^{2}}{\sqrt{2}}$. Generally
two mixing matrices $Z_{+}$ and $Z_{-}$ can be obtained
by making the mass matrix ${\cal M}_{c}$ diagonal in a similar 
way as in the SM to make the mass matrix of quark diagonal i.e.
the product $(Z_{+})^{T}{\cal M}_{C}Z_{-}$ turns to a diagonal matrix:
\begin{equation}
(Z_{+})^{T}{\cal M}_{C}Z_{-}= \left(
\begin{array}{ccccc}
m_{\kappa_{1}^{-}} & 0 & 0 & 0 &0 \\
0 & m_{\kappa_{2}^{-}} & 0 & 0 & 0 \\
0 & 0 & m_{e} & 0 & 0 \\
0 & 0 & 0 & m_{\mu} & 0 \\
0 & 0 & 0 & 0 & m_{\tau}
\end{array}  \right)\; .
\label{dig-chargino}
\end{equation}
We denote the mass eigenstates with $\tilde{\chi}$ as follows:
\begin{eqnarray}
&&-i\lambda_{A}^{\pm} = Z_{\pm}^{1i}\tilde{\chi}_{i}^{\pm}, \hspace{3mm}
\psi_{H^{2}}^{1} = Z_{+}^{2i}\tilde{\chi}_{i}^{+}, \nonumber \\
&&\psi_{H^{1}}^{2} = Z_{-}^{2i}\tilde{\chi}_{i}^{-}, \hspace{6mm}
e_{L} = Z_{-}^{3i}\tilde{\chi}_{i}^{-}, \nonumber \\
&&e_{R} = Z_{+}^{3i}\tilde{\chi}_{i}^{+}, \hspace{8mm}
\mu_{L} = Z_{-}^{4i}\tilde{\chi}_{i}^{-}, \nonumber \\
&&\mu_{R} = Z_{+}^{4i}\tilde{\chi}_{i}^{+}, \hspace{8mm}
\tau_{L} = Z_{-}^{5i}\tilde{\chi}_{i}^{-}, \nonumber \\
&&\tau_{R} = Z_{+}^{5i}\tilde{\chi}_{i}^{+}.
\label{mixing-chargino}
\end{eqnarray}
The four-component fermions are defined as:
\begin{equation}
\kappa_{i}^{+}(i=1, 2, 3, 4, 5)=
\left(  \begin{array}{c}
\tilde{\chi}_{i}^{+} \\
\bar{\tilde{\chi}}_{i}^{-}  \end{array}  
\right),  \label{define-chargino}
\end{equation}
where $\kappa_{1}^{\pm}$, $\kappa_{2}^{\pm}$ are the usual charginos 
and $\kappa_{i}^{\pm}$ ($i=3$, $4$, $5$) correspond to $e$, $\mu$ and
$\tau$ lepton respectively. For convenience,
late on we will call the mixtures of charginos and charged leptons as
`charginos' shortly sometimes.

Due to the trilinear terms with coefficients $\lambda'_{IJK}$ (a lepton
superfield couples to two quark superfields) in Eq.\ (\ref{eq-4}),
the squark mixing is also affected by the lepton number breaking
interactions. Since it is not the main subject of this paper, besides
being discussed elsewhere\cite{cchf}, we outline the effects in
the simplest case in Appendix B briefly. 

According to the above analysis, we have achieved the 
formulations of the mass spectrum of
the neutralinos-neutrinos, charginos-charged leptons, 
neutral Higgs-sneutrinos and charged Higgs-charged sleptons. 
Since the vertices of the interactions are also important,
thus in the next section we will give 
the Feynman rules of the model, which are new to those of
the MSSM with R-parity conserved.

\section{The Feynman rules for the R-parity violating interactions}

We have discussed the various masses of the MSSM with R-parity violation.
Now, we are discussing the Feynman rules for the model that are
new to those in MSSM with R-parity conserved. 
For convenience in loop calculations, we work out here the rules 
only in t$^{\prime}$Hooft-Feynman gauge\cite{s16}, 
which has the gauge fixed terms:
\begin{eqnarray}
{\cal L}_{GF} &=& -\frac{1}{2\xi}\Big(\partial^{\mu}A_{\mu}^{3} + \xi 
M_{Z}C_{W}H_{6}^{0}\Big)^{2} - \frac{1}{2\xi}\Big(\partial^{
\mu}B_{\mu} - \xi M_{Z}S_{W}H_{6}^{0}\Big)^{2} - 
\frac{1}{2\xi}\bigg(\partial^{\mu}A_{\mu}^{1}  \nonumber \\
 & &+\frac{i}{\sqrt{2}}\xi M_{W}\Big(
H_{1}^{+} - H_{1}^{-}\Big)\bigg)^{2} -  
\frac{1}{2\xi}\bigg(\partial^{\mu}A_{\mu}^{2} + 
\frac{1}{\sqrt{2}}\xi M_{W}\Big(H_{1}^{+} + H_{1}^{-}\Big)\bigg)^{2}  
\nonumber \\
 &=& \Bigg\{-\frac{1}{2\xi}(\partial^{\mu}Z_{\mu})^{2} - 
\frac{1}{2\xi}(\partial^{\mu}F_{\mu})^{2} - \frac{1}{\xi}
(\partial^{\mu}W_{\mu}^{+})(\partial^{\mu}W_{\mu}^{-}) \Bigg\} - 
\Bigg\{M_{Z}H_{6}^{0}\partial^{\mu}Z_{\mu}  \nonumber \\
 & & +iM_{W}\Big(H_{1}^{+}
\partial^{\mu}W_{\mu}^{-} - H_{1}^{-}\partial^{\mu}W_{\mu}^{+}\Big)\Bigg\}
 - \Bigg\{ \frac{1}{2}\xi M_{Z}^{2}H_{6}^{0^{2}} - 
\xi M_{W}^{2}H_{1}^{+}H_{1}^{-} 
\Bigg\},
\label{gauge-fixed}
\end{eqnarray}
where $C_{W}=\cos\theta_{W}$ and $H_{6}^{0}$, $H_{1}^{\pm}$ 
are defined as the above. By inserting the expressions
Eq.\ (\ref{gauge-fixed}) into Lagrangian, the desired vertices for the
Higgs bosons are obtained. We assume the relevant parameters are real,
i.e. at this moment only CP being conserved is considered, one may find
that $H_{1}$, $H_{2}$, $H_{3}^{0}$ $H_{4}^{0}$, $H_{5}^{0}$ are scalars
but $H_{6}^{0}$, $H_{7}^{0}$, $H_{8}^{0}$, $H_{9}^{0}$, $H_{10}^{0}$
pseudoscalar.

\subsection{Feynman rules for Higgs (slepton)- gauge boson interactions}

\begin{figure}
\setlength{\unitlength}{1mm}
\begin{picture}(140,180)(30,30)
\put(-10,-10){\includegraphics{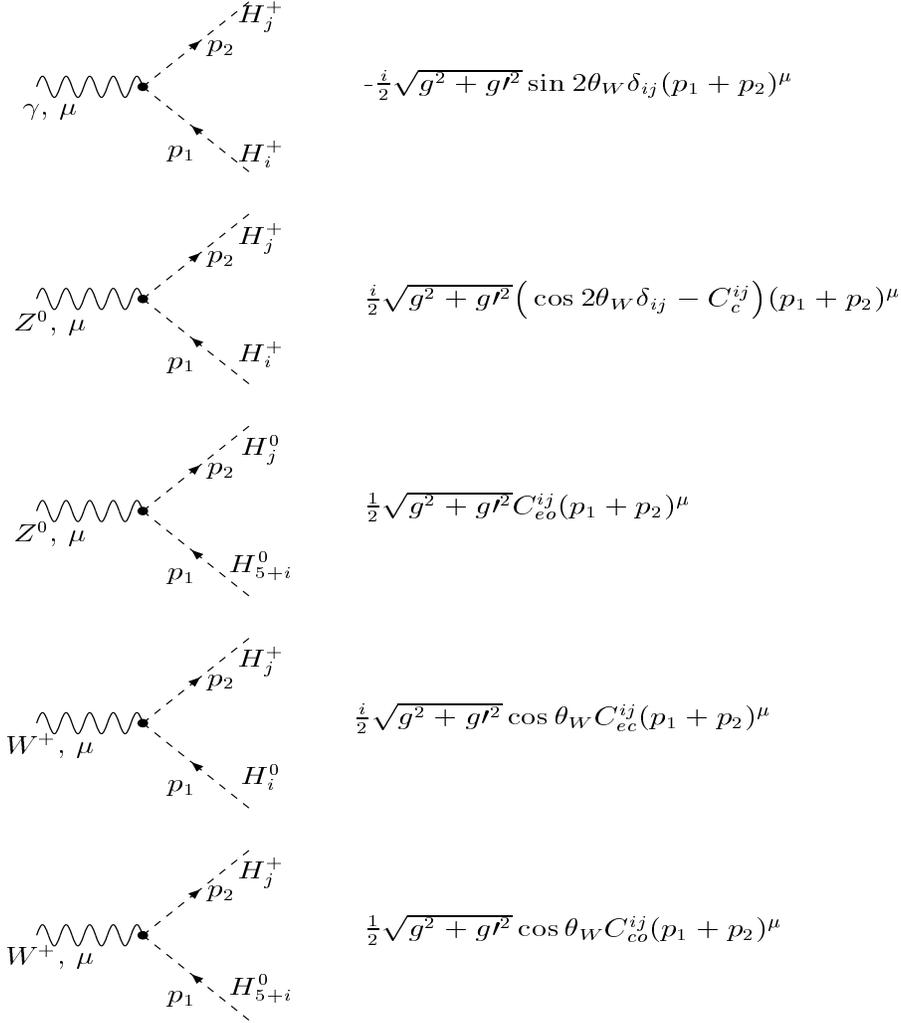}}
\end{picture}
\caption[]{Feynman rules for SSV vertices, the direction of momentum is 
indicated above}
\label{fig1}
\end{figure}

Let us compute the vertices of Higgs (slepton)- gauge bosons in the
model precisely. The interaction terms of Higgs bosons and
gauge bosons are:
\begin{eqnarray}
{\cal L}_{int}^{1} &=& -\sum_{I}({\cal D}_{\mu}\tilde{L}^{I\dag}{\cal 
D}^{\mu}\tilde{L}^{I} - {\cal D}_{\mu}\tilde{R}^{I*}{\cal D}^{\mu}
\tilde{R}^{I}) - {\cal D}_{\mu}H^{1\dag}{\cal D}^{\mu}H^{1} - {\cal 
D}_{\mu}H^{2\dag}{\cal D}^{\mu}H^{2}  \nonumber \\
 &=&\sum_{I}\Bigg\{ \bigg[i\tilde{L}^{I\dag}\Big(g\frac{\tau^{i}}{2}
A_{\mu}^{i} - 
\frac{1}{2}g^{\prime}B_{\mu}\Big)\partial^{\mu}\tilde{L}^{I} +
h.c. \bigg] - \tilde{L}^{I\dag}\Big(g\frac{\tau^{i}}{2}A_{\mu}^{i} 
\nonumber \\
&& - \frac{1}{2}g^{\prime}B_{\mu}\Big) \Big(g\frac{\tau^{j}}{2}A^{j\mu} - 
\frac{1}{2}g^{\prime}B^{\mu}\Big)\tilde{L}^{I} +  \Big( ig^{\prime} 
B_{\mu}\tilde{R}^{I*}\partial^{\mu}\tilde{R}^{I} \nonumber \\
&& +h.c. \Big) - g^{\prime^{2}}\tilde{R}^{I*}\tilde{R}^{I}B_{\mu}B^{\mu} 
\Bigg\} + \Bigg\{ H^{1\dag}\Big(g\frac{\tau^{i}}{2}A_{\mu}^{i} - 
\frac{1}{2}g^{\prime}B_{\mu}\Big)\partial^{\mu}H^{1} \nonumber \\
 & &+ h.c. \Bigg\}
- H^{1\dag}\Big(g\frac{\tau^{i}}{2}A_{\mu}^{i} - 
\frac{1}{2}g^{\prime}B_{\mu}\Big) \Big(g\frac{\tau^{j}}{2}A^{j\mu} -
\frac{1}{2}g^{\prime}B^{\mu}\Big)H^{1}  \nonumber \\
 & &+\Bigg\{ H^{2\dag}\Big(g\frac{\tau^{i}}{2}A_{\mu}^{i} - 
 \frac{1}{2}g^{\prime}B_{\mu}\Big)\partial^{\mu}H^{2} + h.c. \Bigg\}
- H^{2\dag}\Big(g\frac{\tau^{i}}{2}A_{\mu}^{i}  \nonumber \\
 & &- \frac{1}{2}g^{\prime}B_{\mu}\Big) \Big(g\frac{\tau^{j}}{2}A^{j\mu} -
\frac{1}{2}g^{\prime}B^{\mu}\Big)H^{2} \nonumber \\
&=& {\cal L}_{SSV} + {\cal L}_{SVV} + {\cal L}_{SSVV},
\label{interaction1}
\end{eqnarray}
here ${\cal L}_{SSV}$, ${\cal L}_{SVV}$ and ${\cal L}_{SSVV}$, 
are the relevant interaction terms. For the Feynman rules
and convenience in practical applications they are precisely rewritten 
in the physical bases which were obtained in the previous section.
Since the precise formulas of ${\cal L}_{SSV}$, ${\cal L}_{SVV}$ and 
${\cal L}_{SSVV}$ are lengthy, so we put them into Appendix C.
Instead, let us summarize the relevant Feynman rules
in Fig.\ \ref{fig1} $\sim$ \ref{fig4} and
emphasize a few features about them.
First, the presence of the vertices 
$Z_{\mu}H_{i}H_{5+j}^{0}$ $(i$, $j=1$, $2$, 
$3$, $4$, $5)$ and the forbiddance of the vertices 
$Z_{\mu}H_{i}H_{j}^{0}$ and $Z_{\mu}H_{5+i}H_{5+j}^{0}$ 
($i$, $j=1$, $2$, $3$, $4$, $5$) are due to their CP nature.
Second, besides 
the $W_{\mu}^{+}Z^{\mu}H_{1}^{-}$ ($H_{1}^{-}$ is just 
the charged Goldstone 
boson) interaction, there are not vertices $W_{\mu}^{+}Z^{\mu}H_{i}^{-}$
$(i=2$, $3$, $4$, $5$, $6$, $7$, $8)$ at tree level, 
that is the same as the MSSM with R-parity being conserved and 
the general two-Higgs doublet models.

\subsection{Self-couplings of the Higgs bosons (sleptons)}

It is a straightforward calculation
by inserting Eqs.\ (\ref{masseven},
\ref{ngold}, 
\ref{oddhiggs}, \ref{cgold}, \ref{charhiggs})
into Eqs.\ (\ref{eq-6}), to obtain the desired interaction 
terms. Similar to the interactions of gauge-Higgs (slepton) 
bosons, we split the Lagrangian into pieces:
\begin{equation}
{\cal L}_{int}^{S} = {\cal L}_{SSS} + {\cal L}_{SSSS}
\end{equation}
where ${\cal L}_{SSS}$ represents trilinear coupling terms, 
and ${\cal L}_{SSSS}$ four scalar boson coupling terms. The trilinear pieces
are most interesting. If the masses of the scalars are appropriate, 
the decays of one Higgs boson into other two Higgs bosons may occur. 
After tedious computation, we obtain:
\begin{eqnarray}
{\cal L}_{SSS} &=& -\frac{g^{2} + 
g^{\prime^{2}}}{8}A_{even}^{ij}B_{even}^{k}H_{i}H_{j}^{0}H_{k}^{0} -
\frac{g^{2} + g^{\prime^{2}}}{8}A_{odd}^{ij}B_{even}^{k}H_{5+i}H_{5+j}^{0}
H_{k}^{0}  \nonumber \\
 & & -A_{ec}^{kij}H_{k}^{0}
H_{i}^{-}H_{j}^{+} + iA_{oc}^{kij}H_{5+k}^{0}H_{i}^{-}H_{j}^{+}
\label{vertex-sss}
\end{eqnarray}
and
\begin{eqnarray}
{\cal L}_{SSSS} &=& -\frac{g^{2} + 
g^{\prime^{2}}}{32}A_{even}^{ij}A_{even}^{kl}H_{i}H_{j}^{0}H_{k}^{0}H_{l}^{0}
- \frac{g^{2} + 
g^{\prime^{2}}}{32}A_{odd}^{ij}A_{odd}^{kl}H_{5+i} 
H_{5+j}^{0}H_{5+k}^{0}H_{5+l}^{0}  \nonumber \\
& & -\frac{g^{2} + 
g^{\prime^{2}}}{16}A_{even}^{ij}A_{odd}^{kl}
H_{i}H_{j}^{0}H_{5+k}^{0}H_{5+l}^{0} - 
{\cal A}_{ec}^{klij}H_{k}^{0}H_{l}^{0}H_{i}^{-}H_{j}^{+}  \nonumber \\
 & &-{\cal A}_{oc}^{klij}H_{5+k}^{0}H_{5+l}^{0}H_{i}^{-}H_{j}^{+} - 
i {\cal A}_{eoc}^{klij}H_{k}^{0}H_{5+l}^{0}H_{i}^{-}H_{j}^{+} - 
{\cal A}_{cc}^{klij}H_{k}^{-}H_{l}^{+}H_{i}^{-}H_{j}^{+}
\label{vertex-ssss}
\end{eqnarray}
with 
\begin{eqnarray}
A_{even}^{ij} &=& \sum_{\alpha=1}^{5}Z_{even}^{i \alpha}Z_{even}^{j 
\alpha} - 2Z_{even}^{i 2}Z_{even}^{j 2}, \nonumber \\
A_{odd}^{ij}  &=& \sum_{\alpha=1}^{5}Z_{odd}^{i \alpha}Z_{odd}^{j \alpha}
- 2Z_{odd}^{i 2}Z_{odd}^{j 2}, \nonumber \\
B_{even}^{i}  &=& \upsilon_{1}Z_{even}^{i 1} - \upsilon_{2}Z_{even}^{i 2}
+ \sum_{I}\upsilon_{\tilde{\nu}_{I}}Z_{even}^{i I+2} .
\end{eqnarray}
The definitions of $A_{ec}^{kij}$, $A_{oc}^{kij}$, ${\cal A}_{ec}^{klij}$, 
${\cal A}_{oc}^{klij}$, ${\cal A}_{eoc}^{klij}$ and ${\cal A}_{cc}^{ijkl}$ 
are lengthy, so we put them in Appendix C. 
The Feynman rules are summarized in Fig.\ref{fig5} and Fig.\ \ref{fig6}.
Note that the lepton number violations have
led to very complicated form for the ${\cal L}_{SSS}$ 
and ${\cal L}_{SSSS}$.

\subsection{The R-parity violation couplings of Higgs}

In this subsection we compute the R-parity violation 
couplings of Higgs i.e. the Higgs couplings to charginos 
(charged lepton) and neutralinos 
(neutrinos). After spontaneous breaking of the 
gauge symmetry SU(2)$\times$U(1), the gauginos, higgsinos and 
leptons with the
same electric charge may mix as described in Section II. Let us 
proceed to compute the interaction
$S\tilde{\kappa}_{i}^{0}\tilde{\kappa}_{j}^{0}$ 
(Higgs-neutralinos-neutralinos interactions).

The interactions (in two-component spinors)\cite{s13} are:
\begin{eqnarray}
{\cal L}_{S\kappa\kappa} &=& 
i\sqrt{2}g\Big(H^{1\dag}\frac{\tau^{i}}{2}\lambda_{A}^{i}\psi_{H^{1}} - 
\bar{\psi}_{H^{1}}\frac{\tau^{i}}{2}
\bar{\lambda}_{A}^{i}H^{1}\Big) - 
i\sqrt{2}g^{\prime}\Big(\frac{1}{2}H^{1\dag}\psi_{H^{1}}\lambda_{B} - 
\frac{1}{2}\bar{\lambda}_{B}\bar{\psi}_{H^{1}}
H^{1}\Big)  \nonumber \\
 & & + i\sqrt{2}g\Big(H^{2\dag}\frac{\tau^{i}}{2}\lambda_{A}^{i}\psi_{H^{2}} 
- \bar{\psi}_{H^{2}}\frac{\tau^{i}}{2}
\bar{\lambda}_{A}^{i}H^{2}\Big) + 
i\sqrt{2}g^{\prime}\Big(\frac{1}{2}H^{2\dag}\psi_{H^{2}}\lambda_{B} - 
\frac{1}{2}\bar{\lambda}_{B}\bar{\psi}_{H^{2}}
H^{2}\Big)  \nonumber \\
 & & + i\sqrt{2}\tilde{L}^{I\dag}\bigg(g\frac{\tau^{i}}{2}
\lambda_{A}^{i}\psi_{L^{I}} - 
\frac{1}{2}g^{\prime}\lambda_{B}\psi_{L^{I}}\bigg) - 
i\sqrt{2}\tilde{L}^{I}\bigg(g\frac{\tau^{i}}{2}
\bar{\lambda}_{A}^{i}\bar{\psi}_{L^{I}} - 
\frac{1}{2}g^{\prime}\bar{\lambda}_{B}
\bar{\psi}_{L^{I}}\bigg)  \nonumber \\
 & & + i\sqrt{2}g^{\prime}\tilde{R}^{I\dag}\lambda_{B}\psi_{R^{I}} - 
i\sqrt{2}g^{\prime}\tilde{R}^{I}\bar{\lambda}_{B}\bar{\psi}_{R^{I}} - 
\frac{1}{2}l_{I}\varepsilon_{ij}\bigg(\psi_{H^{1}}^{i}
\psi_{L^{I}}^{j}\tilde{R}^{I} + 
\psi_{H^{1}}^{i}\psi_{R^{I}}\tilde{L}_{j}^{I}
 \nonumber \\
 & & + \psi_{R^{I}}\psi_{L^{I}}^{j}H_{i}^{1} + h.c.\bigg) 
-\frac{1}{2}\lambda_{IJK}\varepsilon_{ij}\bigg(\psi_{L^{I}}^{i}
\psi_{L^{J}}^{j}
\tilde{R}^{K} + \psi_{L^{I}}^{i}\psi_{R^{K}}\tilde{L}_{j}^{J} \nonumber \\
&& + \psi_{L^{J}}^{j}\psi_{R^{K}}\tilde{L}_{i}^{I} + h.c.\bigg)\; .
\label{snn-tcom}
\end{eqnarray}
We sketch the derivation for the vertices,
$S\tilde{\kappa}_{i}^{0}\tilde{\kappa}_{j}^{0}$ etc. Starting with the 
Eq.\ (\ref{snn-tcom}), we convert the pieces from the two-component spinor
notation into four-component spinor notation first,
then using the definitions in 
Eq.\ (\ref{define-eneutrino} $\sim$ \ref{define-neutralino})
and Eq.\ (\ref{define-chargino}), we obtain:
\begin{eqnarray}
{\cal L}_{S\kappa\kappa} &=& \frac{\sqrt{g^{2} + g^{\prime^{2}}}}{2}\bigg[
C_{snn}^{ijm}H_{i}\bar{\kappa}_{j}^{0}P_{L}\kappa_{m}^{0}  
+C_{snn}^{ijm*}H_{i}\bar{\kappa}_{j}^{0}P_{R}\kappa_{m}^{0}\bigg]
\nonumber \\
 & & + 
\frac{g}{\sqrt{2}}\bigg[C_{skk}^{ijm}H_{i}\bar{\kappa}_{m}^{+}P_{L} 
\kappa_{j}^{+}
+C_{skk}^{ijm*}H_{i}\bar{\kappa}_{j}^{+}P_{R}\kappa_{m}^{+}
 \bigg]  \nonumber \\
 & & + i\frac{\sqrt{g^{2} + g^{\prime^{2}}}}{2}\bigg[
C_{onn}^{ijm}H_{5+i}\bar{\kappa}_{j}^{0}P_{R}\kappa_{m}^{0} 
-C_{onn}^{ijm*}H_{5+i}\bar{\kappa}_{m}^{0}P_{L}\kappa_{j}^{0}\bigg]
\nonumber \\
 & & +i\frac{g}{\sqrt{2}}\bigg[C_{okk}^{ijm}H_{5+i}
\bar{\kappa}_{m}^{+}P_{L}\kappa_{j}
^{+}
 -C_{okk}^{ijm*}H_{5+i}\bar{\kappa}_{m}^{+}P_{R}\kappa_{j}^{+} \bigg]
\nonumber 
\\
&&+ \sqrt{g^{2} + g^{\prime^{2}}}\bigg[C_{Lnk}^{ijm} 
\bar{\kappa}_{j}^{+}P_{L}\kappa_{m}^{0}H_{i}^{+}
-C_{Rnk}^{ijm}\bar{\kappa}_{j}^{+}P_{R}\kappa_{m}^{0}H_{i}^{+}\bigg] 
\label{vertex-snn}
\end{eqnarray}
and the coefficients $C_{snn}^{ijm}$, $C_{Lnk}^{ijm}$, $C_{Rnk}^{ijm}$ and 
$C_{skk}^{ijm}$, being lengthy, are put in Appendix D 
Here $P_{L,R} = \frac{1 \pm \gamma_{5}}{2}$ are project operators and
the transformation matrices $Z_{\pm}$, $Z_{N}$ are defined in Sect.II.
The corresponding Feynman rules are summarized in Fig.\ \ref{fig7}. 
Note here: as for $\kappa_{i}^{0}$ being a Majorana fermion, 
the useful identity
\begin{equation}
\bar{\kappa}_{j}^{0}(1 \pm \gamma_{5})\kappa_{k}^{0} = 
\bar{\kappa}_{k}^{0}(1 \pm \gamma_{5})\kappa_{j}^{0} ,
\label{majorana-protet}
\end{equation}
holds for the anticommuting four-component Majorana spinors
always, that the $H_{i}\bar{\kappa}_{j}^{0}\kappa_{k}^{0}$
interaction can be rearranged symmetrically 
under the interchange of the indices $j$ and $k$.

Since $\nu_{e}$ $(e)$, $\nu_{\mu}$ $(\mu)$ and 
$\nu_{\tau}$ $(\tau)$ should be 
identified with the lightest three `neutralinos' (`charginos') in the 
model, there must be some fresh and interesting 
phenomena relevant to them, e.g.
$\kappa_{i}^{0}$ $(i=1,2,3,4) \rightarrow \tau H_{j}^{+}$ $(j=2$,$3$,
$\cdots$, $8)$, $\kappa_{i}^{0}$ $(i=1,2,3,4) \rightarrow \nu_{e,\mu,\tau}
H_{j}^{0}$ $(j=1$, $2$, $\cdots$, $5)$ etc may occur, if the 
masses are suitable (the phase space is allowed).
Namely, these interactions without R-parity conservation may 
induce new rare processes\cite{s5,s55,s4,s6,se6,s20}.

\subsection{The R-parity violation couplings of gauge bosons}

In this subsection we focus on the R-parity
violation couplings of the gauge bosons 
($W$, $Z$, $\gamma$) i.e. the couplings of the gauge bosons 
($W$, $Z$, $\gamma$) to the charginos
(charged leptons) and neutralinos (neutrinos). 
Since we identify the three types of charged leptons (neutrinos)
with the three lightest charginos (neutralinos), the restrictions 
relating to them from the present experiments must
be considered carefully. The relevant interactions 
come from the following pieces of Lagrangian:
\begin{eqnarray}
{\cal L}_{int}^{gcn} &=& -i\bar{\lambda}_{A}^{i}\bar{\sigma}^{\mu}{\cal 
D}_{\mu}\lambda_{A}^{i} - i\bar{\lambda}_{B}\bar{\sigma}^{\mu}
{\cal D}_{\mu}\lambda_{B} - i\bar{\psi}_{H^{1}}\bar{\sigma}^{\mu}{\cal 
D}_{\mu}\psi_{H^{1}} - i\bar{\psi}_{H^{2}}\bar{\sigma}^{\mu}
{\cal D}_{\mu}\psi_{H^{2}} - i\bar{\psi}_{L^{I}}\bar{\sigma}^{\mu}{\cal 
D}_{\mu}\psi_{L^{I}}  \nonumber \\
 & &  -i\bar{\psi}_{R^{I}}\bar{\sigma}^{\mu}{\cal D}_{\mu}\psi_{R^{I}}
\label{lang-gcn}
\end{eqnarray}
with
\begin{eqnarray}
{\cal D}_{\mu}\lambda_{A}^{1} &=& \partial_{\mu}\lambda_{A}^{1} - 
gA_{\mu}^{2}\lambda_{A}^{3} + gA_{\mu}^{3}\lambda_{A}^{2} ,\nonumber \\
{\cal D}_{\mu}\lambda_{A}^{2} &=& \partial_{\mu}\lambda_{A}^{2} - 
gA_{\mu}^{3}\lambda_{A}^{1} + gA_{\mu}^{1}\lambda_{A}^{3} ,\nonumber \\
{\cal D}_{\mu}\lambda_{A}^{3} &=& \partial_{\mu}\lambda_{A}^{3} - 
gA_{\mu}^{1}\lambda_{A}^{2} + gA_{\mu}^{2}\lambda_{A}^{1} ,\nonumber \\
{\cal D}_{\mu}\lambda_{B} &=& \partial_{\mu}\lambda_{B} ,\nonumber \\
{\cal D}_{\mu}\psi_{H^{1}} &=& (\partial_{\mu} + igA_{\mu}^{i}
\frac{\tau^{i}}{2} 
- \frac{i}{2}g^{\prime} B_{\mu})\psi_{H^{1}} ,\nonumber \\
{\cal D}_{\mu}\psi_{H^{2}} &=& (\partial_{\mu} + igA_{\mu}^{i}
\frac{\tau^{i}}{2} 
+ \frac{i}{2}g^{\prime} B_{\mu})\psi_{H^{2}} ,\nonumber \\
{\cal D}_{\mu}\psi_{L^{I}} &=& (\partial_{\mu} + igA_{\mu}^{i}
\frac{\tau^{i}}{2} 
- \frac{i}{2}g^{\prime} B_{\mu})\psi_{L^{I}} ,\nonumber \\
{\cal D}_{\mu}\psi_{R^{I}} &=& (\partial_{\mu} + ig^{\prime} 
B_{\mu})\psi_{R^{I}}\; .
\label{con-partial}
\end{eqnarray}
Similar to the couplings in ${\cal L}_{S\kappa\kappa}$, we convert 
all spinors in 
Eq.\ (\ref{lang-gcn}) into four component ones and with
Eq.\ (\ref{define-neutralino}) 
and Eq.\ (\ref{define-chargino}), then we obtain:
\begin{eqnarray}
{\cal L}_{int}^{gcn} 
&=&\bigg\{\sqrt{g^{2}+g^{\prime^{2}}}\sin\theta_{W}\cos\theta_{W}A_{\mu}
\bar{\kappa}_{i}^{+}\gamma^{\mu}\kappa_{i}^{+} - 
\sqrt{g^{2}+g^{\prime^{2}}}Z_{\mu}\bar{\kappa}_{i}^{+}
\bigg[\cos^{2}\theta_{W}\delta_{ij}\gamma^{\mu}  \nonumber \\
 & &+\frac{1}{2}\Big(Z_{-}^{*i 2}Z_{-}^{j 2} + 
\sum_{I=1}^{3}Z_{-}^{*i 2+I}Z_{-}^{j 2+I}\Big)\gamma^{\mu}P_{R} 
\nonumber \\
&& + \Big(\frac{1}{2}Z_{+}^{*i 2}Z_{+}^{j 2} - 
\sum_{I=1}^{3}Z_{+}^{*i 2+I}Z_{+}^{j 2+I}\Big)
\gamma^{\mu}P_{L}\bigg]\kappa_{j}^{+} \bigg\}  \nonumber \\
 & &+\bigg\{ g\bar{\kappa}_{j}^{+}\bigg[ \Big(-Z_{+}^{*i 1}Z_{N}^{j 2} + 
\frac{1}{\sqrt{2}}Z_{+}^{*i 2}
Z_{N}^{j 4}\Big)\gamma^{\mu}P_{L} + \bigg(Z_{N}^{*i 2}Z_{-}^{j 1}
\nonumber \\
 & &+\frac{1}{\sqrt{2}}\Big(Z_{N}^{*i 3}Z_{-}^{j 2} +  
\sum_{I=1}^{3}Z_{N}^{*i 4+I}Z_{-}^{j 2+I}\Big)\bigg)
\gamma^{\mu}P_{R}\bigg]\kappa_{i}^{0}W_{\mu}^{+} + h.c. \bigg\}  \nonumber \\
 & &+\frac{\sqrt{g^{2}+g^{\prime^{2}}}}{2}
\bar{\kappa}_{i}^{0}\gamma^{\mu}\Bigg\{\bigg[Z_{N}^{*i 4}Z_{N}^{j 4} - 
\Big(Z_{N}^{*i 3}Z_{N}^{j 3} + 
\sum_{\alpha=5}^{7}Z_{N}^{*i \alpha}Z_{N}^{j \alpha}\Big)\bigg]P_{L}
\nonumber \\
&& - \bigg[Z_{N}^{*i 4}Z_{N}^{j 4} - \Big(Z_{N}^{*i 3}Z_{N}^{j 3} + 
\sum_{\alpha=5}^{7}Z_{N}^{*i \alpha}Z_{N}^{j \alpha}\Big)\bigg]P_{R}
\Bigg\}\kappa_{j}^{0}Z_{\mu} \;.
\label{vertex-gcn}
\end{eqnarray}
The corresponding Feynman rules are summarized in Fig.\ \ref{fig8}.
Since we identify the three lightest neutralinos (charginos) with three
types of neutrinos (charged leptons), some features about
Eq.\ (\ref{vertex-gcn}) should be emphasized:
\begin{itemize}
\item At tree level for the $\gamma$-$\kappa$-$\kappa$ vertices,
there is no lepton flavor-changing current interaction, that
is the same as
that in the SM and MSSM with R-parity.
\item At tree level for the $Z$-$\kappa$-$\kappa$ vertices, there
are lepton flavor-changing current interactions, that is different
from the MSSM with R-parity.
\item Similar to the $Z$-$\kappa$-$\kappa$ vertices, there are
the vertices such as $W\tau\nu_{e}$, which are forbidden in the MSSM
with R-parity.
\end{itemize}

\subsection{The R-parity violation couplings of quarks and/or squarks}

In this subsection we focus the R-parity violation couplings of 
quarks and/or squarks i.e. pursue the Feynman rules for the interactions 
of quarks and scalar-quarks with charginos (charged leptons) and
neutralinos (neutrinos) e.g. the $\tilde{Q}q\kappa_{i}^{\pm}$ 
vertices. Because the mixing of neutrinos (charged leptons) and 
original neutralinos (charginos), so the vertices will lead to certain
interesting phenomenology, thus it is interesting to write them 
down precisely.
Of the vertices, they can be divided into two categories:
the supersymmetric analogies of the $q\bar{q}W^{\pm}$ and $q\bar{q}Z$
interactions and the supersymmetric analogy of the $q\bar{q}H$
interaction which is proportional to quark mass and 
depends on the properties of the Higgs bosons in the model. 

Let us consider the $\bar{q}q\kappa_{i}^{\pm}$ interactions first.
In two-component spinors they are as the follows:
\begin{eqnarray}
{\cal L}_{\tilde{Q}q\kappa^{\pm}} &=& ig\bigg(C^{IJ}\tilde{Q}_{2}^{I*}
\lambda_{A}^{-}\psi_{Q_{1}}^{J} + C^{IJ*}\tilde{Q}_{1}^{J*}\lambda_{A}^{+}
\psi_{Q_{2}}^{I}\bigg) - 
ig\bigg(C^{IJ*}\tilde{Q}_{2}^{I}\bar{\lambda}_{A}^{-}\bar{\psi}_{Q_{1}}^{J} + 
C^{IJ}\tilde{Q}_{1}^{J}\bar{\lambda}_{A}^{+}\bar{\psi}_{Q_{2}}^{I}\bigg)  
\nonumber \\
 & & - \frac{d^{I}}{2}\bigg(C^{IJ}
\psi_{H^{1}}^{2}\psi_{Q^{1}}^{J}\tilde{D}^{I} 
 + C^{IJ}\psi_{H^{1}}^{2}\tilde{Q}_{1}^{J}\psi_{D}^{I} +
h.c.\bigg) + 
\frac{u^{I}}{2}\bigg(C^{JI*}\psi_{H^{2}}^{1}\psi_{Q^{2}}^{J}\tilde{U}^{I}  
\nonumber \\
 & & + C^{JI*}\psi_{H^{2}}^{1}\psi_{U}^{I}\tilde{Q}_{2}^{J} + h.c.\bigg) 
-\frac{1}{2}\lambda_{IJK}
\bigg(C^{JK}\psi_{L^{I}}^{2}\psi_{Q^{1}}^{K}\tilde{D}^{J} \nonumber \\
&& + C^{JK}\psi_{L^{I}}^{2}\tilde{Q}_{1}^{K}\psi_{D}^{J} + h.c.\bigg)\; .
\label{Qqc-tcom}
\end{eqnarray}
As discussed above, we convert the two-component spinors into
four-component spinors:
\begin{eqnarray}
{\cal L}_{\tilde{Q}q\kappa^{\pm}} &=& 
C^{IJ}\bar{\kappa}_{j}^{+}\Bigg\{\bigg(-gZ_{D_{I}}^{i 1}Z_{-}^{j 1} + 
\frac{d^{I}}{2}Z_{D_{I}}^{i 2}Z_{-}^{j 2} + 
\frac{1}{2}\lambda_{KIJ}^{\prime}Z_{D_{I}}^{i 2}Z_{-}^{j 2+K}\bigg)P_{L} 
\nonumber \\
&& + 
\frac{u^{J}}{2}Z_{+}^{j 2*}Z_{D_{I}}^{i 1}P_{R}
\Bigg\}\psi_{u^{I}}\tilde{D}_{I i}^{+} +  
C^{IJ*}\bar{\kappa}_{j}^{-}\Bigg\{\bigg(-gZ_{U_{J}}^{i 1}Z_{+}^{j 1} + 
\frac{u^{J}}{2}Z_{U_{J}}^{i 2}Z_{+}^{j 2}\bigg)P_{L}  \nonumber \\
 & &-\bigg(\frac{d^{I}}{2}Z_{U_{J}}^{j 1*}Z_{-}^{j 2*} + 
\frac{1}{2}\lambda_{KIJ}^{\prime}Z_{U_{J}}^{j 1*}
Z_{-}^{j 2+K*}\bigg)P_{R}\Bigg\}\psi_{d^{J}}\tilde{U}_{Ji}^{-}
+ h.c.\; .
\label{Qqc-fcom}
\end{eqnarray}
Here $\psi_{u^{I}}$, $\psi_{d^{I}}$ are four-component quark spinors of
the I-th generation. The $\kappa_{j}^{-}$ is a charged-conjugate state
of $\kappa_{j}^{+}$, and $\kappa_{j}^{+}$ is defined by
Eq.\ (\ref{define-chargino}). 

For the $\tilde{Q}q\kappa_{i}^{0}$ interactions, in two-component notation
they are:
\begin{eqnarray}
{\cal L}_{\tilde{Q}q\kappa_{i}^{0}} &=& i\sqrt{2}\tilde{Q}^{I*}\left( 
g\frac{\tau^{3}}{2}\lambda_{A}^{3} + 
\frac{1}{6}g^{\prime}\lambda_{B}\right)
\psi_{Q}^{I} - 
i\sqrt{2}\tilde{Q}^{I}\left(g\frac{\tau^{3}}{2}\bar{\lambda}_{A}^{3} + 
\frac{1}{6}g^{\prime}\bar{\lambda}_{B}\right)
\bar{\psi}_{Q}^{I}  \nonumber \\
 & & -i\frac{2\sqrt{2}}{3}g^{\prime}\tilde{U}^{I*}\lambda_{B}\psi_{U}^{I} + 
i\frac{2\sqrt{2}}{3}g^{\prime}\tilde{U}^{I}\bar{\lambda}_{B}
\bar{\psi}_{U}^{I} + i\frac{\sqrt{2}}{3}g^{\prime}
\tilde{D}^{I*}\lambda_{B}\psi_{D}^{I} - 
i\frac{\sqrt{2}}{3}g^{\prime}\tilde{D}^{I}
\bar{\lambda}_{B}\bar{\psi}_{D}^{I}  \nonumber \\
 & & +\frac{d^{I}}{2}\left\{\psi_{H^{1}}^{1}\psi_{Q^{2}}^{I}\tilde{D}^{I} + 
\psi_{H^{1}}^{1}\psi_{D}^{I}\tilde{Q}_{2}^{I} + h.c.\right\} - 
\frac{u^{I}}{2}\left\{\psi_{H^{2}}^{2}\psi_{Q^{1}}^{I}\tilde{U}^{I} 
+ \psi_{H^{2}}^{2}\psi_{U}^{I}\tilde{Q}_{1}^{I} + h.c.\right\}   \nonumber \\
& & + 
\frac{1}{2}\lambda_{IJJ}^{\prime}
\left\{\psi_{L^{I}}^{1}\psi_{Q^{2}}^{J}\tilde{D}^{J} + 
\psi_{L^{I}}^{1}\psi_{D}^{J}\tilde{Q}_{2}^{J} + h.c.\right\} \; .
\label{Qqn-tcom}
\end{eqnarray}
After converting into four-component notation
straightforward and using the definition for neutralino mass eigenstates,
we have:
\begin{eqnarray}
{\cal L}_{\tilde{Q}q\kappa_{i}^{0}} &=& \kappa_{j}^{0}\left\{ 
\left[\frac{e}{\sqrt{2}\sin\theta_{W}
\cos\theta_{W}}Z_{U^{I}}^{i 1*}\left(\cos\theta_{W}
Z_{N}^{i 2} + \frac{1}{3}\sin\theta_{W}Z_{N}^{j 1}\right) - 
\frac{u^{I}}{2}Z_{U^{I}}^{i 1*}Z_{N}^{j 4*}
\right]P_{L}  \right. \nonumber \\
 & &+\left. \left[\frac{2\sqrt{2}}{3}g^{\prime}
Z_{U^{I}}^{i 2*}Z_{N}^{j 1} - 
\frac{u^{I}}{2}Z_{U^{I}}^{i 1*}Z_{N}^{j 4*}\right]P_{R}\right\}
\psi_{u^{I}}\tilde{U}_{I,i}^{-} + 
\bar{\kappa}_{j}^{0}\left\{\left[
\frac{e}{\sqrt{2}\sin\theta_{W}\cos\theta_{W}}
Z_{D^{I}}^{i 1} \right. \right. \nonumber \\
 & &\left. \left.\left(-\cos\theta_{W}Z_{N}^{i 2} + \frac{1}{3}
\sin\theta_{W}Z_{N}^{j 1}\right) +
\frac{d^{I}}{2}Z_{D^{I}}^{i 2}Z_{N}^{j 3} + 
\frac{1}{2}\lambda_{KIJ}Z_{D^{I}}^{i 2}Z_{N}^{j 4+K} \right]P_{L} \right. 
\nonumber \\
 & & + \left. \left[-\frac{\sqrt{2}}{3}g^{\prime} Z_{D^{I}}^{i 2}
Z_{N}^{j 1} + \frac{d^{I}}{2}Z_{D^{I}}^{i 1*}Z_{N}^{j 3*} + 
\frac{1}{2}\lambda_{KIJ}Z_{D^{I}}^{i 1*}Z_{N}^{j 4+K*}\right]P_{R} 
\right\}\psi_{d^{I}}\tilde{D}_{I i}^{+} + h.c. \;. 
\label{Qqn-fcom}
\end{eqnarray}
Thus the Feynman rules for the concerned interactions may 
be depicted exactly as the last two diagrams in Fig.\ \ref{fig9}.

\section{Various R-parity breaking models and the freedom for
redefining the superfields}

As stated at beginning, we are working in a 
very general model, where the R-parity is broken 
via various lepton-number violations with possible parameters 
explicitly and spontaneously, then an interesting problem 
is raised. Namely, we should realize possible freedom in
representing the model and should fix it properly. In fact, there
is some confusion on the freedom in literature.
In this section we focus the problem carefully.

If the superpotential and the soft SUSY breaking were 
switched off, the MSSM models would turn to have a $U(n+1)$ 
global symmetry, i.e. in the case the `down' type of
Higgs boson $H^{1}$ and the leptons chiral superfields $L^{I} 
(I=e, \mu, \tau; n=3)$ can be composed as a `vector', and under 
a transformation as $(H^{1}, L^{I}) \rightarrow U\cdot 
(H^{1}, L^{I})$, $U\in U(n+1=4)$, the theory would be invariant. 
Due to the invariance, the quantum numbers of leptons 
would become meaningless. Whereas the $U(4)$ symmetry is 
completely broken down when switching 
on the superpotential and the SUSY breaking terms as well, 
then two possibilities happen: a) If the superpotential and the 
soft SUSY breaking terms in the Lagrangian of the model 
conserve each lepton number respectively as a global symmetry, 
the lepton quantum numbers are fixed so the lepton numbers make 
senses. In fact this case is just the MSSM with R-parity. b)
If they are switched on, but break the lepton numbers, 
although the $U(4)$ symmetry is lost, instead a freedom to 
re-define the three lepton superfields and the down type
Higgs is raised in representing the model. Namely if all 
the terms in the superpotential and in the soft SUSY breaking 
terms undergo a $U(4)$ transformation accordingly, i.e. the 
$U(4)$ acts as redefining the lepton and Higgs superfields in the 
model, then superficially the VEVs of the sneutrinos, the mass 
matrices and the relevant couplings are changed accordingly, 
whereas the physics is not changed.
The MSSMs without R-parity may be constructed at beginning with 
very different parameters even very different assumptions naively, 
but they may be equivalent exactly i.e. they are just the same one 
of the models. Indeed in the case b) with such broken 
lepton-number superfields and complicated soft SUSY-breaking terms, 
it is not so straightforward to see the freedom, so
it is an important and non-trivial task. 
Let us examine the problem now.

To bare the task to realize the freedom for the MSSM 
without R-parity in mind, in the section we precisely show 
the equivalence of the models which are related to each other 
by $U(n+1)$ ($n$ is the family number with broken lepton
number) transformations. Namely the $U(n+1)$ transformations
can be understood as a freedom for re-defining the 
relevant superfields. Finally in this section
we propose two suggestions which may be considered as 
`conventions' for possible choices to fix the freedom.
Of the two, one is `to rotate away' the nonzero vacuum
expectation values (VEVs) of all the three generations 
of sneutrinos, this is emphasized by Refs.\cite{se6,s25,se1},
the other one is to `rotate away' the bilinear terms in the
superpotential\cite{s21,s5,s55,s6,s9}. 
Note that in the second case, in general,
the bilinear R-parity violation terms in soft breaking SUSY
terms so nonzero VEVs of the sneutrinos still may exist.

We would emphasize here that based on the whole effective 
Lagrangian we may compute the spectrum (mixing) of the content 
particles of the model precisely, and may have a global view
of the model too. Furthermore, having the precise Lagrangian 
one may easy connect the effective one to a more fundamental 
theory, thus we would not do the problem such as in the 
references\cite{s7} where from the very beginning only the 
`basis-indepedent' parameters are focused on so as to investigate 
the phenomenology of the model. Whereas in the present way, we 
should examine the freedom in defining the fields carefully, and 
make all the parameters being substantial. In fact, when all
possible terms (not only in superpotential but also
in SUSY soft-breaking and D-terms) are involved
the problem is not so transparent to realize the freedom.

Here we take the general case, that the three lepton-numbers 
are broken, as an example to examine the freedom. In fact, 
the freedom is $U(n+1=4)$ ($n$: the number of 
violated lepton-numbers) in defining the superfields 
within the extented MSSM as shown in the follows.
 
The $U(4)$ 
\begin{equation}
X^{\dagger}X = X X^{\dagger} = I,
\label{eq29}
\end{equation}
and $X\in U(4)$ is $4\times 4$ matrix, $I$ is the unit matrix. 
The $U(4)$ matrix `acting' on the `old' Higgs and lepton
superfields $\hat{H}^{1}, \hat{L}^{1}, \hat{L}^{2}, \hat{L}^{3}$ 
is defined as:
\begin{equation}
\left(
\begin{array}{c}
\hat{H}^{a^{1}} \\
\hat{L}^{a^{1}} \\
\hat{L}^{a^{2}} \\
\hat{L}^{a^{3}} \\  \end{array} \right) = X
\left(
\begin{array}{c}
\hat{H}^{1} \\
\hat{L}^{1} \\
\hat{L}^{2} \\
\hat{L}^{3} \\  \end{array} \right),
\label{eq30}
\end{equation}
where $\hat{H}^{a^{1}}, \hat{L}^{a^{1}}, \hat{L}^{a^{2}}, 
\hat{L}^{a^{3}}$ are `new' Higgs-lepton superfields.
As the D-terms in the model are invariant under the $U(4)$
transformation, so we need not consider them at all for
present purpose. As for the superpotential, 
having the $U(4)$ transformation performed, it 
turns into the following form accordingly
by means of the `new' superfields (with a upper-suffix
as in Eq.\ (\ref{eq30})):
\begin{eqnarray}
&W =& \varepsilon_{ij}\mu^{a}\hat{H}_{i}^{a^{1}}\hat{H}_{j}^{2} 
+ \varepsilon_{ij}
\epsilon_{I}^{a}\hat{H}_{i}^{2}\hat{L}_{j}^{a^{I}} + 
\varepsilon_{ij}h_{IJ}^{l}
\hat{H}_{i}^{a^{1}}\hat{L}_{j}^{a^{I}}\hat{R}^{J} \nonumber \\
&&- u_{I}(\hat{H}_{1}^{2}C^{JI*}
\hat{Q}_{2}^{J} - \hat{H}_{2}^{2}\hat{Q}_{1}^{J}\delta^{IJ})\hat{U}^{I} 
 \nonumber \\
& & -h_{IJ}^{d}(\hat{H}_{1}^{a^{1}}\hat{Q}_{2}^{J}\delta^{IJ} - 
\hat{H}_{2}^{a^{1}}
\hat{Q}_{1}^{J}C^{IJ})\hat{D}^{I} + \varepsilon_{ij}
\lambda_{IJK}^{a}\hat{L}_{i}^{a^{I}}
\hat{L}_{j}^{a^{J}}\hat{R}^{K}  \nonumber \\
 & & + \lambda_{IJK}^{a^{\prime}}(\hat{L}_{1}^{a^{K}}
\hat{Q}_{2}^{J}\delta{IJ} - 
\hat{L}_{2}^{a^{K}}\hat{Q}_{1}^{J}C^{IJ})\hat{D}^{I}
 + \lambda_{IJK}^{\prime\prime}\hat{U}^{I}\hat{D}^{J}\hat{D}^{K},
\label{eq31}
\end{eqnarray}
and the `new' coefficients are related to the `old' 
ones through the $U(4)$ matrix elements precisely:
\begin{eqnarray}
&h_{IJ}^{l} =& l_{J}(X_{11}^{*}X_{(I+1)(J+1)}^{*} - 
X_{(I+1)1}^{*}X_{1(J+1)}^{*})  \nonumber \\
& &+ \sum\limits_{KM}
\lambda_{KMJ}(X_{1(K+1)}^{*}X_{(I+1)(M+1)}^{*} \nonumber \\
&& - X_{(I+1)(K+1)}^{*}X_{1(M+1)}^{*}) , \label{eq32} \\
&h_{IJ}^{d} =& d_{J}X_{11}^{*} - 
\sum\limits_{K}\lambda_{KIJ}^{\prime}X_{1(K+1)}^{*} ,
\label{eq33} \\
&\lambda_{IJK}^{a} =& l_{K}X_{(I+1)1}^{*}X_{(J+1)(K+1)}^{*} \nonumber \\
&& + \sum\limits_{MN}\lambda_{MNK}
X_{(I+1)(M+1)}^{*}X_{(J+1)(N+1)}^{*} , \label{eq34} \\
&\lambda_{IJK}^{a^{\prime}} =& -d_{I}X_{(I+1)1}^{*} + \sum\limits_{M}
\lambda_{MKJ}^{\prime}
X_{(I+1)(M+1)}^{*}  ,\label{eq35} \\
&h_{I}^{a} =& l_{I}X_{11}^{*}X_{1(I+1)}^{*} + \sum\limits_{JK}
\lambda_{KJI}X_{1(K+1)}^{*}X_{1(J+1)}^{*}
, \label{eq36} \\
&\mu^{a} =& X_{11}^{*}\mu - \sum\limits_{I}\epsilon_{I}X_{1(I+1)}^{*} , 
\label{eq37} \\
&\epsilon_{I}^{a} =& -X_{(I+1)1}^{*}\mu + 
\sum\limits_{J}X_{(I+1)(J+1)}^{*}\epsilon_{J}. \label{eq38}
\end{eqnarray}
Here $X_{\alpha\beta}^* (\alpha, \beta=1, 2, 3, 4)$ are the 
complex conjugations of $X_{\alpha\beta}$, and $(I, J, K, 
M, N=1, 2, 3)$ always.
As the same way, the soft SUSY-breaking terms
in Lagrangian now correspondingly become:
\begin{eqnarray}
{\cal L}_{soft} &=& - M_{11}^{s^{2}}H_{i}^{a^{1*}}H_{i}^{a^{1}} - 
\sum\limits_{I}M_{1(I+1)}^{s^{2}}
(H_{i}^{a^{1*}}\tilde{L}_{i}^{a^{I}} + 
H_{i}^{a^{1}}\tilde{L}_{i}^{a^{I*}})  \nonumber  \\
 &  & - \sum\limits_{IJ}
M_{(I+1)(J+1)}^{s^{2}}\tilde{L}_{i}^{a^{I*}}\tilde{L}_{i}^{a^{J}} 
-m_{R^{I}}^{2}\tilde{R}^{I*}\tilde{R}^{I} -m_{Q^{I}}^{2} \tilde{Q}_{i}^{I*} 
\tilde{Q}_{i}^{I} \nonumber \\
&& - m_{D^{I}}^{2}
\tilde{D}
 ^{I*} \tilde{D}^{I} - m_{U^{I}}^{2}\tilde{U}^{I*} \tilde{U}^{I} + (m_{1}
\lambda_{B}\lambda_{B}+ m_{2}\lambda_{A}^{i}\lambda_{A}^{i}  \nonumber  \\
 &  &  + m_{3} \lambda_{G}^{a}\lambda_{
 G}^{a} + h.c.) + \{\varepsilon_{ij}B^{a} H_{i}^{a^{1}}H_{j}^{2} + 
\varepsilon_{ij}B_{I}^{a} H_{i}^{2}\tilde{L}_{j}^{a^{I}}  
 \nonumber \\
& &+\varepsilon_{ij}l_{s_{IJ}}^{a}H_{i}^{a^{1}}
\tilde{L}_{j}^{a^{I}}\tilde{R}^{J} 
- d_{s_{IJ}}^{a}(H_{1}^{a^{1}}\tilde{Q}_{2}^{J}\delta^{IJ} 
- H_{2}^{a^{1}}\tilde{
Q}_{1}C^{IJ})\tilde{D}^{I}   \nonumber \\
& & + u_{s_{I}}(-C^{KI*}H_{1}^{2}\tilde{Q}_{2}^{I} + 
H_{2}^{2}\tilde{Q}_{1}^{I})\tilde{U}^{I} + 
\varepsilon_{ij}\lambda_{IJK}^{s_{a}}
\tilde{L}_{i}^{a^{I}}\tilde{L}_{i}^{a^{J}}\tilde{R}^{K}  \nonumber \\
& & +\lambda_{KIJ}^{s_{a}^{\prime}}(\tilde{L}_{1}\tilde{Q}_{2}^{J}\delta^{IJ}
-\tilde{L}_{2}^{a^{K}}\tilde{Q}_{1}^{J}C^{IJ})\tilde{D}^{I} \nonumber \\
&&+\lambda_{IJK}^{s^{\prime\prime}}\tilde{U}^{I}\tilde{D}^{J}\tilde{D}^{K} 
+ h.c. \}. \label{39}
\end{eqnarray}
Here the 'new' soft breaking parameters are defined as
\begin{eqnarray}
&M_{s_{\alpha\beta}}^{2}=& m_{H^{1}}^{2}X_{\alpha 1}^{*}X_{\beta 1}^{*} +
\sum\limits_{I}m_{L^{I}}^{2}X_{\alpha (I+1)}^{*}X_{\beta (I+1)}^{*} + 
\sum\limits_{I\neq J}m_{L^{IJ}}^{2}X_{\alpha 
(I+1)}^{*}X_{\beta (J+1)}^{*}
\nonumber \\
&& + \sum\limits_{I}m_{HL^{I}}^{2}X_{\alpha 1}^{*}X_{\beta (I+1)}^{*}
, \label{eq40} \\
&B^{a} = & B X_{11}^{*} - 
\sum\limits_{I}B_{I}X_{1(I+1)}^{*},\label{eq41} \\
&B_{I}^{a} =& -B X_{(I+1)1}^{*} + \sum\limits_{J}B_{J}
X_{(I+1)(J+1)}^{*}, \label{eq42} \\
&l_{s_{MK}}^{a} =&\sum\limits_{s_{I}}(-X_{(M+1)1}^{*}X_{1(I+1)}^{*}+ 
X_{11}^{*}X_{(M+1)(I+1)}^{*})\delta_{IK} \nonumber \\
& & + \sum\limits_{IJ}\lambda_{IJK}^{s}(X_{(M+1)(I+1)}^{*}X_{1(J+1)}^{*} 
+ X_{1(I+1)}^{*}X_{(M+1)(J+1)}^{*}  , \label{eq43}  \\
& d_{s_{IJ}}^{a} =&d_{s_{I}}X_{11}^{*} - \sum\limits_{K}
\lambda_{KIJ}^{s^{\prime}}X_{1(K+1)}^{*} , \label{eq44}  \\
&\lambda_{IJK}^{s_{a}} = & l_{s_{K}}X_{(I+1)1}^{*}X_{(J+1)(K+1)}^{*} 
\nonumber \\
&& + \sum\limits_{MN}\lambda_{MNK}^{s}X_{(I+1)(M+
1)}^{*}X_{(J+1)(N+1)}^{*} , \label{eq45} \\
&\lambda_{KIJ}^{s_{a}^{\prime}} =& -d_{s_{I}}X_{(K+1)1}^{*} + 
\sum\limits_{M}\lambda_{MJI}^{s^{\prime}}X_{(K+1)(M+1)}^{*}, \label{eq46} \\
&h_{I}^{s_{a}} =& l_{s_{I}}X_{11}^{*}X_{1(I+1)}^{*} + 
\sum\limits_{JK}\lambda_{KJI}X_{1(K+1)}^{*}X_{1(J+1)}^{*}. \label{eq47}
\end{eqnarray}
Now, the $H^{a^{1}}$ and slepton acquire 
vacuum expectation values (VEVs):
\begin{equation}
H^{a^{1}}=
\left( 
\begin{array}{c}
\frac{1}{\sqrt{2}}(\chi_{1}^{a} + \upsilon_{1}^{a} + i\phi_{1}^{a}) \\
H_{2}^{a^{1}}  \end{array}  \right)
\label{eq48}
\end{equation}
\begin{equation}
H^{2}=
\left(
\begin{array}{c}
H_{1}^{2}  \\
\frac{1}{\sqrt{2}}(\chi_{2}^{0} + \upsilon_{2} + i\phi_{2}^{0})  
\end{array}  \right)
\label{eq49}\\[2mm]
\end{equation}
and

\begin{equation}
\tilde{L}^{a^{I}} = 
\left(
\begin{array}{c}
\frac{1}{\sqrt{2}}(\chi_{\tilde{\nu}_{I}}^{a} + 
\upsilon_{\tilde{\nu}_{I}}^{a} + 
i\phi_{\tilde{\nu}_{I}}^{a})  \\
\tilde{L}_{2}^{a^{I}}    \end{array}  \right)
\label{eq50}\\[2mm]
\end{equation}
where 
\begin{eqnarray}
\upsilon_{1}^{a} &=& X_{11}\upsilon_{1} + \sum\limits_{J}X_{1(J+1)}
\upsilon_{\tilde{\nu}_{J}} , \nonumber \\
\upsilon_{\tilde{\nu}_{I}}^{a} &=& X_{(I+1)1}\upsilon_{1} + 
\sum\limits_{J}X_{(I+1)(J+1)}\upsilon_{\tilde{\nu}_{J}}. \label{eq51}
\end{eqnarray}
As before by expanding the scalar potential, the tadpole terms become:
\begin{equation}
V_{tadpole} = t_{1}^{0}\chi_{1}^{0} + t_{2}^{0} \chi_{2}^{0} + 
t_{\tilde{\nu}_{1}}^{0} \chi_{\tilde{\nu}_{1}}^{0} + 
t_{\tilde{\nu}_{2}}^{0} \chi_{\tilde{\nu}_{2}}^{0} + 
t_{\tilde{\nu}_{3}}^{0} \chi_{\tilde{\nu}_{3}}^{0}
\label{eq52}
\end{equation}
where
\begin{eqnarray}
t_{1}^{a} &=& \frac{1}{8}(g^{2} + g^{'2})
\upsilon_{1}^{a}(\upsilon_{1}^{a^{2}} - 
\upsilon_{2}^{2} + \sum_{I}\upsilon_{\tilde{\nu}_{I}}^{a^{2}})
+ \mu^{2}\upsilon_{1}^{a} - B^{a} \upsilon_{2}  \nonumber \\
 & & -\sum_{I} \mu^{a} \epsilon_{I}^{a}\upsilon_{\tilde{\nu}_{I}}^{a}+
M_{s_{11}}^{2}\upsilon_{1}^{a} 
 + \sum\limits_{I}M_{s_{1(I+1)}}^{2}\upsilon_{\tilde{\nu}_{I}}^{a} 
\nonumber \\
t_{2}^{a} &=& -\frac{1}{8}(g^{2} + g^{'2})\upsilon_{2}
(\upsilon_{1}^{a^{2}} - 
\upsilon_{2}^{2} + \sum_{I}\upsilon_{\tilde{\nu}_{I}}^{a^{2}})
+ \mu^{a^{2}}\upsilon_{2} - B^{a} \upsilon_{1}^{a} \nonumber \\
 & &  + \sum_{I}\epsilon_{I}^{a^{2}}\upsilon_{2} 
+\sum_{I} B_{I}^{a}\upsilon_{\tilde{\nu}_{I}}^{a} + 
 m_{H^{2}}^{2}\upsilon_{2},    \nonumber \\
t_{\tilde{\nu}_{I}}^{a} &=& \frac{1}{8}(g^{2} + 
g^{\prime^{2}})\upsilon_{\tilde{\nu}_{I}}^{a}(\upsilon_{1}^{a^{2}} - 
\upsilon_{2}^{2} + 
\sum_{I}\upsilon_{\tilde{\nu}_{I}}^{a^{2}}) + 
\epsilon_{I}^{a}\sum_{J}\epsilon_{J}^{a}
\upsilon_{\tilde{\nu}_{J}}^{a}  \nonumber \\
 & & - \mu^{a}\epsilon_{I}^{a}\upsilon_{1}^{a} + 
B_{I}^{a}\upsilon_{2} + M_{s_{(I+1)(I+1)}}^{2}
\upsilon_{\tilde{\nu}_{I}}^{a} + 
M_{s_{1(I+1)}}^{2}\upsilon_{1}^{a} +  \nonumber \\
 & &\sum\limits_{J \neq I}M_{s_{(I+1)(J+1)}}^{2}
\upsilon_{\tilde{\nu}_{J}}^{a}.
\label{eq53}
\end{eqnarray}
Namely in the new basis $(\phi_{1}^{a}$, $\phi_{2}^{0}$, 
$\phi_{\tilde{\nu}_{1}}^{a}$,
 $\phi_{\tilde{\nu}_{2}}^{a}$, $\phi_{\tilde{\nu}_{3}}^{a})$, the 
mass matrix of the CP-odd Higgs becomes
\begin{equation}
M_{odd}^{a^{2}} =  \left(
\begin{array}{ccccc}
s_{11}^{a} & B^{a} & M_{s_{12}}^{2} - \mu^{a}\epsilon_{1}^{a} & 
M_{s_{13}}^{2} - \mu^{a}\epsilon_{2}^{a} & M_{s_{14}}^{2} - 
\mu^{a}\epsilon_{3}^{a} \\
B^{a} & s_{22}^{a} & -B_{1}^{a} & -B_{2}^{a} &
-B_{3}^{a} \\
M_{s_{12}}^{2} - \mu^{a}\epsilon_{1}^{a} & 
-B_{1}^{a} & s_{33}^{a} &
\epsilon_{1}^{a}\epsilon_{2}^{a} + M_{s_{23}}^{2} & 
\epsilon_{1}^{a}\epsilon_{3}^{a} 
+ M_{s_{24}}^{2} \\
M_{s_{13}}^{2} - \mu^{a}\epsilon_{2}^{a} & -B_{2}^{a} & 
\epsilon_{1}^{a}\epsilon_{2}^{a} + M_{s_{23}}^{2} & s_{44}^{a} & 
\epsilon_{2}^{a}\epsilon_{3}^{a} 
+ M_{s_{34}}^{2}  \\
M_{s_{14}}^{2} - \mu^{a}\epsilon_{3}^{a} & -B_{3}^{a} & 
\epsilon_{1}^{a}
\epsilon_{3}^{a} + M_{s_{24}}^{2} & \epsilon_{2}^{a}\epsilon_{3}^{a} + 
M_{s_{34}}^{2} & s_{55}^{a}
\end{array} \right).
\label{eq54}
\end{equation}
The definitions for the parameters:
\begin{eqnarray}
&&s_{11}^{a} = \frac{g^{2} + g^{\prime^{2}}}{8}(\upsilon_{1}^{a^{2}} - 
\upsilon_{2}^{2} + 
\sum_{I}\upsilon_{\tilde{\nu}_{I}}^{a^{2}}) + \mu^{a^{2}}
+ M_{s_{11}}^{2}  \nonumber \\
 &&= \sum_{I}\mu^{a}\epsilon_{I}^{a}\frac{\upsilon_{\tilde{\nu}_{I}}^{a}}
{\upsilon_{1}^{a}} + 
B^{a}\frac{\upsilon_{2}}{\upsilon_{1}^{a}} -\sum\limits_{I}
M_{s_{\small{1(I+1)}}}^{2}, \nonumber \\
&&s_{22}^{a} = -\frac{g^{2} + g^{\prime^{2}}}{8}(\upsilon_{1}^{a^{2}} - 
\upsilon_{2}^{2} + 
\upsilon_{\tilde{\nu}_{I}}^{a^{2}}) + \mu^{a^{2}} +
\sum_{I}\epsilon_{I}^{a^{2}} + m_{H^{2}}^{2}  \nonumber \\
 &&= B^{a}\frac{\upsilon_{1}^{a}}{\upsilon_{2}} - 
 \sum_{I}B_{I}^{a}\frac{\upsilon_{\tilde{\nu}_{I}}^{a}}
{\upsilon_{2}}, \nonumber \\
&&s_{\small{(I+2)(I+2)}}^{a} = \frac{g^{2} + 
g^{\prime^{2}}}{8}(\upsilon_{1}^{a^{2}} - 
\upsilon_{2}^{2} + \sum_{I}\upsilon_{\tilde{\nu}_{I}}^{a^{2}}) + 
\epsilon_{I}^{a^{2}} + M_{s_{\small{(I+1)(I+1)}}}^{2} \nonumber \\
& &= \mu^{a}\epsilon_{I}^{a}\frac{\upsilon_{1}^{a}}
{\upsilon_{\tilde{\nu}_{e}}^{a}} 
- B_{I}^{a}\frac{\upsilon_{2}}
{\upsilon_{\tilde{\nu}_{I}}^{a}}
- \sum\limits_{J \neq I}(\epsilon_{I}^{a}\epsilon_{J}^{a}
 + M_{s_{\small{(I+1)(J+1)}}}^{2})
\frac{\upsilon_{\tilde{\nu}_{J}}^{a}}{\upsilon_{\tilde{\nu}_{I}}^{a}}
 \;\; . \nonumber \\
\label{eq55}
\end{eqnarray}
Using Eqs.\ (\ref{eq32}) $\sim$ \ (\ref{eq38}) and Eqs.\ (\ref{eq40}) 
$\sim$ \ (\ref{eq47}) 
properly, we find 
\begin{equation}
M_{odd}^{a^{2}} = T_{1}^{-1} M_{odd}^{2} T_{1} 
\label{eq561}
\end{equation}
and the unitary matrix $T_1$ is defined:
\begin{equation}
T_{1} = \left(
\begin{array}{ccccc}
X_{11}^{*} & 0 & X_{21}^{*} & X_{31}^{*} & X_{41}^{*} \\
0 & 1 & 0 & 0 & 0 \\
X_{12}^{*} & 0 & X_{22}^{*} & X_{32}^{*} & X_{42}^{*} \\
X_{13}^{*} & 0 & X_{23}^{*} & X_{33}^{*} & X_{43}^{*} \\
X_{14}^{*} & 0 & X_{24}^{*} & X_{34}^{*} & X_{44}^{*}
\end{array} \right)\;\;.
\label{eq56}
\end{equation}
It is known from Eq.\ (\ref{eq561}) that
$M_{odd}^{a^{2}}$ and 
$M_{odd}^{2}$ have the same eigenvalues and the eigenstates of them
are related by ${\cal X}^\alpha=T_1 {\cal Y}^\alpha$, if ${\cal X}^\alpha$
presents an eigenstate of $M_{odd}^{2}$ and ${\cal Y}^\alpha$ presents
that of $M_{odd}^{a^{2}}$ for the same eigenvalue $m_\alpha^2$.
For the $CP$-even `Higgs', in the same way one can find 
$$M_{even}^{a^{2}} = T_{1}^{-1}M_{even}^{2}T_{1}$$
here $T_{1}$ is the same as that in Eq.\ (\ref{eq56}). So the same
conclusion on the `new' and `old' relation of the eigenvalues 
and the eigenstates as that in the $CP$-odd case is obtained.

As for the mass matrix of charged Higgs, the case is more complicated 
because the right handed sleptons should be included in the
mixing . In the basis 
$\Phi_{c}=(H_{2}^{a^{1*}}$, $H_{1}^{2}$, $\tilde{L}_{2}^{a^{1*}}$,  
$\tilde{L}_{2}^{a^{2*}}$, $\tilde{L}_{2}^{a^{3*}}$, $\tilde{R}^{1}$, 
$\tilde{R}^{2}$, $\tilde{R}^{3})$, one find:
\begin{equation}
M_{c}^{a^{2}} = T_{2}^{-1}M_{c}^{2} T_{2},
\label{eq61}
\end{equation}
and the unitary transformation matrix $T_{2}$ now is 
\begin{equation}
T_{2}= \left(
\begin{array}{cccccccc}
X_{11}^{*} & 0 & X_{21}^{*} & X_{31}^{*} & X_{41}^{*} & 0 & 0 & 0 \\
0 & 1 & 0 & 0 & 0 & 0 & 0 &0 \\
X_{12}^{*} & 0 & X_{22}^{*} & X_{32}^{*} & X_{42}^{*} & 0 & 0 & 0 \\
X_{13}^{*} & 0 & X_{23}^{*} & X_{33}^{*} & X_{43}^{*} & 0 & 0 & 0 \\
X_{14}^{*} & 0 & X_{24}^{*} & X_{34}^{*} & X_{44}^{*} & 0 & 0 & 0 \\
0 & 0 & 0 & 0 & 0 & 1 & 0 & 0 \\
0 & 0 & 0 & 0 & 0 & 0 & 1 & 0 \\
0 & 0 & 0 & 0 & 0 & 0 & 0 & 1 
\end{array}  \right)\; .
\label{eq62}
\end{equation}
The same conclusion is obtained as the above neutral Higgs cases.

Now, let us consider the neutral fermions. In the basis
$(\Phi^{0})^{T} = (-\lambda_{B}$, 
$-i\lambda_{A}^{3}$, $\psi_{H^{a^{1}}}^{1}$, 
$\psi_{H^{2}}^{2}$, $\nu_{e_{L}^{a}}$,
 $\nu_{\mu_{L}^{a}}$, $\nu_{\tau_{L}^{a}})$, 
the mass matrix of neutralino-neutrino is written as
\begin{equation}
{\cal M}_{N}^{a} = \left(
\begin{array}{ccccccc}
2m_{1} & 0 & -\frac{1}{2}g^{\prime}\upsilon_{1}^{a} & \frac{1}{2}g^{\prime}
\upsilon_{2} & -\frac{1}{2}g^{\prime}\upsilon_{\tilde{\nu}_{e}}^{a} &
-\frac{1}{2}g^{\prime}\upsilon_{\tilde{\nu}_{\mu}}^{a} & -\frac{1}{2}g^{\prime}
\upsilon_{\tilde{\nu}_{\tau}}^{a} \\
0 & 2m_{2} & \frac{1}{2}g\upsilon_{1}^{a} & -\frac{1}{2}g\upsilon_{2} & 
\frac{1}{2}g\upsilon_{\tilde{\nu}_{e}}^{a} 
& \frac{1}{2}g\upsilon_{\tilde{\nu}_{\mu}}^{a} & \frac{1}{2}g
\upsilon_{\tilde{\nu}_{\tau}}^{a} \\
-\frac{1}{2}g^{\prime}\upsilon_{1}^{a} & \frac{1}{2}g\upsilon_{1}^{a} & 0 & 
-\frac{1}{2}\mu^{a} & 0  & 0 & 0\\
\frac{1}{2}g^{\prime}\upsilon_{2} & -\frac{1}{2}g\upsilon_{2} 
& -\frac{1}{2}\mu^{a} & 0 & \frac{1}{2}\epsilon_{1}^{a} 
& \frac{1}{2}\epsilon_{2}^{a} & \frac{1}{2}\epsilon_{3}^{a}   \\
-\frac{1}{2}g^{\prime}\upsilon_{\tilde{\nu}_{e}}^{a} & \frac{1}{2}g
\upsilon_{\tilde{\nu}_{e}}^{a} & 0 & \frac{1}{2}\epsilon_{1}^{a} & 0 
& 0 & 0\\
-\frac{1}{2}g^{\prime}\upsilon_{\tilde{\nu}_{\mu}}^{a} & \frac{1}{2}g
\upsilon_{\tilde{\nu}_{\mu}}^{a} & 0 & \frac{1}{2}\epsilon_{2}^{a} & 0 & 
0 & 0\\
-\frac{1}{2}g^{\prime}\upsilon_{\tilde{\nu}_{\tau}}^{a} & \frac{1}{2}g
\upsilon_{\tilde{\nu}_{\tau}}^{a} & 0 & \frac{1}{2}\epsilon_{3}^{a} & 0 
& 0 & 0 
\end{array} \right),
\label{eq63}
\end{equation}
and we find 
\begin{equation}
M_{N}^{a} = T_{3}^{-1}M_{N}T_{3}.
\label{eq64}
\end{equation}
The $T_{3}$ unitary matrix is defined:
\begin{equation}
T_{3} = \left(
\begin{array}{ccccccc}
1 & 0 & 0 & 0 & 0 & 0 & 0 \\
0 & 1 & 0 & 0 & 0 & 0 & 0 \\
0 & 0 & X_{11}^{*} & 0 & X_{21}^{*} & X_{31}^{*} & X_{41}^{*} \\
0 & 0 & 0 & 1 & 0 & 0 & 0 \\
0 & 0 & X_{12}^{*} & 0 & X_{22}^{*} & X_{32}^{*} & X_{42}^{*} \\
0 & 0 & X_{13}^{*} & 0 & X_{23}^{*} & X_{33}^{*} & X_{43}^{*} \\
0 & 0 & X_{14}^{*} & 0 & X_{24}^{*} & X_{34}^{*} & X_{44}^{*} 
\end{array} \right).
\label{eq65}
\end{equation}
Once more the same conclusion on the relations of `new' and `old' eigenvalues
and eigenstates is reached.

As for the mixing of chargino-charged lepton, the mass matrix from the
Lagrangian now can be related as follows: 
\begin{equation}
{\cal M}_{\small{C}}^{a} = T_{4}^{-1}{\cal M}_{\small{C_{T}}} I
\label{eq66}
\end{equation}
with
\begin{equation}
T_{4} = \left(
\begin{array}{ccccc}
1 & 0 & 0 & 0 & 0 \\
0 & X_{11}^{*} & X_{21}^{*} & X_{31}^{*} & X_{41}^{*} \\
0 & X_{12}^{*} & X_{22}^{*} & X_{32}^{*} & X_{42}^{*} \\
0 & X_{13}^{*} & X_{23}^{*} & X_{33}^{*} & X_{43}^{*} \\
0 & X_{14}^{*} & X_{24}^{*} & X_{34}^{*} & X_{44}^{*} 
\end{array} \right)\; .
\label{eq67}
\end{equation}
Since charged fermions are considered here, 
the situation is a little complicated.
We need to make the mass matrix diagonal as the case of SM 
for quarks i.e. first to diagonalize the mass squared matrix
(the combination of the matrix and its conjugate), whereas, owing to
the relation Eq.\ (\ref{eq66}), the same conclusion can be obtained
too as the above.

Furthermore, it is easy to check the interaction terms are the same
no matter to start with what an `old' Lagrangian or a `new' Lagrangian:
as long as the vertices for the model all turn to represent by means of 
their eigenvalues (physical value) and corresponding eigenstates 
(physical states) coordinately, the equivalence for the interactions 
can be seen clearly. Therefore, the $U(4)$ transformation 
Eq.\ (\ref{eq30}) indeed is shown a freedom for
defining the superfields, and the problem
how to fix a model of R-parity violation MSSM emerges.
To solve this problem, we would like to suggest two
`conventions' for choices: a) with the freedom to
rotate away the VEVs for all sneutrinos; b) with the
freedom to rotate away all the bilinear terms of R-parity
violation in superpotential. Note that one can apply the freedom
only once that is to mean one can make either a) or b)
but cannot do both successfully in a general case.
In fact, besides the two choices we suggest here, there are many 
choices to fix the freedom. For instance, one may rotate 
part of the trilinear lepton-superfield terms or the
linear lepton-superfield terms which couple to the 
quark superfields properly in Eq.\ (\ref{eq-4}),
for specific convenience.

Now let us show the convention a) first: in fact,
Refs.\cite{se6,s25,se1} may be considered as the case. 
For convenience, let us define
the angles $\theta, \phi, \xi$ if all the VEVs are real\footnote{If 
the VEVs of sneutrinos are not real but complex values
i.e. there are spontaneous CP violation in the model through
the VEVs, the discussion here is still valid. Only the change
of the discussion is from a rotation into a 
unitary $U(4)$ transformation.}:
\begin{eqnarray}
&&\sin\theta=\frac{v_{\tilde\nu_e}}{\sqrt{v_1^2+
v_{\tilde\nu_e}^2}}\;, \nonumber \\
&&\sin\phi=\frac{v_{\tilde\nu_\mu}}{\sqrt{v_1^2+ 
v_{\tilde\nu_e}^2+v_{\tilde\nu_\mu}^2}}\; , \nonumber \\
&&\sin\xi=\frac{v_{\tilde\nu_\tau}}
{\sqrt{v_1^2+v_{\tilde\nu_e}^2+v_{\tilde\nu_\mu}^2+
v_{\tilde\nu_\tau}^2}} \; .
\label{bas1}
\end{eqnarray}
Indeed, anyone of the R-parity
violation MSSMs with nonzero VEVs of sneutrinos may be rotated
to the one where only the Higgs superfield 
$\hat {H}^1$ has nonzero VEV ($v_1\neq 0$; $v_{\tilde\nu_I}=0$,
with $I=e, \mu, \tau$). The
`rotation matrix' (here the $U(4)$ transformation `degenerates'
just to a rotation) can be decomposed into three rotations as below:
\begin{equation}
X=R_1 \cdot R_2 \cdot R_3 \;.
\label{rot1} 
\end{equation}
Here
$$ R_1= \left(
\begin{array}{cccc}
\cos\xi & 0 & 0 & \sin\xi\\
0 & 1 & 0 & 0 \\  
0 & 0 & 1 & 0 \\  
-\sin\xi & 0 & 0 & \cos\xi 
\end{array} \right) \; ,$$ 
$$ R_2= \left(
\begin{array}{cccc}
\cos\phi & 0 & \sin\phi & 0\\
0 & 1 & 0 & 0 \\  
-\sin\phi & 0 & \cos\phi & 0 \\
0 & 0 & 0 & 1 \\  
\end{array} \right)\; , $$
and 
$$ R_3= \left(
\begin{array}{cccc}
\cos\theta & \sin\theta& 0 & 0 \\
-\sin\theta & \cos\theta & 0 & 0 \\  
0 & 0 & 1 & 0 \\  
0 & 0 & 0 & 1 
\end{array} \right)\; . $$
It is easy further to check that
\begin{eqnarray}
H_{6}^{0} &=&\frac{1}{\upsilon}\Big( \upsilon_{d}\phi_{1}^{a^{0}} 
- \upsilon_{2}
\phi_{2}^{0} \Big) \nonumber \\
&=&\frac{1}{\upsilon}\Big(\upsilon_{1}\phi_{1}^{0} - 
\upsilon_{2}\phi_{2}^{0} + 
  \upsilon_{\tilde{\nu}_{e}}\phi_{\tilde{\nu}_{e}}^{0}
 + \upsilon_{\tilde{\nu}_{\mu}}\phi_{\tilde{\nu}_{\mu}}^{0} + 
 \upsilon_{\tilde{\nu}_{\tau}}\phi_{\tilde{\nu}_{\tau}}^{0}\Big) 
\hspace{1mm},
\end{eqnarray}
and
\begin{eqnarray}
H_{1}^{+}& = &\frac{1}{\upsilon}\Big( \upsilon_{d}H_{2}^{1^{a}} 
- \upsilon_{2}
H_{1}^{2*} \Big) \nonumber \\
& = &\frac{1}{\upsilon}(\upsilon_{1}H_{2}^{1*} - 
\upsilon_{2}H_{1}^{2} + 
\upsilon_{\tilde{\nu}_{e}}\tilde{L}_{2}^{1*} + 
\upsilon_{\tilde{\nu}_{\mu}}\tilde{L}_{2}^{2*} + 
\upsilon_{\tilde{\nu}_{\tau}}\tilde{L}_{2}^{3*}).
\end{eqnarray}
are just the Goldstones for spontaneously breaking the EW gauge symmetry.
The rest parts of the models can be checked without difficulty, but to
shorten the paper we will not show them here precisely.

The second `convention' b), which is suggested above, can be 
`realized' from anyone of the R-parity violation MSSMs by a proper 
rotation which is similar to the above, 
if the coefficients $\epsilon_I$ of the bilinear 
terms in the superpotential are real, otherwise an
according $U(4)$ transformation instead of the rotation 
to complete the purpose. For convenience in various application 
let us present the rotation precisely as below:
\begin{equation}
X'=R'_1 \cdot R'_2 \cdot R'_3 \;.
\label{rot2} 
\end{equation}
Here
$$ R'_1= \left(
\begin{array}{cccc}
\cos\xi' & 0 & 0 & \sin\xi'\\
0 & 1 & 0 & 0 \\  
0 & 0 & 1 & 0 \\  
-\sin\xi' & 0 & 0 & \cos\xi' 
\end{array} \right) \; ,$$ 
$$ R'_2= \left(
\begin{array}{cccc}
\cos\phi' & 0 & \sin\phi' & 0\\
0 & 1 & 0 & 0 \\  
-\sin\phi' & 0 & \cos\phi' & 0 \\
0 & 0 & 0 & 1
\end{array} \right)\; , $$
and 
$$ R'_3= \left(
\begin{array}{cccc}
\cos\theta'  & \sin\theta' & 0 & 0 \\
-\sin\theta' & \cos\theta' & 0 & ) \\ 
0 & 0 & 1 & 0 \\
0 & 0 & 0 & 1
\end{array} \right)\; , $$
with
\begin{eqnarray}
&&\sin\theta'=\frac{\mu }{\sqrt{\epsilon_1^2 +
\mu^2}}\;\;, 
\sin\phi'=\frac{\mu }{\sqrt{\epsilon_1^2 + 
\epsilon_2^2 + \mu^2}}\;, \nonumber \\
&&\sin\xi'=\frac{\mu}{\sqrt{\epsilon_1^2 + 
\epsilon_2^2 + \epsilon_3^2 + \mu^2}}\;.
\label{bas2}
\end{eqnarray}
After the rotation, in superpotential only the term
$\mu'\varepsilon_{ij}{\hat H}_i^1{\hat H}_j^2$
with $\mu'=
\sqrt{\mu^2+\epsilon_1^2+\epsilon_2^2
+\epsilon_3^2}$ is survived and the other
bilinear terms disappear totally.

Before closing this section, we would like to emphasize again: 
if one would like to compare different R-parity violation MSSMs
and to draw any definite conclusion, he must
fix the freedom in defining the four superfields (three leptons and the 
relevant Higgs which has the same quantum numbers as those
of leptons) first, and then carry on the comparisons. Otherwise, 
the obtained surface `differences' can be attributed to a 
different definition on the superfields totally or partly.
For convenience in applications and not only to fix the freedom
for redefining the superfields, we will further `simplify'
the parameterization in various ways elsewhere\cite{cchf}.

\section{Numerical results}

In this section, to be reference results for further studies,
we analyze the masses of neutral Higgs 
and charginos numerically and show their values 
in proper ways. We have obtained the mass matrices 
by setting the three type sneutrinos 
with non-zero vacuum expectation values and $\epsilon_{i} \neq 0$ $(i=1$,
$2$, $3)$. However, the matrices are quite big that may obscure the
typical features. To simplify the `problem' and to deduct
the parameters, we assume only those terms
which related the third generation (only $\tau$-lepton 
number) of lepton number is broken, but those
to the first two generations are not relevant
i.e. the terms relating to
the `first two generation lepton-numbers' disappear correspondly. 
Furthermore through fixing the freedom for redefining the 
superfields as discussed in the previous section,
for the `survived' trilinear terms relevant to the third 
generation leptons in superpotential and SUSY soft breaking 
terms, we will `rotate away' them as possible as one can, that
only the trilinear terms $\varepsilon_{ij}\lambda_{333}
\hat {L}^3_i \hat{L}^3_j\hat{R}^3$ in Eq.\ (\ref{eq-4}) 
and $\varepsilon_{ij}\lambda_{333} \tilde {L}^3_i
\tilde {L}^3_j\tilde {R}^3$ in Eq.\ (\ref{eq-5}) are kept. 
Namely in the Section for the numerical calculation, we restrict
ourselves to compute the case that the VEV of $\tau$-sneutrino
is nonzero, the bilinear terms relevant to $\tau$-family
lepton as well as the two trilinear terms 
$\varepsilon_{ij}\lambda_{333} \hat {L}^3_i \hat{L}^3_j\hat{R}^3$ 
and $\varepsilon_{ij}\lambda_{333} \tilde {L}^3_i
\tilde {L}^3_j\tilde {R}^3$ are present. 

Two reasons to make such an assumption that only
$\tau$-lepton number is violated:
\begin{itemize}
  \item    Under the assumption, we think the main feature will 
  not be lost too much but the mass matrices will turn much simple.
  \item    According to experimental indications, 
  the $\tau$-neutrino may be the heaviest among the three type
  neutrinos, and so far the constraints for the $\tau$ lepton
  rare decays are comparatively loose etc, i.e. the third
  generation of leptons probably are special.
\end{itemize}
In the numerical calculations below, input parameters are chosen as:
$\alpha = \frac{e^{2}}{4\pi}=\frac{1}{128}$, $M_{Z}=91.19$GeV,
$M_{W}=80.23$GeV, $m_{\tau}=1.77$GeV, but for the parameters $m_{1}$, $m_{2}$, 
we assume $m_{1} = m_{2} =250$GeV and the upper limit 
on $\tau$-neutrino mass $m_{\nu_{\tau}} \leq 10$MeV is taken into account 
seriously.

Now let us consider the masses of the charginos first, 
when $\epsilon_{1} = \epsilon_{2} =0$, and 
$\upsilon_{\tilde{\nu}_{e}} = 
\upsilon_{\tilde{\nu}_{\mu}}=0$, the Eq.\ (\ref{chargino-matrix}) becomes:
\begin{equation}
{\cal M}_{C} = \left(
\begin{array}{ccc}
2m_{2} & \frac{e\upsilon_{2}}{\sqrt{2}S_{W}} & 0  \\
\frac{e\upsilon_{1}}{\sqrt{2}S_{W}} & \mu  & 
\frac{l_{3}\upsilon_{\tilde{\nu}_{\tau}}}{\sqrt{2}} \\
\frac{e\upsilon_{\tilde{\nu}_{\tau}}}{\sqrt{2}S_{W}} & \epsilon_{3} & 
\frac{l_{3}\upsilon_{1}}{\sqrt{2}}+\frac{\lambda_{333}
\upsilon_{\tilde{\nu}_{\tau}}}{\sqrt{2}}
\end{array}  \right).
\label{rchargino-matrix}
\end{equation}
Because $m_{\tau}^{2}$ should be identified
as the lightest eigenvalue of the
matrix ${\cal M}_{C}^{\dag}{\cal M}_{C}$, we should take it 
as an eigenvalue away first so as not to conflict
the measurement of $\tau$ lepton mass.
After taking the eigenvalue $m_{\tau}^{2}$ 
away, the survived eigenvalue equation becomes:
\begin{equation}
\lambda^{2} - {\cal A}_{C}\lambda + {\cal B}_{C} = 0,
\end{equation}
and
\begin{eqnarray}
{\cal A}_{C} &=& X^{2} + Y^{2} + 4m_{2}^{2} + \frac{l_{3}^{2}\bigg(
\upsilon_{1}^{2} + \upsilon_{\tilde{\nu}_{\tau}}^{2}\bigg)
+ \lambda_{333}^{2}\upsilon_{\tilde{\nu}_{\tau}}^{2}}{2}
 + \frac{e^{2}\upsilon^{2}}{2S_{W}^{2}},  \nonumber \\
{\cal B}_{C} &=& \frac{2l_{3}^{2}}{m_{\tau}^{2}} 
\Bigg\{m_{2}\upsilon\cos\beta Y
\bigg(1+\frac{\lambda_{333}}{l_{3}}\sin\theta_{\upsilon}\bigg) \nonumber \\
&& +\frac{e^{2}}{4S_{W}^{2}}
\upsilon^{3}\cos^{2}\beta\sin\beta \bigg(\sin^{2}\theta_{\upsilon} - 
\cos^{2}\theta_{\upsilon}\bigg) \Bigg\}^{2},
\label{coeffi3}
\end{eqnarray}
with the parameters $X$, $Y$ are defined by
\begin{eqnarray}
X &=& \epsilon_{3}\cos\theta_{\upsilon} + \mu\sin\theta_{\upsilon}, 
\nonumber \\
Y &=& -\epsilon_{3}\sin\theta_{\upsilon} + \mu\cos\theta_{\upsilon} ,
\label{define-XY}
\end{eqnarray}
Therefore the masses of the other two charginos are expressed as:
\begin{equation}
m_{\kappa_{1,2}^{\pm}}^{2} = \frac{1}{2}\bigg\{ {\cal A}_{C} \mp 
\sqrt{{\cal A}_{C}^{2} - 4{\cal B}_{C}}\bigg\}.
\label{charginomass}
\end{equation}
The auxiliary parameter $l_{3}$ can be fixed by the condition that 
$Det|m_{\tau}^{2} - {\cal M}_{c}^{\dag}{\cal M}_{c}|=0$. 
When the values of $m_{1}$, $m_{2}$,
$\tan\beta$, $\tan\theta_{\upsilon}$ and Y are fixed, the value of X 
will be fixed by the mass of $\tau$-neutrino. Trying 
to take $m_{\nu_{\tau}} = 0.1$ MeV, we plot the mass of the
lightest charginos  versus with Y in Fig.\ \ref{fig10}. The two 
lines in the figure correspond to
$\lambda_{333}=0$ and  $\lambda_{333}=0.5$ respectively. 
In the figure, we
find that the trilinear effect on the chargino masses is small when 
$\tan\beta >> 1$ and $\tan\theta_{\upsilon} < 1$. In Fig.\ \ref{fig10}(c), 
the line correspond to $\lambda_{333}=0.5$ is coincide with the 
line for $\lambda_{333}=0.5$. When the
$\tan\beta \sim 1$ and $\tan\beta > 1$, the difference between 
$\lambda_{333}= 0.5$ and $\lambda_{333}=0$ is large. In the case,
$\sqrt{\upsilon_{1}^{2} + \upsilon_{\tilde{\nu_{\tau}}}^{2}} \sim
\upsilon_{2}$ and $\upsilon_{1} < \upsilon_{\tilde{\nu_{\tau}}}$ and the
effect of $\lambda_{333} \upsilon_{\tilde{\nu_{\tau}}}$ on the lightest
chargino mass cannot  be neglected. For comparison and considering the
results obtained at Super-K for neutrino oscillations, with a smaller 
neutrino mass $m_{\nu_{\tau}} = 10 $eV but the same parameters
being taken, we do the numerical calculation once more. The obtained
curves are different from those in Fig.\ \ref{fig10} by certain amount 
but not qualitatively and we plot them in Fig.\ \ref{fig11}.

Now, as for the mass-matrices of the neutral Higgs, under the same 
assumption, the one for CP-even Higgs is truncated to:
\begin{equation}
{\cal M}_{even}^{2} = \left(
\begin{array}{ccc}
r_{11} & -e_{12} - B  &  
e_{15} - \mu\epsilon_{3}  \\
-e_{12} - B & r_{22} & 
-e_{25} + B_{3}  \\
e_{15} - 
\mu\epsilon_{3} & -e_{25} + B_{3} & r_{33}
\end{array}
\right)
\label{rmatrix-even}
\end{equation}
with
\begin{eqnarray}
r_{11} &=& \frac{g^{2} + g^{\prime^{2}}}{8}(3\upsilon_{1}^{2} -
\upsilon_{2}^{2} 
+ \upsilon_{\tilde{\nu}_{\tau}}^{2}) +
|\mu|^{2} + m_{H^{1}}^{2}  \nonumber \\
 &=& \frac{g^{2}+g^{\prime^{2}}}{4}\upsilon_{1}^{2} + 
\bigg(\mu\epsilon_{3}-m_{HL^{3}}^{2}\bigg)
\frac{\upsilon_{\tilde{\nu}_{\tau}}}{\upsilon_{1}}
+ B\frac{\upsilon_{2}}{\upsilon_{1}} , \nonumber \\
r_{22} &=& \frac{g^{2} + g^{\prime^{2}}}{8}(-\upsilon_{1}^{2} + 
3\upsilon_{2}^{2} - \upsilon_{\tilde{\nu}_{\tau}}^{2}) +
|\mu|^{2} + |\epsilon_{3}|^{2} + m_{H^{2}}^{2} \nonumber \\
 &=& \frac{g^{2}+g^{\prime^{2}}}{4}\upsilon_{2}^{2} + 
B\frac{\upsilon_{1}}{\upsilon_{2}}
- B_{3}\frac{\upsilon_{\tilde{\nu}_{\tau}}}{\upsilon_{2}},  
\nonumber \\
r_{33} &=& \frac{g^{2} + g^{\prime^{2}}}{8}(\upsilon_{1}^{2} - \upsilon_{2}^{2} 
+ 3\upsilon_{\tilde{\nu}_{\tau}}^{2}) +
|\epsilon_{3}|^{2} + m_{L^{3}}^{2}  \nonumber \\
 &=& \frac{g^{2}+g^{\prime^{2}}}{4}\upsilon_{\tilde{\nu}_{\tau}}^{2} + 
\bigg(\mu\epsilon_{3}-m_{HL^{3}}^{2}\bigg)\frac{\upsilon_{1}}{\upsilon_{
\tilde{\nu}_{\tau}}}
- B_{3}\frac{\upsilon_{2}}{\upsilon_{\tilde{\nu}_{\tau}}}. 
\label{r123}
\end{eqnarray}
and $e_{12}$, $e_{15}$, $e_{25}$ are defined in Eq.\ (\ref{eij}).
Whereas the mass matrix of CP-odd Higgs is truncated to:
\begin{equation}
{\cal M}_{odd}^{2} =
\left(
\begin{array}{ccc}
s_{11} & B & -\mu\epsilon_{3}+m_{HL^{3}}^{2}  \\
B & s_{22} & -B_{3} \\
-\mu\epsilon_{3}+m_{HL^{3}}^{2} & -B_{3} & s_{33}
\end{array}
\right)  \label{rmassodd}
\end{equation}
with
\begin{eqnarray}
s_{11} &=& \frac{g^{2} + g^{\prime^{2}}}{8}(\upsilon_{1}^{2} -
\upsilon_{2}^{2} 
+ \upsilon_{\tilde{\nu}_{\tau}}^{2}) + |\mu|^{2}
+ m_{H^{1}}^{2}  \nonumber \\
 &=& \bigg(\mu\epsilon_{3}-m_{HL^{3}}^{2}\bigg)
\frac{\upsilon_{\tilde{\nu}_{\tau}}}{\upsilon_{1}} + 
B\frac{\upsilon_{2}}{\upsilon_{1}}, \nonumber \\
s_{22} &=& -\frac{g^{2} + g^{\prime^{2}}}{8}(\upsilon_{1}^{2} - \upsilon_{2}^{2} 
+ \upsilon_{\tilde{\nu}_{\tau}}^{2}) + |\mu|^{2}
+|\epsilon_{3}|^{2} + m_{H^{2}}^{2}  \nonumber \\
 &=& B\frac{\upsilon_{1}}{\upsilon_{2}} - 
B_{3}\frac{\upsilon_{\tilde{\nu}_{\tau}}}{\upsilon_{2}}, \nonumber 
\\
s_{33} &=& \frac{g^{2} + g^{\prime^{2}}}{8}(\upsilon_{1}^{2} - \upsilon_{2}^{2} 
+ \upsilon_{\tilde{\nu}_{\tau}}^{2}) + 
|\epsilon_{3}|^{2} + m_{L^{3}}^{2} \nonumber \\
 &=& \bigg(\mu\epsilon_{3}-m_{HL^{3}}^{2}\bigg)
\frac{\upsilon_{1}}{\upsilon_{\tilde{\nu}_{\tau}}} - 
B_{3}\frac{\upsilon_{2}}{\upsilon_{
\tilde{\nu}_{\tau}}}.
\label{s123}
\end{eqnarray}
Introducing the following auxiliary variables:
\begin{eqnarray}
X_{s} &=& B ,\nonumber \\
Y_{s} &=& \mu\epsilon_{3}-m_{HL^{3}}^{2},\nonumber \\
Z_{s} &=& B_{3},
\label{xyzs}
\end{eqnarray}
the masses of the neutral Higgs can be expressed by the parameters
$X_{s}$, $Y_{s}$, $Z_{s}$
and $\tan\beta$, $\tan\theta_{\upsilon}$.
For the masses of CP-odd Higgs,
the masses of the two CP-odd Higgs are given by
\begin{equation}
m_{\small{H_{3+2,3}^{0}}}^{2} = \frac{1}{2}({\cal A} \mp \sqrt{{\cal A}^{2} - 4
{\cal  B}} )
\label{massodd1}
\end{equation}
if we define:
\begin{eqnarray}
{\cal A} &=& X_{s} (\frac{\upsilon_{1}}{\upsilon_{2}} + 
\frac{\upsilon_{2}}{\upsilon_{1}}) + Y_{s} (\frac{
\upsilon_{1}}{\upsilon_{\tilde{\nu}_{\tau}}} + 
\frac{\upsilon_{\tilde{\nu}_{\tau}}}{\upsilon_{1}}) - Z_{s}(\frac{\upsilon_{2}}{
\upsilon_{\tilde{\nu}_{\tau}}} + 
\frac{\upsilon_{\tilde{\nu}_{\tau}}}{\upsilon_{2}})  ,\nonumber \\
{\cal B} &=& - Y_{s}Z_{s} (\frac{\upsilon_{1}}{\upsilon_{2}} + 
\frac{\upsilon_{2}}{\upsilon_{1}}) -
X_{s}Z_{s}(\frac{\upsilon_{1}}{\upsilon_{\tilde{\nu}_{\tau}}} + 
\frac{\upsilon_{\tilde{\nu}_{\tau}}}{\upsilon_{1}}) + X_{s}Y_{s}
(\frac{\upsilon_{2}}{\upsilon_{\tilde{\nu}_{\tau}}} + 
\frac{\upsilon_{\tilde{\nu}_{\tau}}}{\upsilon_{2}}) + \nonumber \\
 & & 
X_{s}Y_{s}\frac{\upsilon_{1}^{2}}{\upsilon_{2}\upsilon_{\tilde{\nu}_{\tau}}} - 
X_{s}Z_{s}\frac{\upsilon_{
2}^{2}}{\upsilon_{1}\upsilon_{\tilde{\nu}_{\tau}}} -
Y_{s}Z_{s} \frac{\upsilon_{\tilde{\nu}_{\tau}}^{2}}{\upsilon_{1}
\upsilon_{2}}.
\label{calab}
\end{eqnarray}
In the numerical calculation, we have taken the parameter 
$\sqrt{|X_{s}|} = 500$GeV. In Fig.\ \ref{fig12}, we plot the 
mass of the lightest CP-even Higgs versus the parameter 
$\sqrt{|Y_{s}|}$. The three lines correspond to 
$\sqrt{|Z_{s}|}=60$GeV, $150$GeV and $250$GeV respectively.
From the Fig.\ \ref{fig12}, we find the mass of the lightest 
CP-even Higgs turns small when the parameter $\sqrt{|Y_{s}|}$
turns large. In Fig.\ \ref{fig13}, we plot the 
mass of the lightest CP-even Higgs versus the parameter 
$\sqrt{|Z_{s}|}$. The three lines correspond to 
$\sqrt{|Y_{s}|}=60$GeV, $300$GeV and $400$GeV respectively.
From the Fig.\ \ref{fig13}, we find the mass of the lightest
Higgs turns large, as the parameter $\sqrt{|Z_{s}|}$ changes
large. From the numerical calculations, we can find
certain parameter space that at tree level the lightest Higgs 
mass can be $m_{H_{1}^{0}} \geq 132$GeV, thus for
the supersymmetry model without R-parity one cannot obtain 
such a stringent limit on the lightest Higgs mass as that in the 
MSSM with R-parity. 

As shown above, the results obtained by our numerical and formulation 
analysis, both confirm the difference from the MSSM with R-parity: 
the upper bound of the lightest CP-even Higgs mass of the MSSM without
R-parity is loosened a lot. 

In summary, besides the formal analysis and clarifying
the confusion on the freedom for the redefining the fields,
with the assumption that only $\tau$-lepton number is broken, we have
calculated the mass spectra in the MSSM without R-parity numerically.
From the restriction on the neutrino mass: even 
$m_{\nu_{\tau}} \leq 10$ eV, we cannot rule out the possibilities 
with large $\epsilon_{3}$. The Feynman rules have been derived in the
$\prime$t-Hooft Feynman gauge which are convenient when studying the
phenomenology beyond tree level of the model. Here, we would like 
to point out some references have analyzed the $0\nu \beta\beta$-decay
in the model\cite{s22} and may obtain certain new constraints about
the upper limits on the first generation R-parity violating parameters, 
such as $\epsilon_{1}$ and $\upsilon_{\tilde{\nu}_{e}}$; 
whereas for the other two generations, there are no such 
serious restrictions on the R-parity violating parameters.

\vspace{20mm}
{\Large\bf Acknowledgment} We would like to thank Prof. T. Han and 
Prof. Z.-Y. Zhao for valuable discussions on the topics of the paper. 
One (C.-H. C) of the authors would like to thank Prof. Edmond L. Berger
for warm hospitality at ANL (Argonne) and some suggestions
during his visit. This work was
supported in part by the National Natural Science 
Foundation of China and the Grant No. LWLZ-1298 
of the Chinese Academy of Sciences.

\appendix 

\section{The parameters\label{01}}

\subsection{The parameters appearing in
the mass matrix for CP-even Higgs\label{app}}

The parameters appearing in the matrix elements are defined 
as follows:
\begin{eqnarray}
&&r_{11} = \frac{g^{2} + g^{\prime^{2}}}{8}\bigg(3\upsilon_{1}^{2} - 
\upsilon_{2}^{2} + \sum_{I}\upsilon_{\tilde{\nu}_{I}}^{2}\bigg) +
|\mu|^{2} + m_{H^{1}}^{2}  \nonumber \\
&&\hspace{2mm} = \frac{g^{2} + g^{\prime^{2}}}{4}\upsilon_{1}^{2} + 
 \sum_{I}\bigg(\mu\epsilon_{I} - m_{HL^{I}}^{2}\bigg)
\frac{\upsilon_{\tilde{\nu}_{I}}}{\upsilon_{1}}
+ B\frac{\upsilon_{2}}{\upsilon_{1}},  \nonumber \\
&&r_{22} = \frac{g^{2} + g^{\prime^{2}}}{8}\bigg(-\upsilon_{1}^{2} + 
3\upsilon_{2}^{2} - \sum_{I}\upsilon_{\tilde{\nu}_{I}}^{2}\bigg) +
|\mu|^{2} + \sum_{I}\epsilon_{I}^{2} + m_{H^{2}}^{2} \nonumber \\
&&\hspace{2mm} = \frac{g^{2} + g^{\prime^{2}}}{4}\upsilon_{2}^{2} + 
B\frac{\upsilon_{1}}{\upsilon_{2}}- 
\sum_{I}B_{I}
\frac{\upsilon_{\tilde{\nu}_{I}}}{\upsilon_{2}},  \nonumber \\
&&r_{\small{(I+2)(I+2)}} = \frac{g^{2} +g^{\prime^{2}}}{8}\bigg(\upsilon_{1}^{2} 
-  \upsilon_{2}^{2} +
\sum_{J}\upsilon_{\tilde{\nu}_{J}}^{2} + 2
\upsilon_{\tilde{\nu}_{I}}^{2}\bigg) +  \epsilon_{I}^{2} + m_{L^{I}}^{2} 
\nonumber \\  
&&\hspace{17mm}= \frac{g^{2} +
g^{\prime^{2}}}{4}\upsilon_{\tilde{\nu}_{I}}^{2} + 
\bigg(\mu\epsilon_{I}-m_{HL^{I}}^{2}\bigg)\frac{\upsilon_{1}}
{\upsilon_{\tilde{\nu}_{I}}} -
B_{I}\frac{\upsilon_{2}}{\upsilon_{\tilde{\nu}_{I}}} \nonumber \\
&&\hspace{17mm}-\sum\limits_{J\neq I}\bigg(\epsilon_{I}\epsilon_{J}  +
m_{L^{IJ}}^{2}\bigg)\frac{\upsilon_{\tilde{\nu}_{J}}}{
\upsilon_{\tilde{\nu}_{I}}} \label{r12345}
\end{eqnarray}
and
\begin{eqnarray}
&&e_{12} = \frac{g^{2} + 
g^{\prime^{2}}}{4}\upsilon_{1}\upsilon_{2}, \hspace{25mm}
e_{13} = \frac{g^{2} + 
g^{\prime^{2}}}{4}\upsilon_{1}\upsilon_{\tilde{\nu}_{e}}
+ m_{HL^{1}}^{2}, \nonumber \\
&&e_{14} = \frac{g^{2} + 
g^{\prime^{2}}}{4}\upsilon_{1}\upsilon_{\tilde{\nu}_{\mu}}
+ m_{HL^{2}}^{2}, \hspace{10mm}
e_{15} = \frac{g^{2} + 
g^{\prime^{2}}}{4}\upsilon_{1}\upsilon_{\tilde{\nu}_{\tau}}
+ m_{HL^{3}}^{2}, \nonumber \\
&&e_{23} = \frac{g^{2} + 
g^{\prime^{2}}}{4}\upsilon_{2}\upsilon_{\tilde{\nu}_{e}} ,\hspace{25mm}
e_{24} = \frac{g^{2} + 
g^{\prime^{2}}}{4}\upsilon_{2}\upsilon_{\tilde{\nu}_{\mu}} ,\nonumber \\
&&e_{25} = \frac{g^{2} + 
g^{\prime^{2}}}{4}\upsilon_{2}\upsilon_{\tilde{\nu}_{\tau}} ,\hspace{25mm}
e_{34} = \frac{g^{2} + g^{\prime^{2}}}{4}\upsilon_{\tilde{\nu}_{e}}
\upsilon_{\tilde{\nu}_{\mu}}+ m_{L^{12}}^{2} ,\nonumber \\
&&e_{35} = \frac{g^{2} + g^{\prime^{2}}}{4}\upsilon_{\tilde{\nu}_{e}}
\upsilon_{\tilde{\nu}_{\tau}}+ m_{L^{13}}^{2} ,\hspace{10mm}
e_{45} = \frac{g^{2} + g^{\prime^{2}}}{4}\upsilon_{\tilde{\nu}_{\mu}}
\upsilon_{\tilde{\nu}_{\tau}}+ m_{L^{23}}^{2} .
\label{eij}
\end{eqnarray}

\subsection{\label{app0}The parameters appearing in
the mass matrix for CP-odd Higgs}

The parameters appearing in the matrix elements are defined 
as follows:
\begin{eqnarray}
s_{11} &=& \frac{g^{2} + g^{\prime^{2}}}{8}\bigg(\upsilon_{1}^{2} - 
\upsilon_{2}^{2} + \sum_{I}\upsilon_{\tilde{\nu}_{I}}^{2}\bigg) + \mu^{2}
+ m_{H^{1}}^{2}  \nonumber \\
 &=& \sum_{I}\bigg(\mu\epsilon_{I}+m_{HL^{I}}^{2}\bigg)
\frac{\upsilon_{\tilde{\nu}_{I}}}
 {\upsilon_{1}} + B\frac{\upsilon_{2}}{\upsilon_{1}}
, \nonumber \\
s_{22} &=& -\frac{g^{2} + g^{\prime^{2}}}{8}\bigg(\upsilon_{1}^{2} - 
\upsilon_{2}^{2} + \upsilon_{\tilde{\nu}_{I}}^{2}\bigg) + \mu^{2} +
\sum_{I}\epsilon_{I}^{2} + m_{H^{2}}^{2}  \nonumber \\
 &=& B\frac{\upsilon_{1}}{\upsilon_{2}} - 
\sum_{I}B_{I}\frac{\upsilon_{\tilde{\nu}_{I}}}
{\upsilon_{2}} ,\nonumber \\
s_{(I+2)(I+2)} &=& \frac{g^{2} + g^{\prime^{2}}}{8}\bigg(\upsilon_{1}^{2} 
- \upsilon_{2}^{2} + \sum_{I}\upsilon_{\tilde{\nu}_{I}}^{2}\bigg) + 
\epsilon_{I}^{2} + m_{L^{I}}^{2} \nonumber \\
 &=& \bigg(\mu\epsilon_{I}-m_{HL^{I}}^{2}\bigg)
\frac{\upsilon_{1}}{\upsilon_{\tilde{\nu}_{I}}} 
 - B_{I}\frac{\upsilon_{2}}{\upsilon_{\tilde{\nu}_{I}}}
- \sum\limits_{J\neq I}\bigg(\epsilon_{I}\epsilon_{J} + m_{L^{IJ}}^{2}\bigg)
\frac{\upsilon_{\tilde{\nu}_{J}}}
{\upsilon_{\tilde{\nu}_{I}}} \;.
\label{s12345}
\end{eqnarray}

\subsection{\label{app1}The elements of the charged Higgs mass-matrix}

With the interaction basis $\Phi_{c}=(H_{2}^{1*}$, $H_{1}^{2}$,
$\tilde{L}_{2}^{1*}$, $\tilde{L}_{2}^{2*}$, $\tilde{L}_{2}^{3*}$,
$\tilde{R}^{1}$, $\tilde{R}^{2}$, $\tilde{R}^{3})$ and Eq.\ (\ref{eq-6}),
the elements of symmetric mass-matrix ${\cal M}_{c}^{2}$ 
for the charged Higgs appearing in Eq.\ ( \ref{eq-20}) may be 
individually rwitten as follows:
\begin{eqnarray}
{\cal M}_{c 1,1}^{2} & = & \frac{g^{2}}{4}\upsilon_{1}^{2} - 
\frac{g^{2}-g'^{2}}{8}(
\upsilon_{1}^{2}-\upsilon_{2}^{2} + \sum_{I}\upsilon_{\tilde{\nu}_{I}}^{2}) + 
\mu^{2} + 
\sum_{I}\frac{1}{2}l_{I}^{2}\upsilon_{\tilde{\nu}_{I}}^{2} + m_{H^{1}}^{2}  
\nonumber  \\
   & = & 
\frac{g^{2}}{4}(\upsilon_{2}^{2}-\sum_{I}\upsilon_{\tilde{\nu}_{I}}^{2}) + 
   \sum_{I}\frac{1}{2}l_{I}^{2}\upsilon_{\tilde{\nu}_{I}}^{2}
 -\sum_{I}(\mu\epsilon_{\tilde{\nu}_{I}}-m_{HL^{I}}^{2})
\frac{\upsilon_{\tilde{\nu}_{I}}}
 {\upsilon_{1}}+B\frac{\upsilon_{2}}{\upsilon_{1}}, \nonumber  \\
{\cal M}_{c 1,2}^{2} &=& \frac{g^{2}}{4}\upsilon_{1}\upsilon_{2} + B , 
\nonumber \\
{\cal M}_{c 1,3}^{2} &=& \frac{g^{2}}{4}\upsilon_{1}\upsilon_{\tilde{\nu}_{e}} -
\mu\epsilon_{1}+m_{HL^{1}}^{2}
-\frac{1}{2}l_{1}^{2}\upsilon_{1}\upsilon_{\tilde{\nu}_{e}} - 
\frac{1}{2}\sum\limits_{IJ}(
\lambda_{J1I}-\lambda_{1JI})\upsilon_{\tilde{\nu}_{I}}\upsilon_{\tilde{\nu}_{J}}
, \nonumber \\
{\cal M}_{c 1,4}^{2} &=& \frac{g^{2}}{4}\upsilon_{1}\upsilon_{\tilde{\nu}_{\mu}} -
\mu\epsilon_{2}+m_{HL^{2}}^{2}
-\frac{1}{2}l_{2}^{2}\upsilon_{1}\upsilon_{\tilde{\nu}_{\mu}}- 
\frac{1}{2}\sum\limits_{IJ}(
\lambda_{J2I}-\lambda_{2JI})\upsilon_{\tilde{\nu}_{I}}\upsilon_{\tilde{\nu}_{J}}
, \nonumber \\
{\cal M}_{c 1,5}^{2} &=& \frac{g^{2}}{4}\upsilon_{1}\upsilon_{\tilde{\nu}_{\tau}} 
-\mu\epsilon_{3}+m_{HL^{3}}^{2}
-\frac{1}{2}l_{3}^{2}\upsilon_{1}\upsilon_{\tilde{\nu}_{\tau}}- 
\frac{1}{2}\sum\limits_{IJ}(
\lambda_{J3I}-\lambda_{3JI})\upsilon_{\tilde{\nu}_{I}}\upsilon_{\tilde{\nu}_{J}}
, \nonumber \\
{\cal M}_{c 1,6}^{2} &=& \frac{1}{\sqrt{2}}l_{1}\epsilon_{1}\upsilon_{2} +
l_{s_{1}}\frac{\upsilon_{\tilde{\nu}_{e}}}{\sqrt{2}} , \nonumber \\
{\cal M}_{c 1,7}^{2} &=& \frac{1}{\sqrt{2}}l_{2}\epsilon_{2}\upsilon_{2} +
l_{s_{2}}\frac{\upsilon_{\tilde{\nu}_{\mu}}}{\sqrt{2}} , \nonumber \\
{\cal M}_{c 1,8}^{2} &=& \frac{1}{\sqrt{2}}l_{3}\epsilon_{3}\upsilon_{2} +
l_{s_{3}}\frac{\upsilon_{\tilde{\nu}_{\tau}}}{\sqrt{2}} , \nonumber \\
{\cal M}_{c 2,2}^{2} &=& \frac{g^{2}}{4}\upsilon_{2}^{2} + 
\frac{1}{8}(g^{2}-g'^{2})
(\upsilon_{1}^{2}-\upsilon_{2}^{2}+\sum_{I}\upsilon_{\tilde{\nu}_{I}}^{2}) + 
\mu^{2} +
\sum_{I}\epsilon_{I}^{2} + m_{H^{2}}^{2}   \nonumber  \\
 & = & \frac{g^{2}}{4}(\upsilon_{1}^{2}+\sum_{I}\upsilon_{\tilde{\nu}_{I}}^{2}) 
- 
\sum_{I}B_{I}\frac{\upsilon_{\tilde{\nu}_{I}}}{\upsilon_{2}} + 
B\frac{\upsilon_{1}}{\upsilon_{2}}, \nonumber \\
{\cal M}_{c 2,3}^{2} &=&
\frac{g^{2}}{4}\upsilon_{2}\upsilon_{\tilde{\nu}_{e}} - 
B_{1} ,\nonumber  \\
{\cal M}_{c 2,4}^{2} &=&
\frac{g^{2}}{4}\upsilon_{2}\upsilon_{\tilde{\nu}_{\mu}} - 
B_{2} ,\nonumber  \\
{\cal M}_{c 2,5}^{2} &=&
\frac{g^{2}}{4}\upsilon_{2}\upsilon_{\tilde{\nu}_{\tau}} - 
B_{3} ,\nonumber  \\
{\cal M}_{c 2,6}^{2} &=& \frac{l_{1}}{\sqrt{2}}\mu\upsilon_{\tilde{\nu}_{e}} + 
\frac{
l_{1}}{\sqrt{2}}\epsilon_{1}\upsilon_{1} - 
\frac{1}{\sqrt{2}}\sum\limits_{IJ}\epsilon_{I}
\lambda_{IJ1}\epsilon_{\tilde{\nu}_{J}} ,\nonumber \\
{\cal M}_{c 2,7}^{2} &=& \frac{l_{2}}{\sqrt{2}}\mu\upsilon_{\tilde{\nu}_{\mu}} + 
\frac{
l_{2}}{\sqrt{2}}\epsilon_{2}\upsilon_{1}- 
\frac{1}{\sqrt{2}}\sum\limits_{IJ}\epsilon_{I}
\lambda_{IJ2}\epsilon_{\tilde{\nu}_{J}} ,\nonumber \\
{\cal M}_{c 2,8}^{2} &=& \frac{l_{3}}{\sqrt{2}}\mu\upsilon_{\tilde{\nu}_{\tau}} + 
\frac{
l_{3}}{\sqrt{2}}\epsilon_{3}\upsilon_{1}- 
\frac{1}{\sqrt{2}}\sum\limits_{IJ}\epsilon_{I}
\lambda_{IJ3}\epsilon_{\tilde{\nu}_{J}} ,\nonumber \\
{\cal M}_{c 3,3}^{2} &=& \frac{g^{2}}{4}\upsilon_{\tilde{\nu}_{e}}^{2} - 
\frac{1}{8}(g^{2}-g'^{2})(\upsilon_{1}^{2}-\upsilon_{2}^{2}+\sum_{I}
\upsilon_{\tilde{\nu}_{I}}^{2}) +\epsilon_{1}^{2} + 
\frac{l_{1}^{2}}{2}\upsilon_{1}^{2} + m_{L^{1}}^{2}  \nonumber \\
 & & +\sum\limits_{J}l_{1}(\lambda_{J11}-\lambda_{1J1})\upsilon_{1}
\upsilon_{\tilde{\nu}_{J}} +\frac{1}{2}\sum_{IJM}(\lambda_{J1I} - 
\lambda_{1JI})(
\lambda_{M1I} - \lambda_{1MI})\upsilon_{\tilde{\nu}_{J}}
\upsilon_{\tilde{\nu}_{M}}  \nonumber \\
 & = & \frac{g^{2}}{4}(\upsilon_{2}^{2}-\upsilon_{1}^{2})+
(\mu\epsilon_{1}-m_{HL^{1}})\frac{\upsilon_{1}}{
\upsilon_{\tilde{\nu}_{e}}} - 
B_{1}\frac{\upsilon_{2}}{\upsilon_{\tilde{\nu}_{e}}} + 
\frac{l_{1}^{2}}{2}\upsilon_{1}^{2}  \nonumber \\
 & & -(\epsilon_{1}\epsilon_{2}+m_{L^{12}}^{2})
\frac{\upsilon_{\tilde{\nu}_{\mu}}}
 {\upsilon_{\tilde{\nu}_{e}}} - 
(\epsilon_{1}\epsilon_{3}+m_{L^{13}}^{2})\frac{\upsilon_{\tilde{\nu}_{\tau}}}
{\upsilon_{\tilde{\nu}_{e}}} - 
\frac{g^{2}}{4}(\upsilon_{\tilde{\nu}_{\mu}}^{2}+
\upsilon_{\tilde{\nu}_{\tau}}^{2})  \nonumber \\
 & & + \sum\limits_{J}l_{1}(\lambda_{J11}-\lambda_{1J1})\upsilon_{1}
\upsilon_{\tilde{\nu}_{J}} +\frac{1}{2}\sum_{IJM}(\lambda_{J1I} - 
\lambda_{1JI})(
\lambda_{M1I} - \lambda_{1MI})\upsilon_{\tilde{\nu}_{J}}
\upsilon_{\tilde{\nu}_{M}} , \nonumber \\
{\cal M}_{c 3,4}^{2} &=& \frac{g^{2}}{4}\upsilon_{\tilde{\nu}_{e}}
\upsilon_{\tilde{\nu}_{\mu}} + \epsilon_{1}\epsilon_{2} + \frac{1}{2}
\sum\limits_{IJM}(\lambda_{J1I}-\lambda_{1JI})(\lambda_{M2I}-\lambda_{2MI})
\upsilon_{\tilde{\nu}_{J}}\upsilon_{\tilde{\nu}_{M}}  \nonumber \\
& & + \frac{1}{2}\sum\limits_{J}l_{1}(\lambda_{J12} + \lambda_{J21} - 
\lambda_{1J2}
-\lambda_{2J1})\upsilon_{1}\upsilon_{\tilde{\nu}_{J}}+m_{L^{12}}^{2}
,\nonumber \\
{\cal M}_{c 3,5}^{2} &=& \frac{g^{2}}{4}\upsilon_{\tilde{\nu}_{e}}
\upsilon_{\tilde{\nu}_{\tau}} + \epsilon_{1}\epsilon_{3} + \frac{1}{2}
\sum\limits_{IJM}(\lambda_{J1I}-\lambda_{1JI})(\lambda_{M3I}-\lambda_{3MI})
\upsilon_{\tilde{\nu}_{J}}\upsilon_{\tilde{\nu}_{M}}  \nonumber \\
& & + \frac{1}{2}\sum\limits_{J}l_{1}(\lambda_{J13} + \lambda_{J31} - 
\lambda_{1J3}
-\lambda_{3J1})\upsilon_{1}\upsilon_{\tilde{\nu}_{J}}+m_{L^{13}}^{2}
,\nonumber \\
{\cal M}_{c 3,6}^{2} &=& \frac{1}{\sqrt{2}}l_{1}\mu\upsilon_{2}+\frac{1}{
\sqrt{2}}l_{s_{1}}\upsilon_{1} - \frac{1}{\sqrt{2}}\sum\limits_{I}\epsilon_{I}
\lambda_{I11}\upsilon_{2} + \frac{1}{\sqrt{2}}\sum_{I}(\lambda_{I11}^{s}-
\lambda_{1I1}^{s})\upsilon_{\tilde{\nu}_{I}}, \nonumber  \\
{\cal M}_{c 3,7}^{2} &=& - \frac{1}{\sqrt{2}}\sum\limits_{I}\epsilon_{I}
\lambda_{I12}\upsilon_{2} + \frac{1}{\sqrt{2}}\sum_{I}(\lambda_{I12}^{s}-
\lambda_{1I2}^{s})\upsilon_{\tilde{\nu}_{I}} ,\nonumber \\
{\cal M}_{c 3,8}^{2} &=& - \frac{1}{\sqrt{2}}\sum\limits_{I}\epsilon_{I}
\lambda_{I13}\upsilon_{2} + \frac{1}{\sqrt{2}}\sum_{I}(\lambda_{I13}^{s}-
\lambda_{1I3}^{s})\upsilon_{\tilde{\nu}_{I}} ,\nonumber \\
{\cal M}_{c 4,4}^{2} &=& \frac{g^{2}}{4}\upsilon_{\tilde{\nu}_{\mu}}^{2} - 
\frac{1}{8}(g^{2}-g'^{2})
(\upsilon_{1}^{2}-\upsilon_{2}^{2}+\sum_{I}\upsilon_{\tilde{\nu}_{I}}^{2}) +
\epsilon_{2}^{2} + 
\frac{l_{2}^{2}}{2}\upsilon_{1}^{2} + m_{L^{2}}^{2}  \nonumber \\
 & & + \sum\limits_{J}l_{2}(\lambda_{J22}-\lambda_{2J2})\upsilon_{1}
\upsilon_{\tilde{\nu}_{J}} +\frac{1}{2}\sum_{IJM}(\lambda_{J2I} - 
\lambda_{2JI})(
\lambda_{M2I} - \lambda_{2MI})\upsilon_{\tilde{\nu}_{J}}
\upsilon_{\tilde{\nu}_{M}} \nonumber \\
 & = & \frac{g^{2}}{4}(\upsilon_{2}^{2}-\upsilon_{1}^{2})+
(\epsilon_{2}\mu-m_{HL^{2}}^{2})\frac{\upsilon_{1}}{
\upsilon_{\tilde{\nu}_{\mu}}} - B_{2}\frac{\upsilon_{2}}{
\upsilon_{\tilde{\nu}_{\mu}}} + 
\frac{l_{2}^{2}}{2}\upsilon_{1}^{2}  \nonumber \\
 & & -(\epsilon_{1}\epsilon_{2}+m_{L^{12}}^{2})
\frac{\upsilon_{\tilde{\nu}_{e}}}{
 \upsilon_{\tilde{\nu}_{\mu}}} - 
(\epsilon_{2}\epsilon_{3}+m_{L^{23}}^{2})\frac{\upsilon_{\tilde{\nu}_{\tau}}}{
\upsilon_{\tilde{\nu}_{\mu}}} - 
\frac{g^{2}}{4}(\upsilon_{\tilde{\nu}_{e}}^{2}+
\upsilon_{\tilde{\nu}_{\tau}}^{2}) \nonumber \\
 & & + \sum\limits_{J}l_{2}(\lambda_{J22}-\lambda_{2J2})\upsilon_{1}
\upsilon_{\tilde{\nu}_{J}} +\frac{1}{2}\sum_{IJM}(\lambda_{J2I} - 
\lambda_{2JI})(
\lambda_{M2I} - \lambda_{2MI})\upsilon_{\tilde{\nu}_{J}}
\upsilon_{\tilde{\nu}_{M}}, \nonumber \\
{\cal M}_{c 4,5}^{2} &=& \frac{g^{2}}{4}\upsilon_{\tilde{\nu}_{\mu}}
\upsilon_{\tilde{\nu}_{\tau}} + \frac{1}{2}
\sum\limits_{IJM}(\lambda_{J2I}-\lambda_{2JI})(\lambda_{M3I}-\lambda_{3MI})
\upsilon_{\tilde{\nu}_{J}}\upsilon_{\tilde{\nu}_{M}}  \nonumber \\
& & + \frac{1}{2}\sum\limits_{J}l_{2}(\lambda_{J23} + \lambda_{J32} - 
\lambda_{2J3}
-\lambda_{3J2})\upsilon_{1}\upsilon_{\tilde{\nu}_{J}}+m_{L^{23}}^{2}  
,\nonumber \\
{\cal M}_{c 4,6}^{2} &=& - \frac{1}{\sqrt{2}}\sum\limits_{I}\epsilon_{I}
\lambda_{I21}\upsilon_{2} + \frac{1}{\sqrt{2}}\sum_{I}(\lambda_{I21}^{s}-
\lambda_{2I1}^{s})\upsilon_{\tilde{\nu}_{I}}  ,\nonumber \\
{\cal M}_{c 4,7}^{2} &=& \frac{1}{\sqrt{2}}l_{2}\mu\upsilon_{2}-\frac{1}{
\sqrt{2}}l_{s_{2}}\upsilon_{1} - 
\frac{1}{\sqrt{2}}\sum\limits_{I}\epsilon_{I}
\lambda_{I22}\upsilon_{2} + \frac{1}{\sqrt{2}}\sum_{I}(\lambda_{I22}^{s}-
\lambda_{2I2}^{s})\upsilon_{\tilde{\nu}_{I}} , \nonumber  \\
{\cal M}_{c 4,8}^{2} &=& - \frac{1}{\sqrt{2}}\sum\limits_{I}\epsilon_{I}
\lambda_{I23}\upsilon_{2} + \frac{1}{\sqrt{2}}\sum_{I}(\lambda_{I23}^{s}-
\lambda_{2I3}^{s})\upsilon_{\tilde{\nu}_{I}} ,\nonumber \\
{\cal M}_{c 5,5}^{2} &=& \frac{g^{2}}{4}\upsilon_{\tilde{\nu}_{\tau}}^{2} - 
\frac{1}{8}(g^{2}-g'^{2})
(\upsilon_{1}^{2}-\upsilon_{2}^{2}+\sum_{I}\upsilon_{\tilde{\nu}_{I}}^{2}) +
\epsilon_{3}^{2} + 
\frac{l_{3}^{2}}{2}\upsilon_{1}^{2} + m_{L^{3}}^{2}  \nonumber \\
 & & + \sum\limits_{J}l_{3}(\lambda_{J33}-\lambda_{3J3})\upsilon_{1}
\upsilon_{\tilde{\nu}_{J}} +\frac{1}{2}\sum_{IJM}(\lambda_{J3I} - 
\lambda_{3JI})(
\lambda_{M3I} - \lambda_{3MI})\upsilon_{\tilde{\nu}_{J}}
\upsilon_{\tilde{\nu}_{M}} \nonumber \\
 & = & \frac{g^{2}}{4}(\upsilon_{2}^{2}-\upsilon_{1}^{2})+(\mu\epsilon_{3}
 -m_{HL^{3}}^{2})
\frac{\upsilon_{1}}{
\upsilon_{\tilde{\nu}_{\tau}}} - B_{3}\frac{\epsilon_{3}\upsilon_{2}}{
\upsilon_{\tilde{\nu}_{\tau}}} + 
\frac{l_{3}^{2}}{2}\upsilon_{1}^{2}  \nonumber \\
 & & -(\epsilon_{1}\epsilon_{3}+m_{L^{13}}^{2})
\frac{\upsilon_{\tilde{\nu}_{e}}}{
\upsilon_{\tilde{\nu}_{\tau}}} - 
(\epsilon_{2}\epsilon_{3}+m_{L^{23}}^{2})\frac{\upsilon_{\tilde{\nu}_{\mu}}}{
\upsilon_{\tilde{\nu}_{\tau}}} - 
\frac{g^{2}}{4}(\upsilon_{\tilde{\nu}_{e}}^{2}+
\upsilon_{\tilde{\nu}_{\mu}}^{2}) \nonumber \\
 & & +\sum\limits_{J}l_{3}(\lambda_{J33}-\lambda_{3J3})\upsilon_{1}
\upsilon_{\tilde{\nu}_{J}} +\frac{1}{2}\sum_{IJM}(\lambda_{J3I} - 
\lambda_{3JI})(
\lambda_{M3I} - \lambda_{3MI})\upsilon_{\tilde{\nu}_{J}}
\upsilon_{\tilde{\nu}_{M}}, \nonumber \\
{\cal M}_{c 5,6}^{2} &=& - \frac{1}{\sqrt{2}}\sum\limits_{I}\epsilon_{I}
\lambda_{I31}\upsilon_{2} + \frac{1}{\sqrt{2}}\sum_{I}(\lambda_{I31}^{s}-
\lambda_{3I1}^{s})\upsilon_{\tilde{\nu}_{I}} ,\nonumber \\
{\cal M}_{c 5,7}^{2} &=& - \frac{1}{\sqrt{2}}\sum\limits_{I}\epsilon_{I}
\lambda_{I32}\upsilon_{2} + \frac{1}{\sqrt{2}}\sum_{I}(\lambda_{I32}^{s}-
\lambda_{3I2}^{s})\upsilon_{\tilde{\nu}_{I}} ,\nonumber \\
{\cal M}_{c 5,8}^{2} &=& \frac{1}{\sqrt{2}}l_{3}\mu\upsilon_{\tilde{\nu}_{\tau}} - 
\frac{1}{\sqrt{2}}l_{s_{3}}\upsilon_{1} - 
\frac{1}{\sqrt{2}}\sum\limits_{I}\epsilon_{I}
\lambda_{I33}\upsilon_{2} + \frac{1}{\sqrt{2}}\sum_{I}(\lambda_{I33}^{s}-
\lambda_{3I3}^{s})\upsilon_{\tilde{\nu}_{I}},\nonumber \\
{\cal M}_{c 6,6}^{2} &=& -\frac{g'^{2}}{4}(\upsilon_{1}^{2}-\upsilon_{2}^{2}+
\sum_{I}\upsilon
_{\tilde{\nu}_{I}}^{2})+\frac{1}{2}l_{1}^{2}(\upsilon_{1}^{2} + 
\upsilon_{\tilde{\nu}_{e}}^{2})
+ m_{R^{1}}^{2} \nonumber \\ 
& & -\frac{1}{2}\sum\limits_{J}l_{1}\lambda_{1J1}
\upsilon_{1}\upsilon_{\tilde{\nu}_{J}} + 
\frac{1}{2}\sum\limits_{IJM}\lambda_{IJ1}
\lambda_{IM1}\upsilon_{\tilde{\nu}_{J}}
\upsilon_{\tilde{\nu}_{M}},\nonumber  \\
{\cal M}_{c 6,7}^{2} &=& \frac{1}{2}l_{1}l_{2}
\upsilon_{\tilde{\nu}_{e}}\upsilon_{\tilde{\nu}_{\mu}} 
-\frac{1}{2}\sum\limits_{J}l_{1}
\lambda_{1J2}\upsilon_{1}\upsilon_{\tilde{\nu}_{J}}
-\frac{1}{2}\sum\limits_{J}l_{2}
\lambda_{2J1}\upsilon_{1}\upsilon_{\tilde{\nu}_{J}} \nonumber \\
&& - 
\frac{1}{2}\sum\limits_{IJM}
\lambda_{IJ1}\lambda_{IM2}\upsilon_{\tilde{\nu}_{J}}\upsilon_{\tilde{\nu}_{M}}
+m_{R^{12}}^{2},
\nonumber \\
{\cal M}_{c 6,8}^{2} &=& \frac{1}{2}l_{1}l_{3}
\upsilon_{\tilde{\nu}_{e}}\upsilon_{\tilde{\nu}_{\tau}} 
-\frac{1}{2}\sum\limits_{J}l_{1}
\lambda_{1J3}\upsilon_{1}\upsilon_{\tilde{\nu}_{J}}  
-\frac{1}{2}\sum\limits_{J}l_{2}
\lambda_{3J1}\upsilon_{1}\upsilon_{\tilde{\nu}_{J}}\nonumber \\
&& -\frac{1}{2}\sum\limits_{IJM}
\lambda_{IJ1}\lambda_{IM3}\upsilon_{\tilde{\nu}_{J}}\upsilon_{\tilde{\nu}_{M}}
+m_{R^{13}}^{2},
\nonumber \\
{\cal M}_{c 7,7}^{2} &=& -\frac{g'^{2}}{4}(
\upsilon_{1}^{2}-\upsilon_{2}^{2}+\sum_{I}\upsilon
_{\tilde{\nu}_{I}}^{2})+\frac{1}{2}l_{1}^{2}(\upsilon_{1}^{2} + 
\upsilon_{\tilde{\nu}_{\mu}}^{2})
+ m_{R^{2}}^{2}    \nonumber \\
& & -\frac{1}{2}\sum\limits_{J}l_{2}\lambda_{2J2}
\upsilon_{1}\upsilon_{\tilde{\nu}_{J}} + 
\frac{1}{2}\sum\limits_{IJM}\lambda_{IJ2}
\lambda_{IM2}\upsilon_{\tilde{\nu}_{J}}
\upsilon_{\tilde{\nu}_{M}},\nonumber  \\
{\cal M}_{c 7,8}^{2} &=& \frac{1}{2}l_{2}l_{3}
\upsilon_{\tilde{\nu}_{\mu}}\upsilon_{\tilde{\nu}_{\tau}}-
\frac{1}{2}\sum\limits_{J}l_{1}
\lambda_{2J3}\upsilon_{1}\upsilon_{\tilde{\nu}_{J}}  
-\frac{1}{2}\sum\limits_{J}l_{2}
\lambda_{3J2}\upsilon_{1}\upsilon_{\tilde{\nu}_{J}}\nonumber \\
&& -\frac{1}{2}\sum\limits_{IJM}
\lambda_{IJ2}\lambda_{IM2}\upsilon_{\tilde{\nu}_{J}}\upsilon_{\tilde{\nu}_{M}} 
+ m_{R^{23}}^{2},\nonumber \\
{\cal M}_{c 8,8}^{2} &=& -\frac{g'^{2}}{4}(\upsilon_{1}^{2}-
\upsilon_{2}^{2}+\sum_{I}\upsilon
_{\tilde{\nu}_{I}}^{2})+\frac{1}{2}l_{3}^{2}(\upsilon_{1}^{2} + 
\upsilon_{\tilde{\nu}_{\tau}}^{2})
+ m_{R^{3}}^{2} \nonumber \\
& & -\frac{1}{2}\sum\limits_{J}l_{3}\lambda_{3J3}
\upsilon_{1}\upsilon_{\tilde{\nu}_{J}}+ 
\frac{1}{2}\sum\limits_{IJM}\lambda_{IJ3}
\lambda_{IM3}\upsilon_{\tilde{\nu}_{J}}
\upsilon_{\tilde{\nu}_{M}}.  
\label{eq-21}
\end{eqnarray}
Note here that to obtain Eq.\ (\ref{eq-21}), Eq.\ (\ref{masspara}) is used 
sometimes.

\section{The mixing of the squarks \label{app2}}

In the concerning model, the lepton numbers are broken, and due to the
VEVs of sneutrinos and trilinear terms the mixing of squarks is affected.
In a general case, the matrix of the squarks mixing should be 6$\times$6.
Under our assumptions, we do not consider the
squarks mixing between different generations. From superpotential 
Eq.\ (\ref{eq-2}) and the soft-breaking terms Eq.(5), we find
the up squarks mass matrix of the  I-th generation can be written as:
\begin{equation}
{\cal M}_{U^{I}}^{2} =
\left(  
\begin{array}{cc}
\frac{1}{24}(3g^{2} - g^{\prime^{2}})(\upsilon^{2} - 2\upsilon_{2}^{2}) + 
\frac{u_{I}^{2}}{2}\upsilon_{2}^{2} + m_{Q^{I}}^{2} & 
\frac{1}{\sqrt{2}}(u_{I}\mu \upsilon_{1} - 
u_{I}\sum\limits_{J=1}^{3}\epsilon_{J}
\upsilon_{\tilde{\nu}_{J}} - u_{S_{I}}\upsilon_{2})  \\
\frac{1}{\sqrt{2}}(u_{I}\mu \upsilon_{1} - 
u_{I}\sum\limits_{J=1}^{3}\epsilon_{J}
\upsilon_{\tilde{\nu}_{J}} - u_{S_{I}}\upsilon_{2})  &
\frac{1}{6}g^{\prime^{2}}(\upsilon^{2} - 2\upsilon_{2}^{2}) + 
\frac{u_{I}^{2}}{2}\upsilon_{2}^{2} + m_{U^{I}}^{2}
\end{array} \right)
\label{matrix-upsqu}
\end{equation}
where $I=(1$, $2$, $3)$ is the index of the generations. 
The interaction eigenstates $\tilde{Q}_{1}^{I}$ and $\tilde{U}^{I}$ 
connect to the two physical (mass) 
eigenstates $\tilde{U}_{I}^{i}$ $(i=1$, $2)$ through
\begin{equation}
\tilde{U}_{I}^{i} = Z_{U^{I}}^{i,1}\tilde{Q}_{1}^{I} 
+ Z_{U^{I}}^{i,2}\tilde{U}^{I}.
\label{upsqu-mix}
\end{equation}
and $Z_{U^{I}}$ is determined by the condition:
\begin{equation}
Z_{U^{I}}^{\dag}{\cal M}_{U^{I}}^{2}Z_{U^{I}} = 
{\rm diag}(M_{U_{I}^{1}}^{2}, M_{U_{I}^{2}}^{2})
\end{equation}
In a similar way, we can give the down squarks mass matrix of the  
I-th generation:
\begin{equation}
{\cal M}_{D^{I}}^{2} =
\left(  
\begin{array}{cc}
-\frac{1}{24}(3g^{2} + g^{\prime^{2}})(\upsilon^{2} - 2\upsilon_{2}^{2}) + 
\frac{d_{I}^{2}}{2}\upsilon_{1}^{2} + m_{Q^{I}}^{2} & 
-\frac{1}{\sqrt{2}}(d_{I}\mu\upsilon_{2} - d_{S_{I}}\upsilon_{1})  \\
-\frac{1}{\sqrt{2}}(d_{I}\mu \upsilon_{2} - d_{S_{I}}\upsilon_{1})  &
-\frac{1}{12}g^{\prime^{2}}(\upsilon^{2} - 2\upsilon_{2}^{2}) + 
\frac{d_{I}^{2}}{2}\upsilon_{1}^{2} + m_{D^{I}}^{2}
\end{array} \right)
\label{matrix-downsqu}
\end{equation}
The fields $\tilde{Q}_{2}^{I}$ and $\tilde{D}^{I}$ relate to the two physical 
(mass)
eigenstates $\tilde{D}_{I}^{i}$ $(i=(1$, $2)$:
\begin{eqnarray}
&&\tilde{D}_{I}^{i} = Z_{D^{I}}^{i,1}\tilde{Q}_{2}^{I} +
Z_{D^{I}}^{i,2}\tilde{D}^{I},  \nonumber \\
&&Z_{D^{I}}^{\dag}{\cal M}_{D^{I}}^{2}Z_{D^{I}} = {\rm diag}
(M_{D_{I}^{1}}^{2}, M_{D_{I}^{2}}^{2}).
\end{eqnarray}

\section{The precise formulas of  
${\cal L}_{SSV}$, ${\cal L}_{SVV}$ and ${\cal L}_{SSVV}$}

\subsection{The precise formulas of ${\cal L}_{SSV}$}

\begin{eqnarray}
{\cal L}_{SSV} &=& \frac{i}{2}\sqrt{g^{2} + g^{\prime^{2}}}Z_{\mu} 
\Bigg\{ \partial^{\mu}\phi_{1}^{0}\chi_{1}^{0} - \phi_{1}^{0}
\partial^{\mu}\chi_{1}^{0} - \partial^{\mu}\phi_{2}^{0}\chi_{2}^{0} + 
\phi_{2}^{0}\partial^{\mu}\chi_{2}^{0} \nonumber  \\
 &  & + \sum_{I}\Big(\partial^{\mu}\phi_{\tilde{\nu}_{I}}^{0}
\chi_{\tilde{\nu}_{I}}^{0} - 
\phi_{\tilde{\nu}_{I}}^{0}\partial^{\mu}\chi_{\tilde{\nu}_{I}}^{0}\Big)\Bigg\} + 
\frac{1}{2}g\Bigg\{ 
W_{\mu}^{+}\bigg[\chi_{1}^{0}\partial^{\mu}H_{2}^{1} - 
\partial^{\mu}\chi_{1}^{0}H_{2}^{1} \nonumber \\
 & & -\chi_{2}^{0}\partial^{\mu}H_{1}^{2*} + 
\partial^{\mu}\chi_{2}^{0}H_{1}^{2*} + 
\sum_{I}\chi_{\tilde{\nu}_{I}}^{0}\partial^{\mu}\tilde{L}_{2}^{I} - 
\partial^{\mu}\chi_{\tilde{\nu}_{I}}^{0}\tilde{L}_{2}^{I} \bigg] + h.c.\Bigg\} 
\nonumber \\
 & & +\frac{i}{2}g\Bigg\{W_{\mu}^{+}
\bigg[\phi_{1}^{0}\partial^{\mu}H_{2}^{1*} -
\partial^{\mu}\phi_{1}^{0}H_{2}^{1} 
+ 
\phi_{2}^{0}\partial^{\mu}H_{1}^{2*} -   \partial^{\mu}\phi_{2}^{0}H_{1}^{2*} 
\nonumber \\
&&+ \sum_{I}(\phi_{\tilde{\nu}_{I}}^{0}\partial^{\mu}\tilde{L}_{2}^{I} -
\partial^{\mu}\phi_{\tilde{\nu}_{I}}^{0}\tilde{L}_{2}^{I}) \bigg] - h.c.
\Bigg\} +  
\Bigg\{\frac{1}{2}\sqrt{g^{2} + g^{\prime^{2}}}\Big(\cos 2\theta_{W}Z_{\mu} 
\nonumber \\
 & &- \sin 
2\theta_{W}A_{\mu}\Big)\bigg[\sum_{I}\Big(\tilde{L}_{2}^{I*}\partial^{\mu}
\tilde{L}_{2}^{I} -  \partial^{\mu}\tilde{L}_{2}^{I*}\tilde{L}_{2}^{I}\Big) - 
H_{1}^{2*}\partial^{\mu}H_{1}^{2}  \nonumber \\
 & &+\partial^{\mu}H_{1}^{2*}
H_{1}^{2} + H_{2}^{1*}\partial^{\mu}H_{2}^{1} - 
\partial^{\mu}H_{2}^{1*}H_{2}^{1}\bigg] + 
\Big(2\sin^{2}\theta_{W}Z_{\mu} \nonumber \\
 & &+ 
2\sin\theta_{W}\cos\theta_{W}A_{\mu}\Big)\bigg[\sum_{I}\Big(\tilde{R}^{I*}
\partial^{\mu}\tilde{R}^{I} - 
\partial^{\mu}\tilde{R}^{I*}\tilde{R}^{I}\Big)\bigg] \Bigg\} \nonumber \\
 &=& \frac{i}{2}\sqrt{g^{2} + g^{\prime^{2}}}C_{eo}^{ij}
\Big(\partial^{\mu}H_{5+i}H_{j}^{0} -
H_{5+i}\partial^{\mu}H_{j}^{0}\Big)Z_{\mu} 
 \nonumber \\
 & &+\Bigg\{ \frac{1}{2}g C_{ec}^{ij}
\Big(H_{i}\partial^{\mu}H_{j}^{-}
- \partial^{\mu}H_{i}H_{j}^{-}\Big)W_{\mu}^{+} 
+ h.c. \Bigg\} \nonumber \\
 & & +\Bigg\{ \frac{i}{2}gC_{co}^{ij}
\Big(H_{5+i}\partial^{\mu}H_{j}^{-} - 
\partial^{\mu}H_{5+i}H_{j}^{-}\Big)W_{\mu}^{+}  \nonumber \\
 & &+ h.c. \Bigg\} + \Bigg\{\frac{i}{2}\sqrt{g^{2} + g^{\prime^{2}}}
\bigg[\Big(\cos 2\theta_{W}\delta^{ij}-C_{c}^{ij}\Big)Z_{\mu}
\Big(H_{i}^{-}\partial^{\mu}H_{j}^{+} - \partial^{\mu}H_{i}^{-}H_{j}^{+}\Big) 
\nonumber \\
 & & - \sin 2\theta_{W}A_{\mu}\Big(H_{i}^{-}\partial^{\mu}H_{i}^{+} 
- \partial^{\mu}H_{i}^{-}H_{i}^{+}\Big)\bigg]\Bigg\},
\label{vertex-ssv}
\end{eqnarray}
with
\begin{eqnarray}
&&C_{eo}^{ij}= \sum_{\alpha=1}^{5}Z_{odd}^{i\alpha}Z_{even}^{j\alpha} 
- 2Z_{odd}^{i2}Z_{even}^{j2}, \nonumber \\
&&C_{ec}^{ij}=\sum_{\alpha=1}^{5}Z_{even}^{i\alpha}Z_{c}^{j\alpha} 
- 2Z_{even}^{i2}Z_{c}^{j2}, \nonumber \\
&&C_{co}^{ij}=\sum_{\alpha=1}^{5}Z_{odd}^{i\alpha}Z_{c}^{j\alpha} - 
2Z_{odd}^{i2}Z_{c}^{j2}
, \nonumber \\
&&C_{c}^{ij}=\sum_{\alpha=6}^{8}Z_{c}^{i\alpha}Z_{c}^{j\alpha}.
\label{cdef}
\end{eqnarray}
Here the transform matrices $Z_{even}$, $Z_{odd}$ and $Z_{c}$ are well
defined in Sect.II.

\subsection{The precise formulas of ${\cal L}_{SVV}$}

\begin{eqnarray}
{\cal L}_{SVV} &=& \frac{g^{2} + 
g^{\prime^{2}}}{4}\Big(\upsilon_{1}\chi_{1}^{0} + 
\upsilon_{2}\chi_{2}^{0}+\sum\limits_{I}\upsilon_{\tilde{\nu}_{I}}
\chi_{\tilde{\
nu}_{I}}^{0}\Big)
\Big(Z_{\mu}Z^{\mu} + 2\cos^{2}\theta_{W}W_{\mu}^{-}W_{+}^{\mu}\Big)
\nonumber 
\\
 & &+\Bigg\{\frac{g^{2} + g^{\prime^{2}}}{4}\bigg[\cos\theta_{W}\Big(-1 + 
\cos 2\theta_{W}\Big)Z_{\mu}W_{+}^{\mu}
\Big(\upsilon_{1}H_{2}^{1} - \upsilon_{2}H_{1}^{2*} \nonumber \\
&&+ \sum_{I}\upsilon_{\tilde{\nu}_{I}}\tilde{L}_{2}^{I}\Big) - 
\cos\theta_{W}\sin 2\theta_{W}A_{\mu}W^{+\mu}\Big(\upsilon_{1}H_{2}^{1} - 
\upsilon_{2}H_{1}^{2*} \nonumber \\
 & &+ \sum_{I}\upsilon_{\tilde{\nu}_{I}}\tilde{L}_{2}^{I}\Big)\bigg] + h.c. 
\Bigg\} \nonumber \\
 &=& \frac{g^{2} + g^{\prime^{2}}}{4}C_{even}^{i}
\Big(H_{i}Z_{\mu}Z^{\mu} +2\cos^{2}\theta_{W}H_{i}W_{\mu}^{-}W^{+\mu}\Big) 
\nonumber \\
 & &-  \frac{g^{2} + g^{\prime^{2}}}{2}
S_{W}C_{W}\upsilon\bigg[S_{W}Z_{\mu}W^{+\mu}H_{1}^{-} + 
C_{W}A_{\mu}W^{+\mu}H_{1}^{-} + h.c. \bigg] ,
\label{vertex-svv}
\end{eqnarray}
with
\begin{eqnarray}
&&C_{even}^{i}=Z_{even}^{i1}\upsilon_{1} + Z_{even}^{i2}\upsilon_{2} + 
\sum_{I}Z_{even}^{i I+2}\upsilon_{\tilde{\nu}_{I}} \; . \label{ceven}
\end{eqnarray}

\subsection{The precise formulas of ${\cal L}_{SSVV}$}

The pieces of ${\cal L}_{SSVV}$ is given as
\begin{eqnarray}
{\cal L}_{SSVV} &=& -\frac{g^{2}+g^{\prime^{2}}}{4}
\bigg[\frac{1}{2}\Big(\chi_{1}^{0}\chi_{1}^{0} + \chi_{2}^{0}\chi_{2}^{0} + 
\sum_{I}\chi_{\tilde{\nu}_{I}}^{0}\chi_{\tilde{\nu}_{I}}^{0}\Big)Z_{\mu}Z^{\mu} 
\nonumber \\
 & &+ \cos^{2}\theta_{W}\Big(\chi_{1}^{0}\chi_{1}^{0} +
\chi_{2}^{0}\chi_{2}^{0} + 
\sum_{I}\chi_{\tilde{\nu}_{I}}^{0}\chi_{\tilde{\nu}_{I}}^{0}\Big) 
W_{\mu}^{-}W^{+
\mu} \bigg] - 
\frac{g^{2}+g^{\prime^{2}}}{4}\bigg[\frac{1}{2}\Big(\phi_{1}^{0}\phi_{1}^{0} 
\nonumber \\
 & &+ \phi_{2}^{0}\phi_{2}^{0} + 
\sum_{I}\phi_{\tilde{\nu}_{I}}^{0}\phi_{\tilde{\nu}_{I}}^{0}\Big)Z_{\mu}Z^{\mu} 
+ 
\cos^{2}\theta_{W}\Big(\phi_{1}^{0}\phi_{1}^{0} + \phi_{2}^{0}\phi_{2}^{0} + 
\phi_{\tilde{\nu}_{I}}^{0}\phi_{\tilde{\nu}_{I}}^{0}\Big)W_{\mu}^{-}W^{+\mu} 
\bigg]  \nonumber \\
 & &-\frac{g^{2}+g^{\prime^{2}}}{4}\cos\theta_{W}\bigg[\Big(-1 + 
 \cos 2\theta_{W}\Big)Z_{\mu}W^{+\mu}\Big(\chi_{1}^{0}H_{2}^{1} -
\chi_{2}^{0}H_{1}^{2*} \nonumber \\
 & & + \sum_{I}\chi_{\tilde{\nu}_{I}}^{0}\tilde{L}_{2}^{I}\Big) - 
\cos\theta_{W}\sin 2\theta_{W}A_{\mu}W^{+\mu}\Big(\chi_{1}^{0}H_{2}^{1} -
\chi_{2}^{0}H_{1}^{2*} \nonumber \\
 & & + \sum_{I}\chi_{\tilde{\nu}_{I}}^{0}\tilde{L}_{2}^{I}\Big) + h.c.\bigg] + 
\frac{i(g^{2} + g^{\prime^{2}})}{4}\cos\theta_{W}\bigg[\Big(-1 + 
\cos 2\theta_{W}\Big)Z_{\mu}W^{+\mu}\Big(\phi_{1}^{0}H_{2}^{1} 
\nonumber \\
&& -\phi_{2}^{0}H_{1}^{2*} + 
\sum_{I}\phi_{\tilde{\nu}_{I}}^{0}\tilde{L}_{2}^{I}\Big) - 
\cos\theta_{W}\sin 2\theta_{W}A_{\mu}W^{+\mu}\Big(\phi_{1}^{0}H_{2}^{1} 
\nonumber \\
 & &-\phi_{2}^{0}H_{1}^{2*} + 
\sum_{I}\phi_{\tilde{\nu}_{I}}^{0}\tilde{L}_{2}^{I}\Big) + h.c.\bigg] - 
\frac{1}{4}(g^{2} + 
g^{\prime^{2}})\bigg[\sin^{2}2\theta_{W}A_{\mu}A^{\mu}\nonumber \\
& & \Big(H_{2}^{1*}H_{2}^{1} + H_{1}^{2*}H_{1}^{2} +
\sum_{I}\tilde{L}_{2}^{I*}
\tilde{L}_{2}^{I}\Big)   \nonumber \\
 & &+\cos^{2}2\theta_{W}Z_{\mu}Z^{\mu}\Big(H_{2}^{1*}H_{2}^{1} + 
H_{1}^{2*}H_{1}^{2} + 
\sum_{I}\tilde{L}_{2}^{I*}\tilde{L}_{2}^{I}\Big)  \nonumber \\
 & &-\sin 4\theta_{W}Z_{\mu}A^{\mu}\Big(H_{2}^{1*}H_{2}^{1} + 
H_{1}^{2*}H_{1}^{2} + 
\sum_{I}\tilde{L}_{2}^{I*}\tilde{L}_{2}^{I}\Big)  \nonumber \\
 & & +2\cos^{2}\theta_{W}\Big(H_{2}^{1*}H_{2}^{1} + H_{1}^{2*}H_{1}^{2} + 
\sum_{I}\tilde{L}_{2}^{I*}\tilde{L}_{2}^{I}\Big) \bigg]  \nonumber \\
 & &-\sum_{I}g^{\prime^{2}}\tilde{R}^{I*}\tilde{R}^{I}B_{\mu}B^{\mu}  \nonumber 
\\
 &=& -\frac{1}{4}(g^{2} + 
g^{\prime^{2}})\Big(\frac{1}{2}H_{i}H_{i}Z_{\mu}Z^{\mu} + 
\cos^{2}\theta_{W}H_{i}H_{i}W_{\mu}^{-}W^{+\mu}\Big)  \nonumber \\
 & &-\frac{1}{4}(g^{2} + 
g^{\prime^{2}})\Big(\frac{1}{2}H_{5+i}H_{5+i}Z_{\mu}Z^{\mu} 
+ \cos^{2}\theta_{W}H_{5+i}H_{5+i}
W_{\mu}^{-}W^{+\mu}\Big)  \nonumber \\
 & &+\frac{1}{4}(g^{2} + g^{\prime^{2}})\sin 2\theta_{W}\Bigg\{C_{ec}^{ij}
\bigg[\sin\theta_{W}H_{i}Z_{\mu}W^{+\mu}H_{j}^{-}   \nonumber \\
 & & + \cos\theta_{W}H_{i}A_{\mu}W^{+\mu}H_{j}^{-}\bigg] 
+ h.c. \Bigg\}  \nonumber \\
& &-\frac{i}{4}(g^{2} + g^{\prime^{2}})\sin 2\theta_{W}\Bigg\{C_{co}^{ij} 
\bigg[\sin\theta_{W}H_{5+i}Z_{\mu}W^{+\mu}H_{j}^{-}\nonumber \\
 & &  + \cos\theta_{W}H_{5+i}A_{\mu}W^{+\mu}H_{i}^{-}\bigg] - h.c. \Bigg\}  
\nonumber \\
 & &-\frac{1}{4}(g^{2} + 
g^{\prime^{2}})\Bigg\{2\cos^{2}\theta_{W}\Big(\delta_{ij}- 
C_{c}^{ij}\Big)H_{i}^{-}
H_{j}^{+}W_{\mu}^{-}W^{+\mu} \nonumber \\
 & &+\bigg[\cos^{2}2\theta_{W}\delta_{ij} -
C_{c}^{ij}\Big(4\sin^{3}\theta_{W} - 
 \cos^{2}2\theta_{W}\Big)\bigg]H_{i}^{-}H_{j}^{+}Z_{\mu}Z^{\mu}  \nonumber \\
 & & +\sin^{2}2\theta_{W}\delta_{ij}H_{i}^{-}H_{j}^{+}A_{\mu}A^{\mu} + 
\bigg[\sin 4\theta_{W}
\delta_{ij} \nonumber \\
 & &-C_{c}^{ij}\Big(\sin 4\theta_{W} + 
8\sin^{2}\theta_{W}\cos \theta_{W}\Big)\bigg]Z_{\mu}A^{\mu}H_{i}^{-}H_{j}^{+} 
\Bigg\} \hspace{1mm},
\label{vertex-ssvv}
\end{eqnarray}
where the $C_{eo}^{ij}$, $C_{co}^{ij}$ and $C_{c}^{ij}$ are defined in Eq.\ 
(\ref{cdef}).

\section{The complementary expressions of the couplings 
in ${\cal L}_{SSS}$ and ${\cal L}_{SSSS}$}

In this appendix, we give precise expressions of the couplings that 
appear in the ${\cal L}_{SSS}$ and ${\cal L}_{SSSS}$.
The method has been described clearly in text, the results are:
\begin{eqnarray}
A_{ec}^{kij} &=& 
\frac{g^{2}+g^{\prime^{2}}}{4}\bigg(\upsilon_{1}
Z_{even}^{k,1}Z_{c}^{i,1}Z_{c}^{j,1} + 
\upsilon_{2}Z_{even}^{k,2}Z_{c}^{i,2}Z_{c}^{j,2} + 
\sum\limits_{I=1}^{3}\upsilon_{\tilde{\nu}_{I}}Z_{even}^{k,2+I}Z_{c}^{i,2+I}
Z_{c}^{j,2+I}\bigg)  \nonumber \\
 & & +\sum\limits_{I=1}^{3}\bigg(\frac{
g^{\prime^{2}} - g^{2}}{4} + 
l_{I}^{2}\bigg)\bigg(\upsilon_{1}Z_{even}^{k,1}Z_{c}^{i,2+I}Z_{c}^{j,2+I} 
+ \upsilon_{\tilde{\nu}_{I}}Z_{even}^{k,2+I}
Z_{c}^{i,1}Z_{c}^{j,1}\bigg)  \nonumber \\
 & &+\sum\limits_{I=1}^{3}\bigg(\frac{g^{2}}{4} - 
\frac{1}{2}l_{I}^{2}\bigg)\bigg\{
\upsilon_{1}Z_{even}^{k,2+I}\Big(Z_{c}^{i,2+I}Z_{c}^{j,2} + 
Z_{c}^{i,2}Z_{c}^{j,2+I}\Big) + 
\upsilon_{\tilde{\nu}_{I}}Z_{even}^{k,1}
\Big(Z_{c}^{i,2+I}Z_{c}^{j,2}  \nonumber \\
& & + Z_{c}^{i,2}Z_{c}^{j,2+I}\Big)\bigg\} + \frac{g^{2} - 
g^{\prime^{2}}}{4}\bigg(
\upsilon_{1}Z_{even}^{k,1}Z_{c}^{i,2}Z_{c}^{j,2} + 
\upsilon_{2}Z_{even}^{k,2}Z_{c}^{i,1}Z_{c}^{j,1}  \nonumber \\
 & & +\sum\limits_{I=1}^{3}\upsilon_{\tilde{\nu}_{I}}
Z_{even}^{k,2+I}Z_{c}^{i,2}Z_{c}^{j,2} +
\upsilon_{2}Z_{even}^{k,2}Z_{c}^{i,2+I}Z_{c}^{j,2+I}\bigg)  \nonumber \\
 & & +\sum\limits_{I=1}^{3}\bigg[\bigg(l_{I}^{2} - 
\frac{g^{\prime^{2}}}{2}\bigg)
\upsilon_{\tilde{\nu}_{I}}Z_{even}^{k,2+I}
Z_{c}^{i,5+I}Z_{c}^{j,5+I} + \bigg(l_{I}^{2} - \frac{g^{\prime^{2}}}{2}\bigg)
\upsilon_{1}Z_{even}^{k,1}Z_{c}^{i,5+I}Z_{c}^{j,5+I}  \nonumber \\
 & &+\frac{g^{\prime^{2}}}{2}\upsilon_{2}
Z_{even}^{k,2}Z_{c}^{i,5+I}Z_{c}^{j,5+I}\bigg] +  
\frac{g^{2}}{4}\bigg(\upsilon_{\tilde{\nu}_{I}}Z_{even}^{k,2} + 
\upsilon_{2}Z_{even}^{k,2+I}\bigg)\bigg(Z_{c}^{i,2+I}Z_{c}^{j,2}  \nonumber \\
 & & +Z_{c}^{i,2}Z_{c}^{j,2+I}\bigg) + 
\frac{g^{2}}{4}\bigg(\upsilon_{1}Z_{even}^{k,2} + 
\upsilon_{2}Z_{even}^{k,1}\bigg)\bigg(Z_{c}^{i,1}Z_{c}^{j,2}  \nonumber \\
 & & +Z_{c}^{i,2}Z_{c}^{j,1}\bigg) + \frac{1}{\sqrt{2}}l_{I}
\epsilon_{3}Z_{even}^{k,1}\bigg(Z_{c}^{i,4}Z_{c}^{j,2}   \nonumber \\
 & & +Z_{c}^{i,2}Z_{c}^{j,4} \bigg) + \frac{1}{\sqrt{2}}
\sum\limits_{I=1}^{3}l_{I}\epsilon_{I}Z_{even}^{k,2}\bigg(
Z_{c}^{i,5+I}Z_{c}^{j,1} + Z_{c}^{i,1}Z_{c}^{j,5+I} \bigg)  \nonumber \\
A_{oc}^{kij} &=& \sum\limits_{I=1}^{3}\Bigg\{\bigg(\frac{g^{2}}{4} - 
l_{I}^{2}\bigg)
\bigg[\upsilon_{\tilde{\nu}_{I}}Z_{odd}^{k,1}\Big(Z_{c}^{i,1}
Z_{c}^{j,2+I} - Z_{c}^{i,2+I}Z_{c}^{j,1}\Big) \nonumber \\
&& + \upsilon_{1}Z_{odd}^{k,2+I}\Big(Z_{c}^{i,1}Z_{c}^{j,2+I} - 
Z_{c}^{i,2+I}Z_{c}^{j,1}\Big)\bigg]  \nonumber \\
 & & + \frac{g^{2}}{4}\Big(\upsilon_{\tilde{\nu}_{I}}Z_{odd}^{k,2} + 
\upsilon_{2}Z_{odd}^{k,2+I}\Big)\Big(Z_{c}^{i,2+I}Z_{c}^{j,2} - 
Z_{c}^{i,2}Z_{c}^{j,2+I}\Big)
+ \frac{g^{2}}{4}\Big(\upsilon_{\tilde{\nu}_{I}}Z_{odd}^{k,2}  \nonumber \\
 & & +\upsilon_{2}Z_{odd}^{k,2+I}\Big)\Big(Z_{c}^{i,2+I}Z_{c}^{j,2} - 
Z_{c}^{i,2}Z_{c}^{j,2+I}\Big)  \nonumber \\
 & & \frac{1}{\sqrt{2}}l_{I}\epsilon_{I}Z_{odd}^{k,1}\Big(- 
Z_{c}^{i,5+I}Z_{c}^{j,2} 
+ Z_{c}^{i,2}Z_{c}^{j,5+I}\Big)  \nonumber \\
 & & -\frac{1}{\sqrt{2}}l_{I}\epsilon_{I}Z_{odd}^{k,2}\Big( 
Z_{c}^{i,5+I}Z_{c}^{j,1} 
- Z_{c}^{i,1}Z_{c}^{j,5+I}\Big) \Bigg\} \nonumber  \\
{\cal A}_{ec}^{klij} &=& \frac{g^{2} + 
g^{\prime^{2}}}{8s_{W}c_{W}}\bigg(Z_{even}^{k,1}
Z_{even}^{l,1}Z_{c}^{i,1}Z_{c}^{j,1} + 
Z_{even}^{k,2}Z_{even}^{l,2}Z_{c}^{i,2}Z_{c}^{j,2} + 
\sum\limits_{I=1}^{3}Z_{even}^{k,2+I}Z_{even}^{l,2+I}
Z_{c}^{i,2+I}Z_{c}^{j,2+I}\bigg) \nonumber \\
 & &+\sum\limits_{I=1}^{3}\bigg(\frac{g^{\prime^{2}} - g^{2}}{8} + 
\frac{1}{2}l_{I}^{2}\bigg)
\bigg(Z_{even}^{k,1}Z_{even}^{l,1}Z_{c}^{i,2+I}Z_{c}^{j,2+I} +
Z_{even}^{k,2+I}Z_{even}^{l,2+I}Z_{c}^{i,1}Z_{c}^{j,1}\bigg)  \nonumber \\
 & &+\frac{g^{\prime^{2}} - 
g^{2}}{8}\bigg[Z_{even}^{k,1}Z_{even}^{l,1}Z_{c}^{i,2}Z_{c}^{j,2} 
+ Z_{even}^{k,2}Z_{even}^{l,2}Z_{c}^{i,1}Z_{c}^{j,1}  \nonumber \\
 & & +\sum\limits_{I=1}^{3}\Big(Z_{even}^{k,2+I}Z_{even}^{l,2+I}
Z_{c}^{i,2}Z_{c}^{j,2} 
+  Z_{even}^{k,2}Z_{even}^{l,2}
Z_{c}^{i,2+I}Z_{c}^{j,2+I}\Big)\bigg]  \nonumber  \\
 & & +\frac{g^{2}}{4}\bigg[Z_{even}^{k,1}Z_{even}^{l,2}
\Big(Z_{c}^{i,2}Z_{c}^{j,1} + 
Z_{c}^{i,1}Z_{c}^{j,2}) + 
\sum\limits_{I=1}^{3}Z_{even}^{k,1}Z_{even}^{l,2+I}(Z_{c}^{i,2+I}Z_{c}^{j,1}  
\nonumber \\
 & &+Z_{c}^{i,1}Z_{c}^{j,2+I}\Big) + 
\sum\limits_{I=1}^{3}Z_{even}^{k,2}Z_{even}^{l,2+I}
\Big(Z_{c}^{i,2+I}Z_{c}^{j,2} + 
Z_{c}^{i,2}Z_{c}^{j,2+I}\Big) \bigg]  \nonumber  \\
 & & 
-\sum\limits_{I=1}^{3}\bigg[\frac{l_{I}^{2}}{2}Z_{even}^{k,1}Z_{even}^{l,2+I}
\Big(Z_{c}^{i,2+I}Z_{c}^{j,1} + 
Z_{c}^{i,1}Z_{c}^{j,2+I}\Big) - \frac{g^{\prime^{2}}}{4}\Big(
Z_{even}^{k,1}Z_{even}^{l,1}Z_{c}^{i,5+I}Z_{c}^{j,5+I}  \nonumber \\
 & &- Z_{even}^{k,2}Z_{even}^{l,2}Z_{c}^{i,5+I}Z_{c}^{j,5+I}
+ Z_{even}^{k,2+I}Z_{even}^{l,2+I}Z_{c}^{i,5+I}Z_{c}^{j,5+I}\Big)  \nonumber  \\
 & & +\frac{l_{I}^{2}}{2}
 Z_{even}^{k,1}Z_{even}^{l,1}Z_{c}^{i,5+I}Z_{c}^{j,5+I} \bigg]  \nonumber  \\
{\cal A}_{oc}^{klij} &=& \frac{g^{2} 
+g^{\prime^{2}}}{8}\Big(Z_{odd}^{k,1}Z_{odd}^{l,1}Z_{c}^{i,1}
Z_{c}^{j,1} + Z_{odd}^{k,2}Z_{odd}^{l,2}
Z_{c}^{i,2}Z_{c}^{j,2} \nonumber \\ & 
&+\sum\limits_{I=1}^{3}Z_{odd}^{k,2+I}
Z_{odd}^{l,2+I}Z_{c}^{i,2+I}Z_{c}^{j,2+I}\Big) + 
\sum\limits_{I=1}^{3}\bigg(\frac{g^{\prime^{2}} - g^{2}}{8} + 
\frac{1}{2}l_{I}^{2}\bigg)
\bigg(Z_{odd}^{k,1}Z_{odd}^{l,1}Z_{c}^{i,2+I}Z_{c}^{j,2+I}  \nonumber \\
 & &+Z_{odd}^{k,2+I}Z_{odd}^{l,2+I}Z_{c}^{i,1}Z_{c}^{j,1}\bigg) + 
\frac{g^{\prime^{2}} - 
g^{2}}{8}\bigg\{Z_{odd}^{k,1}Z_{odd}^{l,1}Z_{c}^{i,2}Z_{c}^{j,2}  \nonumber \\
 & &+Z_{odd}^{k,2}Z_{odd}^{l,2}
Z_{c}^{i,1}Z_{c}^{j,1} + 
\sum\limits_{I=1}^{3}\Big(Z_{odd}^{k,2+I}Z_{odd}^{l,2+I}Z_{c}^{i,2}Z_{c}^{j,2}  
\nonumber \\
 & & +Z_{odd}^{k,2}Z_{odd}^{l,2}Z_{c}^{i,2+I}Z_{c}^{j,2+I}\Big)\bigg\} +
\frac{g^{2}}{4}\Bigg\{Z_{odd}^{k,1}Z_{odd}^{l,2}\Big(Z_{c}^{i,2}Z_{c}^{j,1} + 
Z_{c}^{i,1}Z_{c}^{j,2}\Big)  \nonumber \\
 & &+ \sum\limits_{I=1}^{3}\bigg[Z_{odd}^{k,1}
Z_{odd}^{l,3}\Big(Z_{c}^{i,3}Z_{c}^{j,1} 
+ Z_{c}^{i,1}Z_{c}^{j,3}\Big) + 
Z_{odd}^{k,2}Z_{odd}^{l,3}\Big(Z_{c}^{i,3}Z_{c}^{j,2} + 
Z_{c}^{i,2}Z_{c}^{j,3}\Big)\bigg] \Bigg\} \nonumber  \\
 & & 
-\sum\limits_{I=1}^{3}\Bigg\{\frac{l_{I}^{2}}{2}Z_{odd}^{k,1}Z_{odd}^{l,2+I}
\Big(Z_{c}^{i,2+I}Z_{c}^{j,1} + Z_{c}^{i,1}Z_{c}^{j,2+I}\Big) - 
\frac{g^{\prime^{2}}}{4}\Big(Z_{odd}^{k,1}
Z_{odd}^{l,1}Z_{c}^{i,5+I}Z_{c}^{j,5+I}
\nonumber \\
 & & -Z_{odd}^{k,2}Z_{odd}^{l,2}Z_{c}^{i,5+I}Z_{c}^{j,5+I}
+ Z_{odd}^{k,2+I}Z_{odd}^{l,2+I}Z_{c}^{i,5+I}Z_{c}^{j,5+I}\Big)  \nonumber  \\
 & & 
+\frac{l_{I}^{2}}{2}Z_{odd}^{k,1}Z_{odd}^{l,1}
Z_{c}^{i,5+I}Z_{c}^{j,5+I}\Bigg\}  
\nonumber  \\
{\cal A}_{eoc}^{klij} &=& \sum\limits_{I=1}^{3}\bigg[\Big(\frac{g^{2}}{8} + 
\frac{1}{2}l_{I}^{2}\Big)\Big(Z_{even}^{k,2+I}Z_{odd}^{l,1} + 
Z_{even}^{k,1}Z_{odd}^{l,2+I}\Big)
\Big(Z_{c}^{i,1}Z_{c}^{j,2+I} - Z_{c}^{i,2+I}Z_{c}^{j,1}\Big) \nonumber \\
 & &-\frac{g^{2}}{8}\Big(Z_{even}^{k,2}Z_{odd}^{l,2+I} + 
Z_{even}^{k,2+I}Z_{odd}^{l,2}\Big)
\Big(Z_{c}^{i,2+I}Z_{c}^{j,2} - Z_{c}^{i,2}Z_{c}^{j,2+I}\Big)\bigg] 
 \nonumber \\
 & & -\frac{g^{2}}{8}\Big(Z_{even}^{k,1}Z_{odd}^{l,2} + 
Z_{even}^{k,2}Z_{odd}^{l,1}\Big)
\Big(Z_{c}^{i,1}Z_{c}^{j,2} - 
Z_{c}^{i,2}Z_{c}^{j,2}\Big)  \nonumber \\
{\cal A}_{cc}^{ijkl} &=&\frac{g^{2} + g^{\prime^{2}}}{8}\bigg[
\sum\limits_{m,n=1}^{5}Z_{c}^{i,m}Z_{c}^{j,m}Z_{c}^{k,n}Z_{c}^{l,n} + 
2\sum\limits_{I=1}^{3}Z_{c}^{i,2+I}Z_{c}^{j,2+I}\Big(Z_{c}^{k,1}Z_{c}^{l,1} - 
Z_{c}^{k,2}Z_{c}^{l,2}\Big) \nonumber \\
 & & -2Z_{c}^{i,1}Z_{c}^{j,1}Z_{c}^{k,2}Z_{c}^{l,2} \bigg] +
\frac{g^{\prime^{2}}}{2}\bigg[\sum\limits_{I=1}^{3}Z_{c}^{i,5+I}Z_{c}^{j,5+I}
 \Big(- Z_{c}^{k,5+I}Z_{c}^{l,5+I} - Z_{c}^{k,2+I}Z_{c}^{l,2+I}\Big)
\nonumber \\
 & & 
-\sum\limits_{I=1}^{3}Z_{c}^{i,5+I}Z_{c}^{j,5+I}Z_{c}^{k,1}Z_{c}^{l,1}\bigg] + 
\sum\limits_{I=1}^{3}l_{I}^{2}Z_{c}^{i,5+I}Z_{c}^{j,5+I}Z_{c}^{k,1}Z_{c}^{l,1} 
\nonumber 
\end{eqnarray}
where the mixing matrices $Z_{even}$, $Z_{odd}$ and $Z_{c}$ are 
defined in Eq.\ (\ref{masseven}), Eq.\ (\ref{oddhiggs})
and Eq.\ (\ref{charhiggs}) respectively.

\section{The coefficients in the R-parity violation couplings of Higgs}

The precise expressions for the coefficients in the R-parity violation 
couplings of Higgs: $C^{ijm}_{snn}$, $C^{ijm}_{Lnk}$, $C^{ijm}_{Rnk}$ 
and $C^{ijm}_{snk}$ are given as the follows:
\begin{eqnarray}
&&C_{snn}^{ijm}=\Big(\cos\theta_{W}Z_{N}^{j2} -
\sin\theta_{W}Z_{N}^{j1}\Big)
\Big(\sum_{\alpha=1}^{5}Z_{even}^{i\alpha}Z_{N}^{m2+\alpha} 
- 2Z_{even}^{i2}Z_{N}^{m4}\Big), \nonumber \\
&&C_{skk}^{ijm}=\Big(Z_{even}^{i 1}Z_{+}^{j 1}Z_{-}^{m,2} + 
Z_{even}^{i 2}Z_{+}^{j 2}Z_{-}^{m1} + 
\sum_{\alpha=3}^{5}Z_{even}^{i \alpha}Z_{+}^{j 1}Z_{-}^{m \alpha}\Big)
\nonumber \\
&&\hspace{10mm} +\frac{1}{2g}\bigg\{\sum_{I=1}^{3}l_{I}\Big(
Z_{even}^{i I+2}Z_{+}^{j I+2}Z_{-}^{m 2} - 
Z_{even}^{i 1}Z_{+}^{j I+2}Z_{-}^{m I+2}\Big) \nonumber \\
&&\hspace{10mm} + \sum\limits_{IJK}\Big(
Z_{even}^{i I+2}Z_{+}^{j J+2}Z_{-}^{m K+2} - 
Z_{even}^{i K+2}Z_{+}^{j J+2}Z_{-}^{m I+2}\Big)
\bigg\}, \nonumber \\
&&C_{onn}^{ijm}=\Big(\cos\theta_{W}Z_{N}^{j 2} - 
\sin\theta_{W}Z_{N}^{j 1}\Big)\Big(
\sum_{\alpha=1}^{5}Z_{odd}^{i \alpha}Z_{N}^{m 2+\alpha} 
- 2Z_{odd}^{i 2}Z_{N}^{m 4}\Big), \nonumber \\
&&C_{okk}^{ijm}=\Big(Z_{odd}^{i 1}Z_{+}^{j 1}Z_{-}^{m 2} + 
Z_{odd}^{i 2}Z_{+}^{j 2}
Z_{-}^{m 1} + 
\sum_{\alpha=1}^{3}Z_{odd}^{i 2+\alpha}
Z_{+}^{j 1}Z_{-}^{m 2+\alpha}\Big)\nonumber \\
&&\hspace{10mm} 
+\frac{i}{2g}\bigg\{\sum\limits_{I}l_{I}
\Big(Z_{odd}^{i 2+I}Z_{+}^{j 2+I}Z_{-}^{m 2} 
- Z_{odd}^{i 1}Z_{+}^{j 2+I}Z_{-}^{m 2+I}\Big)   \nonumber \\
&&\hspace{10mm}+
\sum\limits_{IJK}\lambda_{IJK}\Big(Z_{odd}^{i 2+I}Z_{+}^{j 2+K}
Z_{-}^{m J+2} 
- Z_{odd}^{i J+2}Z_{+}^{j 2+K}Z_{-}^{m 2+I}\Big)\bigg\} ,\nonumber \\
&&C_{Lnk}^{ijm}=\bigg[Z_{c}^{i 1}\bigg(\frac{1}{\sqrt{2}}
\Big(\cos\theta_{W}Z_{-}^{j 2}Z_{N}^{m 2} + \sin\theta_{W}Z_{-}^{j 2}
Z_{N}^{m,1}\Big) - 
\cos\theta_{W}Z_{-}^{j 1}Z_{N}^{m 3}\bigg)  \nonumber \\
 & &\hspace{10mm} + 
 \sum_{\alpha=3}^{5}Z_{c}^{i \alpha}\bigg(\frac{1}{\sqrt{2}}
 \Big(\cos\theta_{W}Z_{-}^{j \alpha}Z_{N}^{m 2} + 
\sin\theta_{W}Z_{-}^{j \alpha}Z_{N}^{m 1}\Big) - 
\cos\theta_{W}Z_{-}^{j 1}Z_{N}^{m 2+\alpha}\bigg) \bigg]  \nonumber \\
 & &\hspace{10mm} + \frac{1}{2\sqrt{g^{2}+g^{\prime^{2}}}}\bigg\{
 \sum\limits_{I=1}^{3}l_{I}\Big(Z_{c}^{i 5+I}Z_{-}^{j 2+I}Z_{N}^{m 3} 
- Z_{c}^{i 5+I}Z_{-}^{j 2}Z_{N}^{m 4+I}\Big) \nonumber \\
&&\hspace{10mm} + 
\sum\limits_{IJK}\lambda_{IJK}\Big(Z_{c}^{i 5+I}Z_{-}^{j 2+J}Z_{N}^{m 4+K} 
- Z_{c}^{i 5+I}Z_{-}^{j 2+K}Z_{N}^{m 4+J}\Big)\bigg\} , \nonumber \\
&&C_{Rnk}^{ijm}=\bigg[Z_{c}^{i 2}\bigg(\frac{1}{\sqrt{2}}
\Big(\cos\theta_{W}Z_{+}^{*j 2}Z_{N}^{*m 2} + 
\sin\theta_{W}Z_{+}^{*j 2}Z_{N}^{*m 1}\Big)  \nonumber \\
 & &\hspace{10mm}+\cos\theta_{W}Z_{+}^{*j 1}Z_{N}^{*m 4}\bigg) 
+ \sqrt{2}\sin\theta_{W}\sum_{I=1}^{3}Z_{c}^{i 5+I}Z_{+}^{*j 2+I}
Z_{N}^{*m 1}\bigg]  \nonumber \\
 & &\hspace{10mm}+\frac{1}{2\sqrt{g^{2}+g^{\prime^{2}}}}\bigg\{
\sum\limits_{I=1}^{3}l_{I}\Big(Z_{c}^{i 2+I}Z_{N}^{*m 3}Z_{+}^{*j 2+I} - 
Z_{c}^{i 1}
Z_{N}^{*m 3}Z_{+}^{*j 2+I}\Big) \nonumber \\
&&\hspace{10mm} 
+\sum\limits_{IJK}\lambda_{IJK}\Big(Z_{c}^{i 2+J}
Z_{N}^{*m 4+I}Z_{+}^{*j 2+K} - 
Z_{c}^{i 2+I}
Z_{N}^{*m 4+J}Z_{+}^{*j 2+K}\Big)\bigg\}\; . 
\label{def-coup}
\end{eqnarray}

\begin{figure}
\setlength{\unitlength}{1mm}
\begin{picture}(140,180)(30,30)
\put(-10,-10){\includegraphics{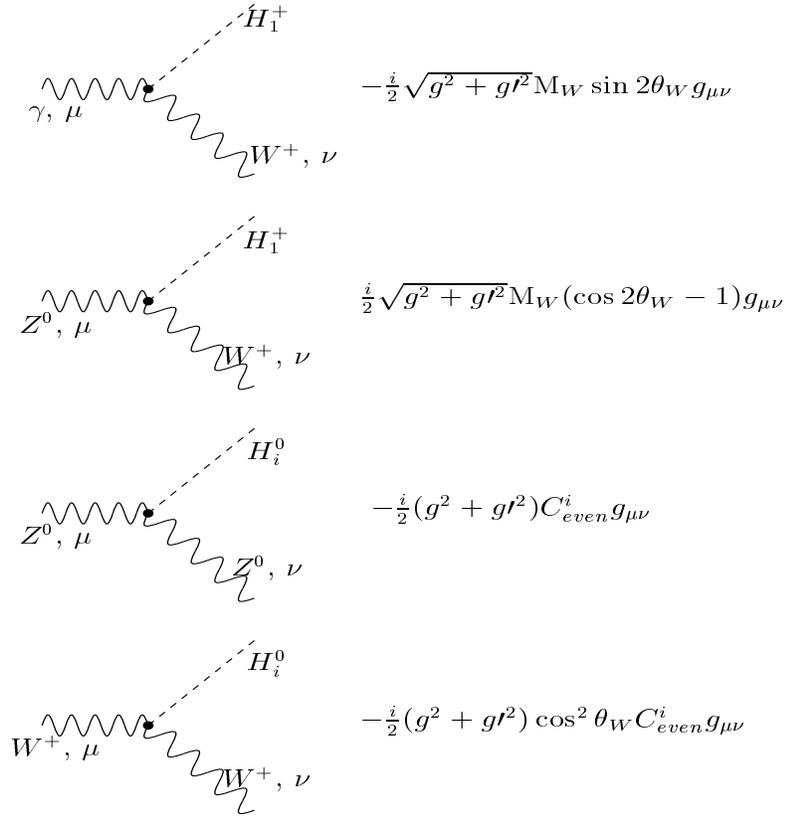}}
\end{picture}
\caption[]{Feynman rules for SVV vertices}
\label{fig2}
\end{figure}

\begin{figure}
\setlength{\unitlength}{1mm}
\begin{picture}(140,180)(30,30)
\put(-10,-10){\includegraphics{fig.3}}
\end{picture}
\caption[]{Feynman rules for SSVV vertices. Part(I)}
\label{fig3}
\end{figure}

\begin{figure}
\setlength{\unitlength}{1mm}
\begin{picture}(140,180)(30,30)
\put(-10,-10){\includegraphics{fig.4}}
\end{picture}
\caption[]{Feynman rules for SSVV vertices. Part(II)}
\label{fig4}
\end{figure}

\begin{figure}
\setlength{\unitlength}{1mm}
\begin{picture}(140,180)(30,30)
\put(-10,-10){\includegraphics{fig.5}}
\end{picture}
\caption[]{Feynman rules for the self-coupling of Higgs. Part(I)}
\label{fig5}
\end{figure}

\begin{figure}
\setlength{\unitlength}{1mm}
\begin{picture}(140,180)(30,30)
\put(-10,-10){\includegraphics{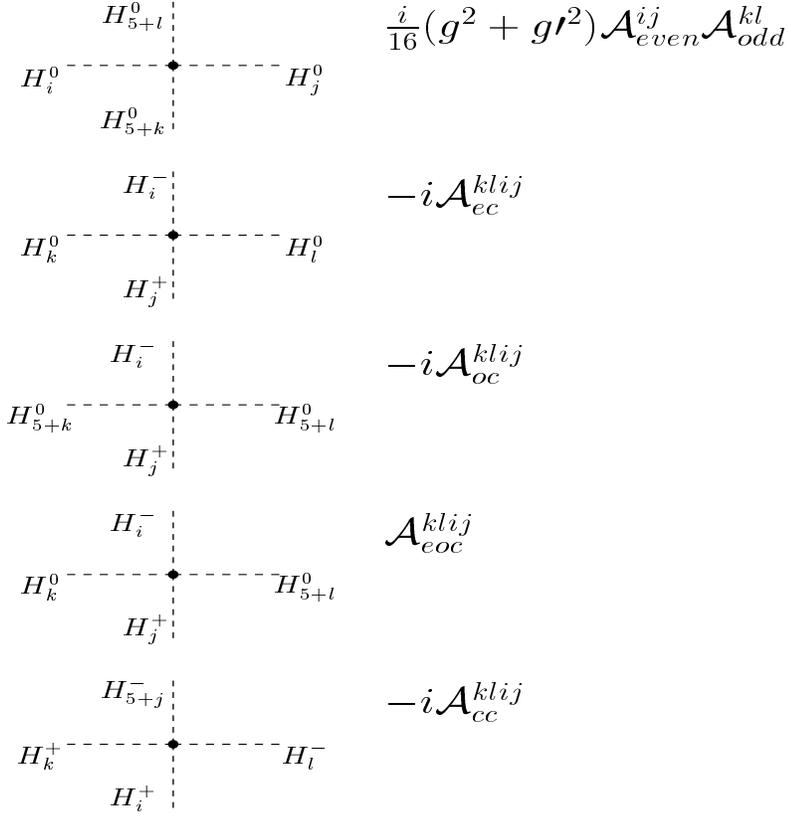}}
\end{picture}
\caption[]{Feynman rules for the self-coupling of Higgs. Part(II)}
\label{fig6}
\end{figure}

\begin{figure}
\setlength{\unitlength}{1mm}
\begin{picture}(140,180)(30,30)
\put(-10,-10){\includegraphics{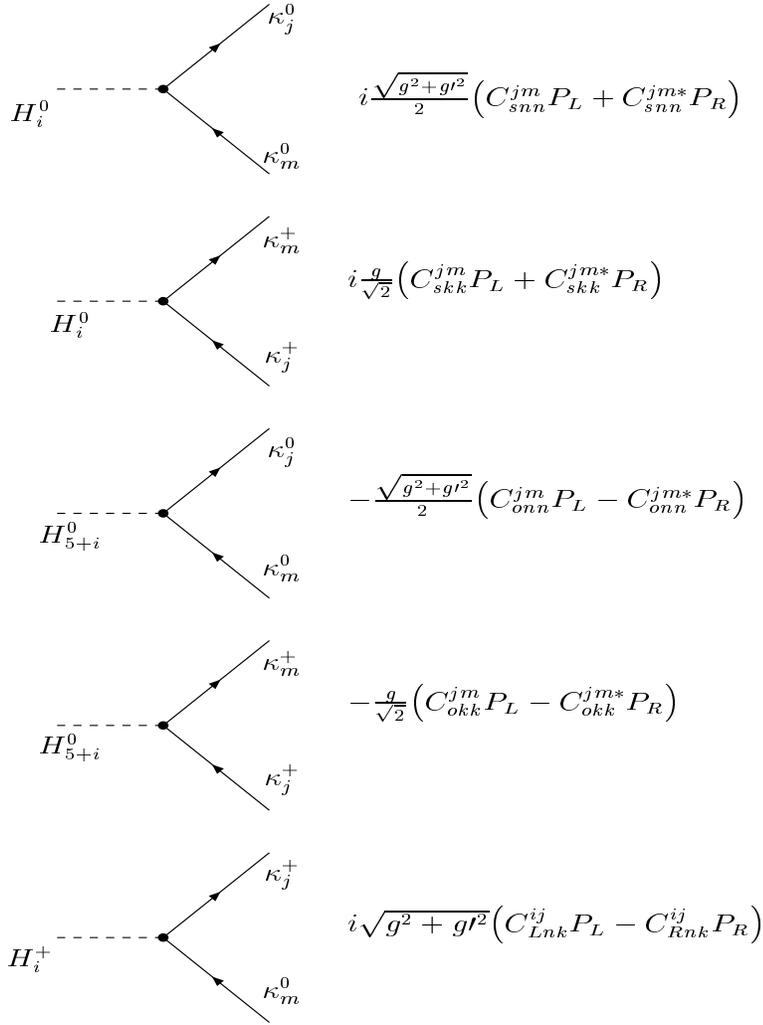}}
\end{picture}
\caption[]{Feynman rules for the coupling of Higgs with 
charginos or neutralinos.}
\label{fig7}
\end{figure}

\begin{figure}
\setlength{\unitlength}{1mm}
\begin{picture}(140,180)(30,30)
\put(-10,-10){\includegraphics{fig.8}}
\end{picture}
\caption[]{Feynman rules for the coupling of gauge bosons with 
charginos or neutralinos.}
\label{fig8}
\end{figure}

\begin{figure}
\setlength{\unitlength}{1mm}
\begin{picture}(140,180)(30,30)
\put(-10,-10){\includegraphics{fig.9}}
\end{picture}
\caption[]{Feynman rules for the coupling of quarks, squarks with charginos or 
neutralinos.}
\label{fig9}
\end{figure}

\begin{figure}
\setlength{\unitlength}{1mm}
\begin{picture}(140,200)(30,30)
\put(-10,0){\includegraphics{fig.10}}
\end{picture}
\caption[]{The mass of the lightest chargino versus Y. 
The parameters are assigned as $m_{1}=m_{2}=250$GeV, $m_{\nu_{\tau}}
= 0.1$MeV
and (a)$\tan\beta = 20$, $\tan\theta_{\upsilon} = 5$; (b)$\tan\beta = 2$, 
$\tan\theta_{\upsilon} = 5$; 
(c)$\tan\beta = 20$, $\tan\theta_{\upsilon} = 0.5$; (d)$\tan\beta = 2$, 
$\tan\theta_{\upsilon} = 0.5$. The dot lines
correspond to $\lambda_{333}=0.5$, the solid lines correspond to 
$\lambda_{333}=0.$}
\label{fig10}
\end{figure}

\begin{figure}
\setlength{\unitlength}{1mm}
\begin{picture}(140,200)(30,30)
\put(-10,0){\includegraphics{fig.11}}
\end{picture}
\caption[]{The mass of the lightest chargino versus Y. 
The parameters are assigned as $m_{1}=m_{2}=250$GeV, $m_{\nu_{\tau}}
= 0.01$MeV
and (a)$\tan\beta = 20$, $\tan\theta_{\upsilon} = 5$; (b)$\tan\beta = 2$, 
$\tan\theta_{\upsilon} = 5$; 
(c)$\tan\beta = 20$, $\tan\theta_{\upsilon} = 0.5$; (d)$\tan\beta = 2$, 
$\tan\theta_{\upsilon} = 0.5$. The dot lines
correspond to $\lambda_{333}=0.5$, the solid lines correspond to 
$\lambda_{333}=0.$}
\label{fig11}
\end{figure}

\begin{figure}
\setlength{\unitlength}{1mm}
\begin{picture}(140,200)(30,30)
\put(-10,0){\includegraphics{fig.12}}
\end{picture}
\caption[]{The mass of the lightest CP-even Higgs versus $\sqrt{|Y_{s}|}$. The parameters are 
assigned as $\sqrt{|X_{s}|}=500$GeV 
and (a)$\tan\beta = 20$, $\tan\theta_{\upsilon} = 5$; (b)$\tan\beta = 2$, 
$\tan\theta_{\upsilon} = 5$; 
(c)$\tan\beta = 20$, $\tan\theta_{\upsilon} = 0.5$; (d)$\tan\beta = 2$, 
$\tan\theta_{\upsilon} = 0.5$. The dot-dash lines
correspond to $\sqrt{|Z_{s}|}=250$GeV, the dash lines correspond to 
$\sqrt{|Z_{s}|}=150$GeV, and dot lines correspond
to $\sqrt{|Z_{s}|}=60$GeV}
\label{fig12}
\end{figure}

\begin{figure}
\setlength{\unitlength}{1mm}
\begin{picture}(140,200)(30,30)
\put(-10,0){\includegraphics{fig.13}}
\end{picture}
\caption[]{The mass of the lightest CP-even Higgs versus $\sqrt{|Z_{s}|}$. The parameters are 
assigned as $\sqrt{|X_{s}|}=500$GeV 
and (a)$\tan\beta = 20$, $\tan\theta_{\upsilon} = 5$; (b)$\tan\beta = 2$, 
$\tan\theta_{\upsilon} = 5$; 
(c)$\tan\beta = 20$, $\tan\theta_{\upsilon} = 0.5$; (d)$\tan\beta = 2$, 
$\tan\theta_{\upsilon} = 0.5$. The dot-dash lines
correspond to $\sqrt{|Y_{s}|}=400$GeV, the dash lines correspond to 
$\sqrt{|Y_{s}|}=300$GeV, and dot lines correspond
to $\sqrt{|Y_{s}|}=60$GeV}
\label{fig13}
\end{figure}

\end{document}